\documentclass[12pt]{article}
\usepackage[utf8]{inputenc}
\usepackage{geometry}
\usepackage{amsmath}
\usepackage{graphicx}
\usepackage[colorlinks=true, allcolors=blue]{hyperref}
\usepackage{hyperref}
\usepackage[title]{appendix}
\usepackage{mathrsfs}
\usepackage{amsfonts}
\usepackage{multirow} 
\usepackage{booktabs} 
\usepackage{caption}  
\usepackage{threeparttable} 
\usepackage{algorithm}
\usepackage{algorithmicx}
\usepackage{adjustbox} 
\usepackage{algpseudocode}
\usepackage{listings}
\usepackage{xcolor}
\usepackage{enumitem}
\usepackage{chngcntr}
\usepackage{booktabs}
\usepackage{lipsum}
\usepackage{subcaption}
\usepackage{authblk}
\usepackage[T1]{fontenc}    
\usepackage{csquotes}       
\usepackage{diagbox}
\usepackage{cite}
\usepackage{longtable, booktabs, multirow, adjustbox} 
\newcommand{\bI}{{\bf I}}
\newcommand\bc{\boldsymbol c}

\newcommand\bv{\boldsymbol v}
\newcommand\bV{\boldsymbol V}

\newcommand\bW{\boldsymbol W}

\newcommand\bT{\boldsymbol T}

\newcommand\bPhi{\boldsymbol{\Phi}}
\newcommand\bSigma{\boldsymbol{\Sigma}}

\newcommand\bR{\boldsymbol{R}}

\numberwithin{equation}{section}

\newcommand{\beqn}{\begin{equation}}
\newcommand{\eeqn}{\end{equation}}
\newcommand{\beqnarr}{\begin{eqnarray}}
\newcommand{\eeqnarr}{\end{eqnarray}}
\newcommand{\baling}{\begin{alignat}{1}}
\newcommand{\ealing}{\end{alignat}}

\definecolor{blackcolour}{rgb}{0, 0, 0}       
\definecolor{codegreen}{rgb}{0, 0.6, 0}       
\definecolor{magenta}{rgb}{1.0, 0, 1.0}       
\definecolor{codegray}{rgb}{0.5, 0.5, 0.5}    
\definecolor{codepurple}{rgb}{0.5, 0, 0.5}    
\definecolor{backcolour}{rgb}{0.95, 0.95, 0.92} 

\lstdefinestyle{mystyle}{
    backgroundcolor=\color{backcolour},   
    commentstyle=\color{codegreen},
    keywordstyle=\color{magenta},
    numberstyle=\tiny\color{codegray},
    stringstyle=\color{codepurple},
    basicstyle=\ttfamily\footnotesize,
    breakatwhitespace=false,         
    breaklines=true,                 
    captionpos=b,                    
    keepspaces=true,                 
    numbers=left,                    
    numbersep=5pt,                  
    showspaces=false,                
    showstringspaces=false,
    showtabs=false,                  
    tabsize=2
}

\title{Ensemble Kalman Filter for Data Assimilation coupled with low-resolution computations techniques applied in Fluid Dynamics}
\author[a,b]{Paul Jeanney: \href{mailto:paul.jeanney@arup.com}{\texttt{paul.jeanney@arup.com}}}
\author[a]{Ashton Hetherington: \href{mailto:ashton.ian@upm.es}{\texttt{ashton.ian@upm.es}}}
\author[c]{Shady E. Ahmed: \href{mailto:shady.ahmed@okstate.edu}{\texttt{shady.ahmed@okstate.edu}}}
\author[b]{David Lanceta: \href{mailto:david.lanceta@arup.com}{\texttt{david.lanceta@arup.com}}}
\author[b]{Susana Saiz: \href{mailto:susana.saiz@arup.com}{\texttt{susana.saiz@arup.com}}}
\author[a]{José Miguel Perez: \href{mailto:josemiguel.perez@upm.es}{\texttt{josemiguel.perez@upm.es}}}
\author[a]{Soledad Le Clainche: \href{mailto:soledad.leclainche@upm.es}{\texttt{soledad.leclainche@upm.es}}}

\affil[a]{ETSI Aeronáutica y del Espacio, Universidad Politécnica de Madrid, Plaza Cardenal Cisneros, 3, 28040 Madrid, Spain.}
\affil[b]{Arup, C. de Alfonso XI, 12, Retiro, 28014 Madrid, Spain.}
\affil[c]{Advanced Computing, Mathematics, and Data Division, Pacific Northwest National Laboratory, Richland, WA 99354, USA.}

\date{\today}

\begin{document}

\maketitle

\begin{abstract}
  This paper presents an innovative Reduced-Order Model (ROM) for merging experimental and simulation data using Data Assimilation (DA) to estimate the "True" state of a fluid dynamics system, leading to more accurate predictions. Our methodology introduces a novel approach by implementing the Ensemble Kalman Filter (EnKF) within a reduced-dimensional framework, grounded in a robust theoretical foundation and applied to fluid dynamics. To address the substantial computational demands of DA, the proposed ROM employs low-resolution (LR) techniques to drastically reduce computational costs. This innovative approach involves downsampling datasets for DA computations, followed by an advanced reconstruction technique based on low-cost Singular Value Decomposition (lcSVD). The lcSVD method, a key innovation in this paper, has never been applied to DA before and offers a highly efficient way to enhance resolution with minimal computational resources. Our results demonstrate significant reductions in both computation time and RAM usage through these LR techniques without compromising the accuracy of the estimations. For instance, in a turbulent test case, for a data compression rate ($C_{R,u_b}$) of 15.9, the LR approach can achieve a speed-up of 13.7 and a RAM compression of 90.9\% while maintaining a low Relative Root Mean Square Error ($RRMSE$) of 2.6\%, compared to 0.8\% in the high-resolution (HR) reference. Furthermore, we highlight the effectiveness of the EnKF in estimating and predicting the state of fluid flow systems based on limited observations and given low-fidelity numerical data. This paper highlights the potential of the proposed DA method in fluid dynamics applications, particularly for improving computational efficiency in CFD and related fields. Its ability to balance accuracy with low computational and memory costs makes it especially suitable for large-scale and real-time applications, such as environmental monitoring or engineering design. This method will be incorporated into ModelFLOWs-app\footnote{The website of the software is available at \url{https://modelflows.github.io/modelflowsapp/}.}.

\end{abstract}
\vspace{0.3cm} 

\hrule\par
\vspace{0.3cm} 

\section{Introduction}

\small
Computational Fluid Dynamics (CFD) is an essential tool in fluid mechanics, using numerical analysis and algorithms to solve and analyze fluid flows. It has broad applications across various industries, including aerospace, automotive, environmental engineering and many more. A major challenge in CFD simulations lies in the accurate representation of boundary and initial conditions, which are essential for producing reliable results. These conditions are often derived from experimental data, which may be difficult to obtain with the required precision, spatial, and temporal resolutions \cite{boundariesCFD}. Inaccurate data can cause errors in simulation outputs, particularly when modeling complex systems. Obtaining experimental data that accurately captures the dynamics of fluid flows is often time-consuming and costly, and may not fully reflect real-world scenarios \cite{expCFD}.

This challenge is common across various fields, including aerodynamics, combustion modeling, and renewable energy systems, where the accuracy of simulations is strongly tied to how well initial and boundary conditions are defined. In aerodynamics, for example, uncertainties in the boundary layer can significantly impact predictions of drag and lift in aircraft design \cite{boundaryAircraft}. In combustion processes, inaccurate initial conditions related to fuel injection and flame dynamics lead to inaccurate predictions of heat release rates and pollutant formation \cite{boundaryComb}. In wind energy, the accuracy of inflow conditions plays a critical role in determining turbine performance \cite{boundariesWind}. Additionally, in urban environments, accurate boundary and initial conditions are crucial for simulating airflow and pollutant dispersion, given the complex interactions between buildings and air currents \cite{boundariesUrban1, boundariesUrban2, boundariesUrban3}. To address these limitations, Data Assimilation (DA) techniques have been developed which can be particularly effective in enhancing the resolution in some areas of CFD simulations, improving the accuracy of boundary conditions for high-fidelity simulations of turbulent flows \cite{DA5} or for large-scale systems with direct measurements \cite{DAlargeScale}.

DA integrates experimental observations with numerical models to accurately estimate the system's "True" state. By combining real-time measurement data with simulations, DA enhances the predictive capabilities of CFD models \cite{DA_CFD_LES}. Among the various DA methods, Ensemble Kalman Filter (EnKF) \cite{DA} is particularly effective. The EnKF uses an ensemble of simulations to represent the uncertainty in the system and iteratively updates these simulations with new observational data, providing a robust framework for merging data from different sources. This approach is particularly effective in fields such as fluid dynamics and CFD, where it is crucial to model complex, uncertain systems. For instance, Nino Ruiz et al. \cite{DA1} applied the EnKF in inverse covariance matrix estimation, demonstrating how the ensemble of estimations captures the uncertainties inherent in the system and how new observational data can refine the simulation outputs. Similarly, Mons et al. \cite{DA2} explored its use in reconstructing unsteady viscous flows in CFD, emphasizing the EnKF's ability to improve predictions by merging simulation data with real-time observations. Additionally, Colburn et al. \cite{DA3} showed its effectiveness in wall-bounded fluid systems by iteratively updating simulations based on observations to reduce uncertainty. Villanueva et al. \cite{DA4} further illustrated this by applying the EnKF for state estimation in urban settings, highlighting how sequential data assimilation in real-time can enhance the accuracy of complex system modeling. Recent studies \cite{DA5, DA6} have also explored advanced applications of DA in turbulent flow simulations, demonstrating the method's versatility and robustness across different scales and physical domains.

Despite the advantages of DA, particularly the EnKF, its application with high-resolution (HR) datasets meets significant computational challenges. High-resolution simulations, necessary for capturing detailed interactions in industrial problems, require substantial computational resources \cite{CFDcost, CFDcost2}. Performing DA on such HR data further exacerbates these demands, making real-time or near-real-time applications impractical \cite{DAcost}. To mitigate these challenges, our research introduces innovative Reduced-Order Model (ROM) based on low-resolution (LR) techniques designed to optimize computational efficiency without compromising the accuracy of DA.

Our methodology involves a three-step process: (1) downsampling the dataset to reduce its size, (2) performing DA computations on this reduced dataset (downsampled), and (3) employing low-cost reconstruction methods to restore the dataset to its original or required resolution. Specifically, for the first time, to the author's knowledge, we implement Super-Resolution using low cost Singular Value Decomposition (lcSVD) \cite{ashton2020low}, which is an extension of Singular Value Decomposition (SVD) \cite{Sirovich87} capable to reconstruct three-dimensional databases fields from sensors for the reconstruction phase. This process ensures that only the most essential data are retained, significantly reducing the amount of data that needs to be processed during the DA computations. The advantage of using SVD lies in the capabilities of this method to extract flow patterns and remove spatial redundancies or filter out noise. This makes it a suitable tool to enhance data resolution in an automatic and effective way \cite{MODELFLOW}. This method is particularly useful in applications where the data is inherently noisy or where the resolution of the measurements is limited.

Our research demonstrates the effectiveness of these LR techniques in reducing computational needs, including both computation time and RAM usage. We validate our approach through a series of benchmark test cases shown in section \ref{testcases}, showcasing the potential of our methodology to deliver accurate predictions. These cases studied highlight the ability of the EnKF to effectively integrate observational data with numerical models, resulting in improved state estimations and predictions across a range of physical systems. The EnKF is well-suited for handling the non-linear dynamics and uncertainties inherent in fluid dynamics applications. By incorporating LR techniques, we make the EnKF more practical for large-scale and real-time applications.

In summary, our work highlights the potential of combining advanced DA techniques with computational optimizations to address the complex and resource-intensive nature of CFD. This integrated approach not only improves the accuracy of simulations but also makes them more accessible and practical for real-time decision-making in various applications. The implications of this research extend to numerous domains where accurate fluid state estimation and efficient data processing are critical, emphasizing the need for ongoing advancements in CFD and DA methodologies.

This article is organized as follows: Section \ref{ENKF} details the implementation of EnKF, highlighting its theoretical background and practical application for improving the accuracy of simulations, then section \ref{svd} introduces the LR techniques designed to optimize computational efficiency, including downsampling methods and low-cost reconstruction technique via lcSVD, with the methodology described subsequently in section \ref{methodo}. Section \ref{testcases} describes the various test cases used to validate the robustness, effectiveness, and precision of the developed methods. Section \ref{results} presents the results of our methodology, showcasing the significant reductions in computational time and memory usage achieved by the LR techniques, alongside the effectiveness of the EnKF in improving state estimations. Finally, the paper concludes in section \ref{conclusion} by summarizing the key findings, discussing the implications for industrial fluid dynamics applications, and presenting the  directions for future research.

\section{Ensemble Kalman Filter}\label{ENKF}

DA integrates observational data into computational models to improve the accuracy of predictions. The EnKF is a widely-used DA method, particularly effective for high-dimensional, nonlinear systems like those found in CFD. This section details the theoretical background and mathematical formulations of the EnKF as applied to CFD. The formulation closely follows the generic EnKF strategy outlined in Ahmed et al. \cite{shady}, applied here on high-dimensional CFD systems.

Traditionally, DA techniques have been tailored to numerical weather predictions, and recently they have been adopted in many other application areas. Interested readers are referred to the monograph by Lewis et al. \cite{lewis} for a wide coverage of DA algorithms while accessible Python implementations can be found in Ahmed et al. \cite{shady}. EnKF is an extension to the classical Kalman filtering approach, tailored for nonlinear dynamics and high-dimensional systems. EnKF maintains an ensemble of possible system states and updates them with observational data to improve accuracy. The method is particularly useful for high-dimensional, nonlinear systems like those found in CFD. It is a widely used algorithm that leverages the collective information from multiple computations to approximate the uncertainty in both the forecasted (background) and analyzed states.

Unlike traditional Kalman filters, which require the direct computation and propagation of the covariance matrices, the EnKF employs a statistical approach. The method uses a Monte Carlo approach to sample the state of a system \cite{evensen, burgers, evensen2}. It maintains an ensemble of possible states, which are evolved according to the model dynamics. This ensemble is then updated with observational data to correct the model predictions, reducing uncertainties in the forecast. 
Specifically, it uses an ensemble of possible state vectors to represent the uncertainty in the state estimate. 

\subsection{Analysis Covariance Matrix}

The background covariance matrix, denoted by \(\mathbf{B}\), plays a crucial role in quantifying the uncertainty associated with the predicted state, often referred to as the background state. This uncertainty typically arises from errors in the initial conditions or imperfections in the model. In EnKF, instead of computing \(\mathbf{B}\) directly, it is approximated using the ensemble of background state vectors, \(\mathbf{u}_b^{(i)}\), where \(E\) represents the number of ensemble members:

\begin{equation}\label{EnKF1}
\mathbf{B} \approx \frac{1}{E-1} \sum_{i=1}^E \left(\mathbf{u}_b^{(i)} - \overline{\mathbf{u}_{\mathbf{b}}}\right)\left(\mathbf{u}_b^{(i)} - \overline{\mathbf{u}_{\mathbf{b}}}\right)\top,
\end{equation}

Here, \(\mathbf{u}_b^{(i)}\) represents the \(i\)-th member of the ensemble, which is one possible realization of the state at a given time based on the model and initial conditions. $()^\top$ represents the transpose operation. The term \(\overline{\mathbf{u}_{\mathbf{b}}}\) denotes the ensemble mean, calculated as:

\[
\overline{\mathbf{u}_{\mathbf{b}}} = \frac{1}{E} \sum_{i=1}^E \mathbf{u}_b^{(i)}.
\]

The ensemble mean serves as the best estimate of the background state, while the background covariance matrix \(\mathbf{B}\) measures the spread of the ensemble members around this mean, thereby providing a quantitative measure of the uncertainty. The larger the spread, the greater the uncertainty in the background state. This matrix is instrumental in determining how much weight the model’s predictions should have relative to new observations during the analysis step. Essentially, it dictates our trust in our model’s predictions, with a larger \(\mathbf{B}\) indicating less confidence in the background state.

Similarly, the analysis covariance matrix, denoted by \(\mathbf{P}\), represents the uncertainty in the state after the assimilation of observational data. Following the same statistical principles, the EnKF approximates \(\mathbf{P}\) using the ensemble of analyzed state vectors, \(\mathbf{u}_a^{(i)}\):

\begin{equation}\label{EnKF2}
\mathbf{P} \approx \frac{1}{E-1} \sum_{i=1}^E \left(\mathbf{u}_a^{(i)} - \overline{\mathbf{u}_{\mathbf{a}}}\right)\left(\mathbf{u}_a^{(i)} - \overline{\mathbf{u}_{\mathbf{a}}}\right)\top.
\end{equation}

In this equation, \(\mathbf{u}_a^{(i)}\) represents the \(i\)-th member of the ensemble after the analysis step, which is the updated state vector that has incorporated observational data. The ensemble mean for the analysis step, \(\overline{\mathbf{u}_{\mathbf{a}}}\), is defined as:

\[
\overline{\mathbf{u}_{\mathbf{a}}} = \frac{1}{E} \sum_{i=1}^E \mathbf{u}_a^{(i)}.
\]

Just as with the background covariance matrix, \(\mathbf{P}\) measures the spread of the ensemble members around the mean of the analyzed state. A smaller analysis covariance matrix suggests that the observations have significantly reduced the uncertainty in the state estimate. Thus, \(\mathbf{P}\) is crucial for understanding the effectiveness of the data assimilation process—how well the observational data have constrained the state estimate. In practice, \(\mathbf{P}\) is used in subsequent forecast steps, continuing the cycle of prediction and correction.

\subsection{Initializing the Ensemble}

The EnKF process begins by generating an initial ensemble of state vectors at a given time, denoted by \(\left\{\widehat{\mathbf{u}}_b^{(i)}(t_k) \mid 1 \leq i \leq E\right\}\). Each ensemble member is drawn from a distribution centered around the best-known estimate of the state at time \(t_k\), which we denote as \(\widehat{\mathbf{u}}_b(t_k)\). Considering the uncertainty, this initial ensemble reflects our current knowledge of the system, with each member representing a possible state the system could be in. The distribution from which the ensemble members are sampled is typically Gaussian, with a mean of \(\widehat{\mathbf{u}}_b(t_k)\) and a covariance matrix \(\widehat{\mathbf{B}}_k\):

\[
\widehat{\mathbf{u}}_b^{(i)}(t_k) \sim \mathcal{N}\left(\widehat{\mathbf{u}}_b(t_k), \widehat{\mathbf{B}}_k\right).
\]

As the ensemble size \(E\) increases, the ensemble mean and covariance matrix \(\widehat{\mathbf{B}}_k\) better approximate the True state and its associated uncertainty. This initialization step is crucial because the accuracy of the EnKF relies heavily on the representativeness of the ensemble. If the ensemble is too small or poorly chosen, the approximations made in subsequent steps may not accurately reflect the True state of the system.

\subsection{Forecast Step}

Once the initial ensemble is established, the EnKF algorithm advances each ensemble member forward in time using the model \(M\). This step is known as the forecast step, and it predicts the state of the system at a future time \(t_{k+1}\). The forecast for each ensemble member is computed as follows:

\[
\mathbf{u}_b^{(i)}(t_{k+1}) = M\left(\widehat{\mathbf{u}}_b^{(i)}(t_k); \boldsymbol{\theta}\right) + \boldsymbol{\zeta}_p^{(i)}(t_{k+1}),
\]

In this equation, \(M(\cdot)\) is the model operator that evolves the state from time \(t_k\) to \(t_{k+1}\), considering the model parameters \(\boldsymbol{\theta}\). The term \(\boldsymbol{\zeta}_p^{(i)}(t_{k+1})\) represents the process noise, which is drawn from a Gaussian distribution with zero mean and a covariance matrix \(\mathbf{Q}_{k+1}\). This process noise accounts for the uncertainties and errors inherent in the model itself, such as approximations, discretization errors, or unknown external influences.

After all ensemble members have been propagated to time \(t_{k+1}\), the ensemble mean and the corresponding background covariance matrix at this new time are computed. The ensemble mean serves as the best estimate of the state at \(t_{k+1}\), and it is calculated as:

\[
\mathbf{u}_b(t_{k+1}) \approx \overline{\mathbf{u}_{\mathbf{b}}}(t_{k+1}) = \frac{1}{E} \sum_{i=1}^E \mathbf{u}_b^{(i)}(t_{k+1}).
\]

The background covariance matrix at time \(t_{k+1}\), which represents the uncertainty in the forecasted state, is given by the equation (\ref{EnKF1}).

This step ensures that the forecasted state and its uncertainty are properly accounted for, setting the stage for the next step, where observational data will be assimilated.

\subsection{Observation Perturbation and Kalman Gain Calculation}

To incorporate observational data into the state estimate, the EnKF generates an ensemble of perturbed observations. These perturbed observations, denoted by \(\{\mathbf{w}^{(i)}(t_{k+1}) \mid 1 \leq i \leq E\}\), are created by adding random Gaussian noise with zero mean and a covariance matrix \(\mathbf{R}_{k+1}\) to the actual observations \(\mathbf{w}(t_{k+1})\). The observation \(\mathbf{w}(t)\) is related to the True state of the system $\mathbf{u(t)}$ using the observation operator \(h(\cdot)\) as follows:

\[
  \mathbf{w}(t_{k+1}) = h(uTrue(t_{k+1})) + \text{noise}.
\]
This noise reflects the uncertainties in the measurement process and allows the EnKF to account for the fact that observations are not perfect:

\[
\mathbf{w}^{(i)}(t_{k+1}) \sim \mathcal{N}\left(\mathbf{w}(t_{k+1}), \mathbf{R}_{k+1}\right).
\]

Once the ensemble of perturbed observations is created, the Kalman gain matrix \(\mathbf{K}_{k+1}\) is computed. The Kalman gain plays a critical role in determining how much weight should be given to the observations versus the model predictions. It is calculated using the background covariance matrix \(\mathbf{B}_{k+1}\), and the observation covariance matrix \(\mathbf{R}_{k+1}\):

\[
\mathbf{K}_{k+1} = \mathbf{B}_{k+1} \mathbf{D}_h\left(\mathbf{u}_b(t_{k+1})\right)^\top\left(\mathbf{D}_h\left(\mathbf{u}_b(t_{k+1})\right) \mathbf{B}_{k+1} \mathbf{D}_h\left(\mathbf{u}_b(t_{k+1})\right)\top + \mathbf{R}_{k+1}\right)^{-1}.
\]

The matrix \(\mathbf{D}_h(\mathbf{u}_b(t_{k+1}))\) is the Jacobian of the observation operator \(h(\cdot)\) with respect to the state, evaluated at the background state \(\mathbf{u}_b(t_{k+1})\). In practice, the observation operator \(h(\cdot)\) is often nonlinear, and the Jacobian matrix \(\mathbf{D}_h(\mathbf{u}_b(t_{k+1}))\) is obtained by linearizing \(h(\cdot)\) around the background state \(\mathbf{u}_b(t_{k+1})\). This linearization can be done either analytically, if \(h(\cdot)\) has an explicit form, or numerically using finite difference approximations, where small perturbations are applied to the state \(\mathbf{u}_b(t_{k+1})\), and the resulting changes in \(h(\mathbf{u}_b(t_{k+1}))\) are used to approximate the Jacobian.

The Kalman gain essentially determines how the forecast should be adjusted to match the observations. A higher Kalman gain indicates that the observations should have a larger influence on the state update, reflecting either high confidence in the observations or high uncertainty in the model forecast.

\subsection{Analysis Step}

In the analysis step, each member of the forecast ensemble is updated by incorporating the perturbed observations using the Kalman gain matrix. The update for each ensemble member is given by:

\[
\mathbf{u}_a^{(i)} = \mathbf{u}_b^{(i)}(t_{k+1}) + \mathbf{K}_{k+1}\left(\mathbf{w}^{(i)}(t_{k+1}) - h\left(\mathbf{u}_b^{(i)}(t_{k+1})\right)\right).
\]

This equation represents the core of the EnKF. It updates the predicted state \(\mathbf{u}_b^{(i)}(t_{k+1})\) by adding a correction term based on the difference between the perturbed observation \(\mathbf{w}^{(i)}(t_{k+1})\) and the predicted observation \(h(\mathbf{u}_b^{(i)}(t_{k+1}))\). The Kalman gain \(\mathbf{K}_{k+1}\) determines the magnitude of this correction. After applying the analysis step to all ensemble members, the mean of the updated ensemble represents the best estimate of the state after incorporating the new observations, while the analysis covariance matrix \(\mathbf{P}_{k+1}\) quantifies the uncertainty in this updated estimate.

The ensemble mean gives the final state estimate at time \(t_{k+1}\):

\[
\mathbf{u}_a(t_{k+1}) \approx \overline{\mathbf{u}_{\mathbf{a}}}(t_{k+1}) = \frac{1}{E} \sum_{i=1}^E \mathbf{u}_a^{(i)}.
\]

Finally, the analysis covariance matrix is recalculated to reflect the updated uncertainty in the state estimate after the assimilation of new observations. The updated covariance matrix, derived from the analyzed ensemble, captures the revised relationship among the ensemble members and their deviation from the new ensemble mean. This is expressed as shown in equation (\ref{EnKF2}).

\section{Low-Resolution techniques}\label{svd}
\small

The main innovation of our DA applied to fluid dynamics resides in the use of LR techniques to alleviate the computational costs of DA when applied to large scale problems. We use LR techniques to bypass the computational bottleneck of the DA workflow. These techniques involve two primary steps: downsampling the dataset to reduce its size and then reconstructing the original data using lcSVD approach, introduced by Hetherington et al. \cite{ashton2020low}. Both methods are presented in section \ref{DS} and \ref{reconstruct} respectively, and the reconstruction performance with different parameters is presented in section \ref{results}.

\subsection{Data organisation}\label{DS}

Downsampling is the process of reducing the dimensionality of the dataset by selecting a subset of data points. This step is crucial for making large datasets more manageable, as it significantly reduces the computational resources required for further processing.

This downsampling organisation method functions as follows: 
a group of K time-varying samples, known as snapshots, are organized in matrix form as

\begin{equation}\label{eq:SnapMatrix}
  \boldsymbol{V}_1^K = \left[
    \boldsymbol{v}_1, \boldsymbol{v}_2, \ldots, \boldsymbol{v}_k, 
    \boldsymbol{v}_{k+1}, \ldots, \boldsymbol{v}_{K-1}, \boldsymbol{v}_K
  \right],
\end{equation}

where $\boldsymbol{V}_1{ }^K$ is known as the snapshot matrix and each one of the $K$ temporal samples $\boldsymbol{v}_k$ is known as snapshot. Each snapshot is formed by a vector with dimension $J= N_{\text {comp }} \times N_x \times N_y$ for two-dimensional problems and $J=N_{\text {comp }} \times$ $N_x \times N_y \times N_z$ for three-dimensional cases, where $N_{\text {comp }}$ correspond to the number of components of the dataset (i.e., in a dataset formed by the pressure vector and the streamwise and normal velocity components $N_{\text {comp }}=3$ ), and $N_x, N_y$ and $N_z$ correspond to the spatial grid points distributed along the streamwise, normal and spanwise directions, respectively. When several components form the dataset, each one of these components is concatenated into columns.

It is possible to generate a modified reduced snapshot matrix $\overline{\boldsymbol{V}}_1^K$ with dimension $\bar{J} \times \bar{K}$, where $\bar{J}<J$ and $\bar{K}<K$. This dimension reduction can be carried out in several ways. For instance, we adopt a uniform downsampling of the dataset, where 1 every $p$ points is retained. Another alternative is to use non-uniform downsampling, based on sensor or sparse measurements along the whole spatial domain.

\begin{equation}
  \bar{\bV}_1^{K} = [\bar{\bv}_{1}, \bar{\bv}_{2}, \dots, \bar{\bv}_{k}, \bar{\bv}_{k+1}, \dots, \bar{\bv}_{K-1}, \bar{\bv}_{K}],
  \label{eq:SnapMatrixRed}
\end{equation}
where $\overline{\boldsymbol{v}}_k$ represents a modified reduced snapshot, $\overline{\boldsymbol{v}}_k \in \mathbb{R}^{\bar{J}}$.

\subsection{Data Reconstruction} \label{reconstruct}

Once the input data has been downsampled, key DA computations can be performed efficiently on the reduced data. After that, a reconstruction to the original dimensionality is often desired. This reconstruction is critical for restoring the data to its original resolution, ensuring that the key features of the dataset are preserved while maintaining computational efficiency. The method used for data reconstruction in our methodology is lcSVD.

\subsubsection{Singular Value Decomposition (SVD) and Proper Orthogonal Decomposition (POD)\label{sec:SVD}}

Singular Value Decomposition (SVD) \cite{Sirovich87} is a mathematical technique widely used for low-rank data approximations, effectively eliminating redundancies and filtering out noise or artifacts. This technique also facilitates the extraction of Proper Orthogonal Decomposition (POD) modes. POD, introduced by Lumley \cite{Lumley}, is a method employed to capture coherent structures in turbulent flows. Although SVD and POD are often considered equivalent in the literature, SVD represents one of the primary approaches for calculating POD modes \cite{LeClaincheetalJFM20}.

SVD allows the decomposition of spatio-temporal data $\bv(x,y,z,t)$ into orthogonal spatial modes, commonly referred to as SVD or POD modes, coupled with corresponding temporal coefficients, as expressed by:
\begin{equation}
\bv(x,y,z,t)\simeq \sum_{n=1}^N \bc_n(t)\bPhi_n(x,y,z),\label{eq:SnapMatrix0pod}
\end{equation}
where $\bPhi_n(x,y,z)$ denotes the spatial modes, and $\bc_n(t)$ represents the temporal coefficients.

The snapshot matrix $\bV_1^k$, given by eq. (\ref{eq:SnapMatrix}), is factorized using the SVD algorithm as:
\begin{equation}
\bV_1^{K}\simeq\bW\,\bSigma\,\bT^\top,\label{eq:svd}
\end{equation}
where $()^\top$ denotes the matrix transpose, $\bW$ and $\bT$ are the matrices containing in columns the spatial SVD (or POD) modes and corresponding temporal coefficients (the columns of these matrices are orthonormal), and $\bSigma$ is a diagonal matrix containing the  singular values $\sigma_1,\cdots,\sigma_{N}$, with $N\leq min(J,K)$ as the number of SVD modes retained, which are ranked (by the singular values) in decreasing order in the previous factorization. It is remarkable that $\bW^\top\bW = \bT^\top\bT= \bI$ are unit matrices of dimension $N\times N$. 

The number of SVD modes $N$ is determined by a tolerance $\varepsilon_{svd}$, such that:
\begin{equation}
\sigma_{N+1}/\sigma_{1}\leq \varepsilon_{svd}.\label{eq:TOLsvd}
\end{equation}
This criterion reduces the dimensionality of the data from $J$ to $N$, referred to as the spatial complexity \cite{LeClaincheVega17}. For experimental data, the tolerance can be set to match the noise level or uncertainty, as smaller singular values generally correspond to noise, while dominant features are represented by larger singular values. For turbulent flow analyses, this threshold can relate to the scale of coherent structures captured in the flow. Other criteria for selecting $N$, such as those outlined in \cite{PCA} within the context of Principal Component Analysis (PCA), may also be applied.

\subsubsection{Low-cost singular value decomposition (lcSVD)\label{sec:lowcostSVD}}

The lcSVD technique, as originally proposed by Hetherington et al. \cite{ashton2020low}, implemented in our test scenarios, builds upon the standard SVD approach to effectively handle reduced datasets while reconstructing their full-scale solution, as detailed in Ref. \cite{Lupod}. This adaptation significantly lowers the computational expense associated with SVD, enabling its application to large datasets while requiring minimal computational resources.

This approach operates by applying SVD to the reduced snapshot matrix $\bar{\bV}_1^{K} \in \mathbb{R}^{\bar J\times \bar K}$ (Eq. \ref{eq:SnapMatrixRed}), followed by utilizing the decomposed data to reconstruct the original snapshot matrix $\bV_1^{K} \in \mathbb{R}^{J\times K}$ (Eq. \ref{eq:SnapMatrix}).

It is also feasible to reduce just one dimension of the snapshot matrix, referred to as a semi-reduced snapshot matrix. This matrix can take the form $\bar{\bV}_{1}^{K,{J\bar K}}\in \mathbb{R}^{J\times \bar K}$ or $\bar{\bV}_{1}^{K,{\bar J K}}\in \mathbb{R}^{\bar J\times  K}$, depending on whether the reduction is performed along the columns (spatial dimension) or rows (temporal dimension).

The following steps summarize the algorithm:
\begin{itemize}
    \item \textit{Step 1: SVD on the reduced data.} Perform SVD on the reduced snapshot matrix (Eq. \ref{eq:SnapMatrixRed}):
    \begin{equation}
    \bar{\bV}_1^{K}\simeq\bar{\bW}\,\bar{\bSigma}\,\bar{\bT}^\top,\label{eq:redSVD}
    \end{equation}
    where $\bar{\bW}^\top\bar{\bW} = \bar{\bT}^\top\bar{\bT}=\bI$ are unit matrices of dimension $\bar{N}\times \bar{N}$. The number of retained SVD modes, $\bar{N}$, is determined using a tolerance criterion (Eq. \ref{eq:TOLsvd}).

    \item \textit{Step 2: SVD mode normalization.} The matrix $\bar{\bSigma}$ can become ill-conditioned when retaining small singular values, potentially causing round-off errors that render the modes in $\bar{\bW}$ slightly non-orthogonal. To address this, QR factorization is applied to re-orthonormalize the modes:
    \begin{equation}
        \bar{\bW}=\bar{\bW} (\bR_{\bar N}^W)^{-1},
    \end{equation}
    where $\bR_{\bar N}^W \in \mathbb{R}^{\bar N\times \bar K}$. Only $\bar N$ modes are retained.

    \item \textit{Step 3: Normalization of temporal coefficients.} Similarly to the step 2, SVD temporal coefficients calculated in $\bar \bT$ may be slighty non-orthogonality. QR factorization is also applied here:
    \begin{equation}
        \bar{\bT}=\bar{\bT} (\bR_{\bar N}^T)^{-1},
    \end{equation}
    with $\bR_{\bar N}^T\in \mathbb{R}^{\bar N\times \bar K}$. Additionally, signs of temporal coefficients can vary based on computational settings, so these are adjusted using:
    \begin{equation}
        \bar{\bT}=\bar{\bT} \text{ sign}(\text{diag} (\bar{\bSigma})),
    \end{equation}
    where sign($\cdot$) and diag($\cdot$) denote the sign function and the diagonal of a matrix, respectively. To prevent conflicts and potential loss of information when determining the signs in $\bar{\bSigma}$, it is advisable to utilize $\bar{\bW}^\top  \bar{\bV}_1^{K} \bar{\bT}$ instead of directly working with $\bar{\bSigma}$. However, this approach may vary depending on the coding language employed in the implementation.

    \item \textit{Step 4: Recover SVD modes.} The SVD modes expanded to their full spatial dimensions $\bW$ from eq. (\ref{eq:svd}) are reconstructed as:

    \begin{equation}
        \bW\simeq \bW^{rec}= (\bar{\bV}_{1}^{K,{J\bar K}}) \bar{\bT} (\bar{\bSigma})^{-1},\label{eq:Wrec}
    \end{equation}
    where $\bW^{rec} \in \mathbb{R}^{J \times \bar N}$.

    \item \textit{Step 5: Recover temporal coefficients.} Temporal coefficients are recovered as:
    \begin{equation}
        \bT\simeq\bT^{rec}\simeq (\bar{\bV}_{1}^{K,{\bar J K}})^\top \bar{\bW} (\bar{\bSigma})^{-1},\label{eq:Trec}
    \end{equation}
    where $\bT^{rec} \in \mathbb{R}^{K \times \bar N}$.

    \item \textit{Step 6: Reconstruct the full dataset.} Using the reconstructed modes and coefficients, along with the singular values, the original dataset is recovered as:
    \begin{equation}
\bV_1^{K}\simeq \bV_1^{K, rec}= \bW^{rec}\,\bar \bSigma\,(\bT^{rec})^\top.\label{eq:svdRec}
\end{equation}

\end{itemize}

In our test cases, when using low-resolution computations with lcSVD to reconstruct the data, the total number of modes ($N_{modes}$) selected will be fixed at 20\%:

\[
  N_{modes} = N_{s, u_b} \times K \times 0.20.
  \label{eq:totalModes}
\]

Since lcSVD works with a high-dimensional dataset of size $N_{s, u_b} \times K$ time dimension, the total number of modes that can be extracted is determined by the number of degrees of freedom. This percentage is chosen to ensure that the most important features of the dataset are retained while reducing the computational cost of the reconstruction process.

\section{Methodology}\label{methodo}

\begin{figure}[H]
  \centering
  \includegraphics[width=0.90\textwidth]{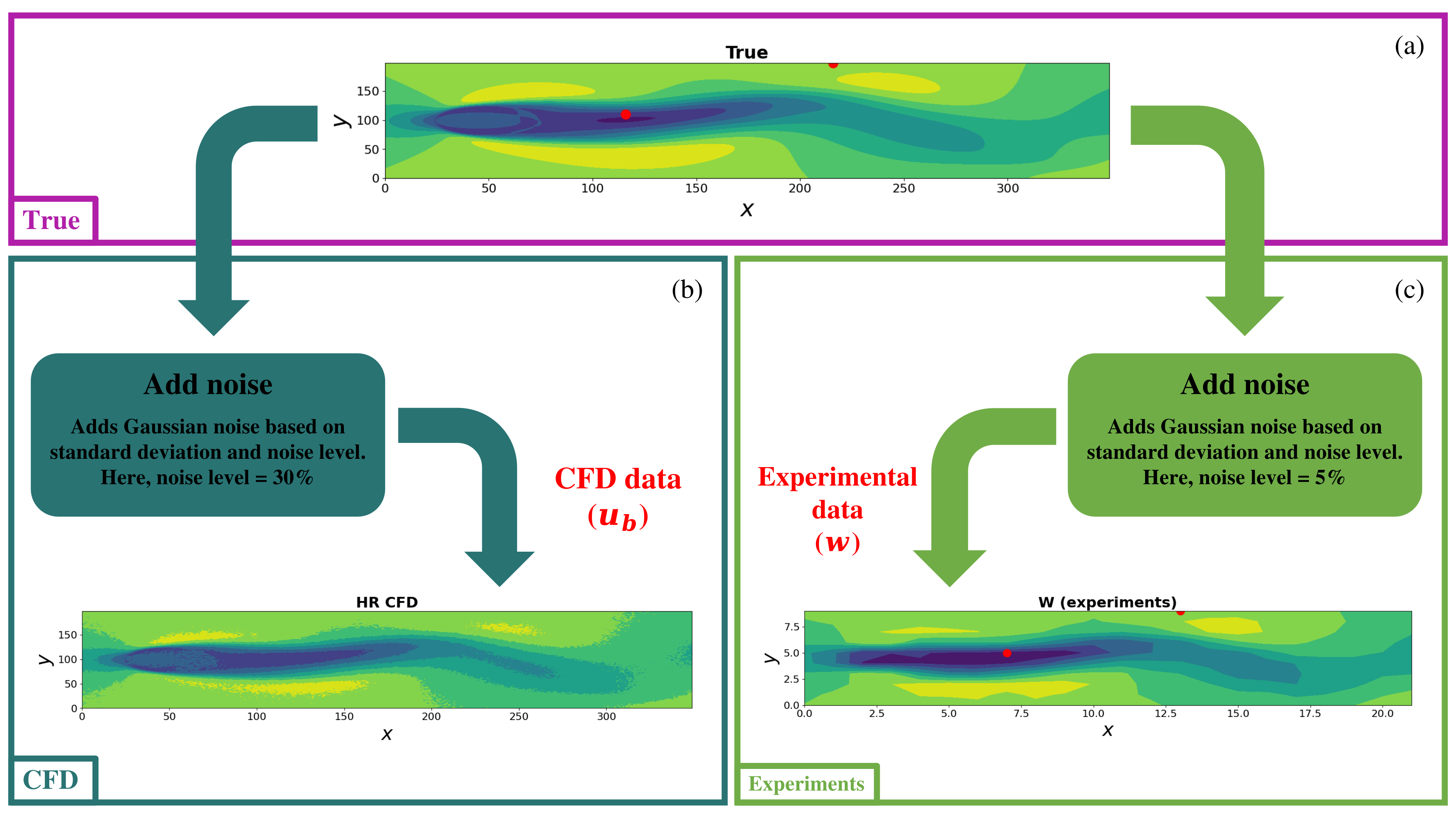}
  \caption{Schematic representation of the methodology to model a (b) CFD database ($u_b$) and an (c) experimental database ($w$), starting from the (a) original database.}
  \label{synthScheme}
\end{figure}

\begin{figure}[H]
  \centering
  \includegraphics[width=0.95\textwidth]{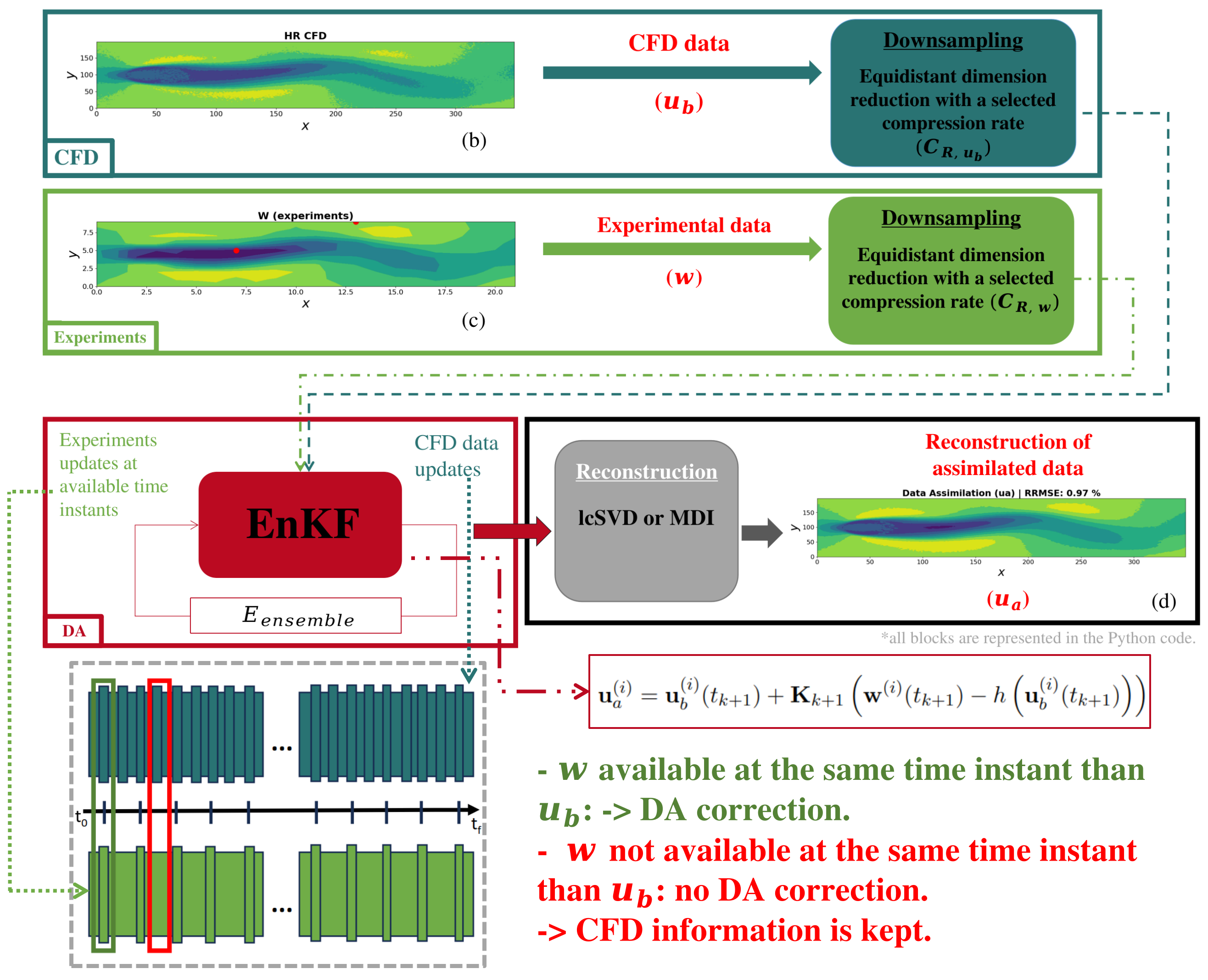}
  \caption{Methodology scheme of the twin-experiment coupling DA code with low-resolution computations, where (b) CFD database ($u_b$) and (c) an experimental database ($w$), resulting in (d) the reconstruction of assimilated data ($u_a$).}
  \label{mainScheme}
\end{figure}

This section outlines the overall computational setup and workflow for assessing the performance of data assimilation with LR techniques. We follow a twin-experiment approach to mimic realistic situations. In particular, combinations of available ground truth datasets with downsampling techniques and noise injection are used to represent low-fidelity CFD simulations and experimental measurements. The procedure described in this section is shown schematically in figures \ref{synthScheme} and \ref{mainScheme}.

The first step involves loading the True data, which serves as the ground truth for this study. This data is typically derived from high-fidelity simulations or detailed experimental measurements of fluid flow phenomena. The True data provides a reference standard that can be used to validate data assimilation and LR techniques.
This dataset encompasses the full temporal and spatial resolution necessary to describe the fluid dynamics under study accurately.

To generate a CFD-like dataset, Gaussian noise is added to the True data, as shown in figure \ref{synthScheme}, to represent the low fidelity CFD dataset, which allows us to demonstrate the robustness of the method when the noise increases. Then, the True data is downsampled to create a lower-resolution version (figure \ref{mainScheme}). The downsampling process reduces the spatial resolution through equidistant dimension reduction, which can be quantified by the compression rate:

\begin{equation}
  C_{R} = \frac{J}{N_{s}},
  \label{eq:compRate}
\end{equation} with $J$ containing the spatial data and the dimensional information of the dataset generated by the sensors, and $N_s$ is the number of sensors, with $N_{s,u_b}$ the number of sensors for $u_b$, and $N_{s,w}$ the number of sensors for $w$. $C_{R, ub}$ and $C_{R, w}$ are the compression rate applied on $u_b$ and $w$ respectively. The number of grid points along each spatial dimension is decreased to reach a LR dataset, referred to as the background CFD data ($u_b$) in this work. The compression rate will be a tuned parameter during the study. Section \ref{DS} describes the fundamentals to understand the downsampling method used in this article.

To create an experimental-like dataset ($w$), Gaussian noise is added to the True data to mimic the effects of measurement noise. The downsampling process further reduces the number of data points in the dataset compared to the one applied on $u_b$, resulting in a more substantial decrease in data resolution to mimic the grid limitations of experimental data. Its own time dimension is also created. The data assimilation code considers the experiments updates at available time steps. 

With the CFD and experimental datasets prepared, the next step involves assimilating the LR data. A more detailed explanation of DA and particularly EnKF is presented in section \ref{ENKF}. The role of EnKF here is to make the best approximation possible of the True state from the low-fidelity and LR CFD and experimental datasets. This result, referred to as $u_a$, represents the assimilated data (at a low-resolution in this stage).

The final step involves reconstructing the LR assimilated data into the original HR data using lcSVD \cite{ashton2020low} described in section \ref{svd}.

All the DA computations presented in this paper were performed using Python. The calculations were executed in serial on a computing cluster, with 500 GB of RAM, ensuring the needs of RAM memory access for our study. This study does not incorporate parallel computing, which is intended to accelerate computations and handle large datasets more efficiently; however, it will be implemented in future work to enhance performance and scalability. 

Adding noise to CFD and experimental data is essential for simulating the uncertainties and fluctuations typically present in the estimation of the True state. This allows testing the robustness and reliability of computational models under conditions that mimic real-world scenarios. Noise is systematically introduced into the dataset by scaling random fluctuations to match a specified noise level. The method involves perturbing each element of the input tensor with a randomly generated value. The magnitude of this perturbation is determined by both the intrinsic variability of the data and the user-defined noise level, denoted by \( \eta \).  

The standard deviation \( \sigma_T \) of the tensor \( \mathbf{T} \) reflects the typical amplitude of natural variations in the dataset. This serves as a baseline for the noise magnitude. The standard deviation of the noise, denoted as \( \sigma_N \), is defined as:

\[
\sigma_N = \eta \cdot \sigma_T,
\]

meaning that, on average, the noise will not introduce bias but will increase variability around the original values. The noise \( \boldsymbol{\epsilon} \) is sampled from a zero-mean Gaussian distribution with variance \( \sigma_N^2 \), and the noisy tensor \( \mathbf{T}_{\text{noisy}} \) is computed as:

\[
\mathbf{T}_{\text{noisy}} = \mathbf{T} + \boldsymbol{\epsilon}, \quad \boldsymbol{\epsilon} \sim \mathcal{N}\left(0, \sigma_N^2\right).
\]

This equation indicates that the noise is directly proportional to the standard deviation of the original tensor and the noise level. The method ensures that the added noise reflects the inherent variability of the data while being scaled appropriately by the noise level. The tuned parameter for this section will be indeed the noise level. The tensors, after adding noise, are introduced directly as inputs for the EnKF. 

To evaluate and compare the performance of the HR and LR methods, we utilize the Relative Root Mean Squared Error ($RRMSE$) and the Mean Absolute Error ($MAE$) as key quantitative metrics.

The $RRMSE$ is defined as:
\begin{equation} 
  RRMSE=\frac{\|\bV_1^{K}-\bV_1^{K,rec}\|_2}{\| \bV_1^{K}\|_2},
  \label{eq:rrmse}
\end{equation}
where $\| \cdot \|_2$ corresponds to the L2-norm, $\bV_1^{K}$ the original snapshot matrix (eq. \ref{eq:SnapMatrix}), and $\bV_1^{K,rec}$ the reconstruction of the original dataset (eq. \ref{eq:svdRec}). $RRMSE$ provides a measure of the average magnitude of the deviation between the original and reconstructed datasets, with a strong emphasis on larger discrepancies.

The $MAE$ is given by:
\begin{equation} 
MAE = \frac{1}{N} \sum_{i=1}^N | \bV_{j_1j_2j_3k}-\bV_{j_1j_2j_3k}^{rec} | \label{eq:meanAbsError}
\end{equation}
where N is the total number of data points, $j_1$, $j_2$ and $j_3$ are the indexes for the velocity component forming the database, $x$ and $y$ coordinates, respectively, while $k$ is used to index the snapshot. The $|.|$ operand represents the absolute value. The mean absolute error provides a measure of the average deviation between the original and reconstructed datasets, offering insights into the overall accuracy of the reconstruction process.

The computational speed-up, denoted as \( S \), is defined as the ratio between HR and LR computation time:

\begin{equation} 
S = \frac{t_{\text{HR}}}{t_{\text{LR}}},
\label{eq:speed-up}  
\end{equation} 

where \( t_{\text{HR}} \) and \( t_{\text{LR}} \) represent the total computational time required for the HR and LR methods, respectively. Each computation time is calculated using the mean of ten independent computations to avoid the CPU time instabilities from the computer. 

In addition, the RAM usage differences between HR and LR computations will be analyzed to evaluate the efficiency of the proposed methods. To quantify these differences, we define the RAM compression rate \( R_{\text{comp}} \) as:  
\begin{equation} 
  R_{\text{comp}} = \frac{R_{\text{HR}} - R_{\text{LR}}}{R_{\text{HR}}},
  \label{eq:RAMComp}  
\end{equation} 
where \( R_{\text{HR}} \) represents is the HR RAM size, and \( R_{\text{LR}} \) is the LR RAM size.

\section{Test Cases} \label{testcases}
\small
To show the accuracy of the DA method applied to fluid dynamics, and the low computational cost (compared to HR computations) of the presented method, two different datasets have been used to show the adaptivity of our method. 
The test cases studied are two flow past cylinder datasets, consisting of a numerical laminar three-dimensional at Re = 100 and an experimental turbulent three-dimensional at Re = 2600 with a two-dimensional database. The third test case is a numerical turbulent boundary layer, a turbulent jet Large Eddy Simulation (LES) at $Re \approx 10^6$. All test cases consist in fluid dynamics experiments or numerical simulations, and are governed by the Navier-Stokes equations. 

For both circular cylinder test cases (sections \ref{testcase1} and \ref{testcase3}), these equations for a viscous, incompressible and Newtonian flow are used: 

\begin{equation}
  \begin{gathered}
  \overrightarrow{\boldsymbol{V}} \cdot \nabla=0, \\
  \frac{\partial u}{\partial t}+(\overrightarrow{\boldsymbol{V}} \cdot \nabla) \overrightarrow{\boldsymbol{V}}=-\nabla p+\frac{1}{R e} \Delta \overrightarrow{\boldsymbol{V}},
  \end{gathered}
\end{equation}

where $\overrightarrow{\boldsymbol{V}}$ is the velocity vector, $p$ is the static pressure, and $Re$ is the Reynolds number. These equations are non-dimensionalized using the characteristic length $L$ and time $L/U$, where $U$ is the characteristic or free stream velocity for each case. In the turbulent jet case (\ref{testcase2}), the temperature equation to be obtained from the transport of a passive scalar will also be taken into account as follows,

\begin{equation}
  \rho c_p\left(\frac{\partial T}{\partial t}+(\overrightarrow{\boldsymbol{V}} \cdot \nabla) T\right)=k \nabla^2 T+\phi,
\end{equation}

where $\rho$ and $T$ are the density and temperature of the fluid, respectively, $c_p$ is the specific heat capacity at constant pressure, $k$ is the thermal conductivity coefficient, and $\phi$ represents the dissipation term which accounts for the conversion of kinetic energy into thermal energy due to the viscous effects within the fluid.

Concerning the EnKF, an ensemble size of $E = 25$ is selected for the circular cylinder case at $Re = 100$ in section \ref{testcase1}. Small ensemble sizes are often sufficient for systems with smoother dynamics, like laminar flows, where the uncertainties are more limited compared to highly chaotic or turbulent systems. This balance ensures computational efficiency while maintaining the ability to represent the system's variability adequately \cite{shady}. Subsequently, an ensemble size of $E = 45$ is tested to assess the impact of increasing the ensemble size on both the accuracy of the solution and the computational resource demands on the turbulent circular cylinder case at $Re = 2600$ in section \ref{testcase3} and for the Jet LES respectively in section \ref{testcase2}, which also captures the flow’s increased variability and unsteady nature. Studies in similar turbulent regimes indicate that ensemble sizes of 40 to 60 are sufficient to capture the chaotic structures in the wake while maintaining computational feasibility \cite{turbulentDA}.

\subsection{Circular cylinder dataset at $Re = 100$}\label{testcase1}
 
The first dataset, circular cylinder, is extracted from Ref. \cite{LeClaincheVega17}. This test case has been selected because it is generally used as a benchmark problem to validate methodologies in fluid dynamics. The dynamics of the flow are closely related to the concept of the Reynolds number, defined with the cylinder diameter $D$. The flow is steady for low Reynolds numbers. 
From Re $\approx$ 46, a Hopf bifurcation that produces an unsteady flow, which is driven by a von Karman vortex street \cite{vK vortex}. These oscillations remain two-dimensional until Re $\approx$ 189 where a second bifurcation occurs and the flow develops into a three-dimensional flow for some specific wavelengths in the spanwise direction \cite{3dim floquet}. The flow becomes fully three-dimensional for some specific wave numbers at Re > 180, and fully turbulent for high Reynolds values.

This is a numerical dataset: numerical simulations were carried out using the open-source solver Nek5000 \cite{nek5000} to solve the incompressible Navier-Stokes equations defining the behavior of the flow. This solver uses spectral elements methods as spatial discretization.

\begin{figure}[H]
  \centering
  \includegraphics[width=1\textwidth]{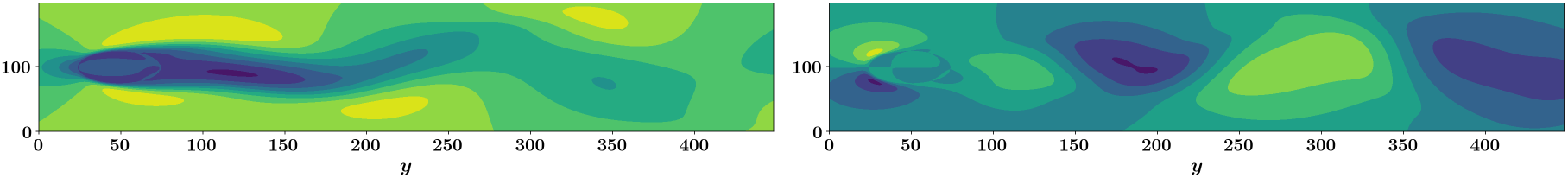}
  \caption{Streamwise (left) and normal (right) velocities of a representative snapshot of the two-dimensional $Re = 100$ cylinder dataset from Ref. \cite{LeClaincheVega17}.}
  \label{Re=100}
\end{figure}

The boundary conditions configured in the simulation for the cylinder surface are Dirichlet for velocity $(u = v = w = 0)$ and homogeneous Neumann for pressure. The conditions in the inlet, upper and lower boundaries of the domain are the same: $u = 1$, $v = w = 0$ for the streamwise, normal and spanwise velocities, respectively, and homogeneous Neumann condition for pressure. The conditions in the outlet are Dirichlet for pressure and homogeneous Neumann for velocity. The domain of the computational simulations is composed of 600 rectangular elements, each one of these is discretized using the polynomial order 9. The dimensions of the computational domain are non-dimensionalized with the 15 diameter of the cylinder. The size of the domain in the normal direction is constant $L_y = 15D$, and extends in the streamwise direction from $L_x = 15D$ upstream of the cylinder to $L_x = 50D$ downstream. As mentioned, the two-dimensional cylinder dataset represents the saturated flow around a circular cylinder with Reynolds number $Re = 100$. The dataset analyzed is composed of $N_t = 150$ snapshots equidistant in time with an interval of $\Delta t = 0.2$. The dimensions are $N_x = 350$ points in the streamwise direction and $N_y = 199$ points in the normal direction. The two variables ($N_{comp} = 2$) used in the reconstruction of the dataset correspond to the streamwise velocity $U$, and normal velocity $V$. Figure \ref{Re=100} shows a representative snapshot of this database. 

\subsection{Turbulent circular cylinder at $Re = 2600$}\label{testcase3}

This dataset is extracted from Ref. \cite{VKIcylinder}. It consists of an experimental database representing the turbulent flow around a circular cylinder with a diameter of $D = 5 \mathrm{mm}$ and a length of $L = 20 \mathrm{cm}$ under transient conditions, featuring a varying free stream velocity.

The experiment was carried out in the low-speed wind tunnel of the Von Karman Institute, utilizing a time-resolved Particle Image Velocimetry (TR-PIV) system from Dantec Dynamics to measure the velocity fields. Seeding particles of approximately $1.5 \, \mu\mathrm{m}$ in diameter were introduced into the flow using a Laskin nozzle and illuminated by a Nd:YLF laser providing $20 \, \mathrm{mJ/pulse}$ at $1 \, \mathrm{kHz}$. The illuminated flow was captured by a SpeedSense camera at $3 \, \mathrm{kHz}$ with a resolution of $1280 \times 488 \, \mathrm{pixels}$.

The dataset consists of the study of a turbulent flow passing a three-dimensional cylinder at $Re = 2600$, where the vortex shedding frequency goes to $f = 303 \, \mathrm{Hz}$, corresponding to a Strouhal number of around $St = f d / U_\infty \approx 0.19$.

The dimensions of the experimental domain are $N_x = 301$ points in the streamwise direction and $N_y = 111$ in the normal direction. This dataset consists of two velocity components ($N_{\mathrm{comp}} = 2$), the streamwise velocity $U$ and normal velocity $V$, which were measured during $5200$ snapshots over $4.5 \, \mathrm{s}$. Each snapshot is taken with time $\Delta t = 0.33 \, \mathrm{s}$.

\begin{figure}[H]
  \centering
  \includegraphics[width=1\textwidth]{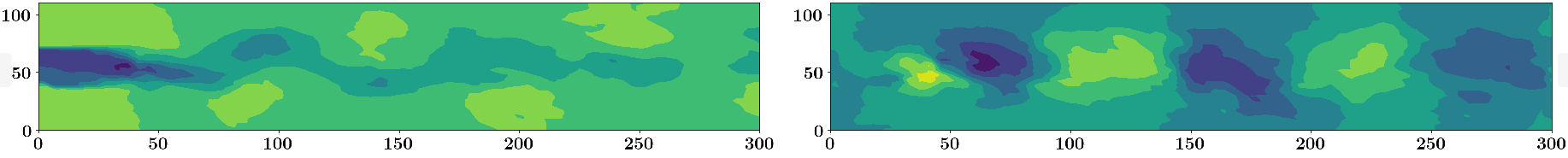}
  \caption{Streamwise (left) and normal (right) velocities of a representative snapshot of the turbulent $Re = 2600$ cylinder dataset from Ref. \cite{VKIcylinder}.}
  \label{vkicylinder}
\end{figure}

\subsection{Turbulent jet large eddy simulation}\label{testcase2}

This next dataset is a numerical database showing an isothermal subsonic jet issued from a round nozzle of exit diameter $D = 50$ mm, see Ref. \cite{JetLES} for more details. This simulation was conducted using the LES compressible flow solver CharLES, which solves the spatially-filtered compressible Navier-Stokes equations on unstructured grids using a finite-volume method and third-order Runge-Kutta time integration. CharLES is a high-fidelity CFD solver that excels in simulating complex fluid flows, including those encountered in jet engines. Using LES, CharLES accurately models turbulent aerothermal interactions and unsteady phenomena, which are critical in understanding engine performance during both design and off-design conditions \cite{mdpi2020,CharLES}. The simulation used approximately sixteen million control volumes and was run for 2000 acoustic time units $(tc_\infty/D)$.

\begin{figure}[H]
  \centering
  \includegraphics[width=1\textwidth]{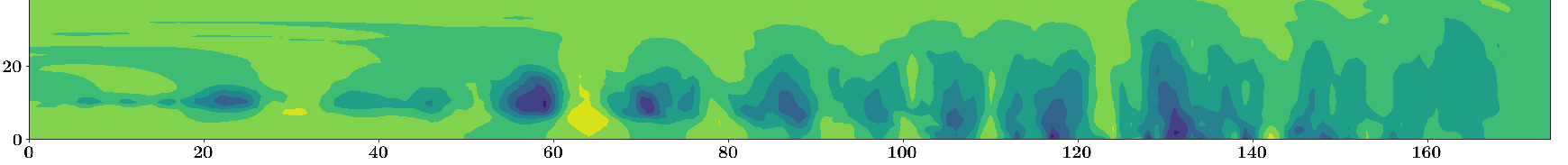}
  \caption{Streamwise velocity of a representative snapshot of the turbulent jet large eddy simulation dataset from Ref. \cite{JetLES}.}
  \label{Jetles}
\end{figure}

This LES dataset consists of 5000 snapshots of the jet's streamwise velocity $(N_{comp} = 1)$ sampled every $\Delta t = 0.2$ acoustic time units on a structured cylindrical output grid that approximately mimics the underlying LES resolution and extends a distance of $30D$ in the streamwise direction and $6D$ in the radial direction, respectively. The simulation settings were set to match the experimental operational conditions, which are defined in terms of the nozzle pressure ratio, $NPR = P_t/P_\infty = 1.7$, and nozzle temperature ratio, $NPR = P_t/P_\infty = 1.15$, where the subscripts $t$ and $\infty$ refer to the stagnation (total) and free-stream (ambient) conditions, respectively. The jet is isothermal $(T /T_\infty = 1.0)$, and the jet Mach number is $M = U/c = 0.9$, where $U$ is the mean (time-averaged) jet exit streamwise velocity and $c$ is the speed of sound. With these conditions, the Reynolds number is $Re = \rho U D/ \mu \approx 10^6$. Each $N_t = 5000$ snapshots contains $N_x = 175$ points in the horizontal plane and $N_y = 39$ points in the vertical plane. Figure \ref{Jetles} shows a representative snapshot of the flow field.

The round nozzle geometry, with output centered at $(x, r) = (0, 0)$, is explicitly included in the axisymmetric computational domain, which stretches from $-10D$ to $50D$ in the streamwise $(x)$ direction and expands from $20D$ to $40D$ in the radial direction. A very slow co-flow at Mach number $M_\infty = 0.009$ is imposed outside the nozzle in the simulation to prevent spurious re-circulation and facilitate flow entrainment.

The Vreman sub-grid model \cite{Vreman} is used to account for the physical effects of unresolved turbulence on the resolved flow, with constant coefficient set to $c_{SGS} = 0.07$. A constant turbulent Prandtl number $Pr = c_{\rho}\mu/k = 0.9$ is selected to complete the energy equation, while the laminar Prandtl number $Pr = c_{\rho}\mu/k = 0.72$. To avoid spurious reflections at the downstream boundary of the computational domain, a damping function \cite{BCLES, Sponge_layers} is applied in the outflow buffer zone as a source term in the governing equations. In addition, the numerical operators are switched to lower-order dissipative discretization in the sponge zone for $x/D > 31$ and $r/D > 7$, to damp turbulent structures, acoustic waves, etc.

All solid surfaces are regarded as adiabatic no-slip walls. In the preliminary parametric study, the far-field noise at $50D$ from the nozzle output was calculated for three different FW-H (Ffowcs Williams–Hawkings) surfaces which consisted of a cylindrical surface of radius $0.65D$ up $x/D = 0$, continued by a conical surface that extended to $x/D = 30$ with different spreading rates of $0.11$, $0.14$ and $0.17$. The jet spreading rate estimates are used as a basis to select the slopes \cite{JetExperiments, JetNozzles}.

\section{Results} \label{results}
\small
This section presents the results and performances of the DA method, integrated with low-cost techniques, applied to each test case.

First, the DA method is applied to the test cases described in section \ref{testcases} without utilizing LR techniques. In this scenario, computations are performed using HR data. A synthetic CFD data ($u_b$) of the same dimensions as the original data is generated. Additionally, synthetic experimental data ($w$) with reduced dimensions is created to simulate data collection from sensors. Random noise (from 5\% to 50\%) is introduced to both $u_b$ and $w$ to emphasize the effectiveness of the data assimilation techniques, resulting in the analysis, denoted as $u_a$.

Subsequently, the HR computational results with 5\% and 50\% of noise are compared to those obtained using the same DA method, but with the incorporation of LR techniques. 5\% represent a standard value for experimental measurements, and at 50\% for high noise comparison. The performance of the lcSVD reconstruction method is evaluated, demonstrating how these LR techniques provide superior results in terms of computational efficiency and memory usage, without compromising the quality of $u_a$. The accuracy of the LR results will be evaluated by comparing them with the HR results, comparing the True state estimation ($RRMSE$ in eq. \ref{eq:rrmse} and $MAE$ in eq. \ref{eq:meanAbsError}), as well as computational resources, specifically computation time and RAM usage. 

Table \ref{tab:dims} displays the original dataset shape of each test case presented in section \ref{testcases}, a reorganized version of each dataset where the dimensions are reshaped to $J \times K$, with $J$ containing the spatial data and the dimensional information of the dataset generated by the sensors with dimensions $N_s$, and $K$ the time dimension. The last column of this table contains the spatial data compression rate $C_R$ comparing the original number of spatial points and the number of sensors as shown in eq. \ref{eq:compRate}.

The results are presented with varying downsampling configurations, denoted by the compression rate $C_{R, u_b}$ (eq. \ref{eq:compRate}) to illustrate the relationship between computational time and solution accuracy using this LR technique. Computations are performed by taking subsets of the domain $J \times K$.
For $C_{R, w}$, we select approximately 1/315 and 1/100 of $J \times K$ points to represent the constraints typically encountered with experimental data points for test cases \ref{testcase1} and \ref{testcase3} respectively. For the JetLES test case \ref{testcase2}, we use approximately 1/250 of $J \times K$ points to show the variation of performances with a low number of points for the experiments ($N_{s,w} \times K = 27 \times 151$).

The dataset shape is $N_x \times N_y \times K$ for two-dimensional data with a singular velocity component, $N_{comp} \times N_x \times N_y \times K$ for two-dimensional data with more than one velocity component. As represented in table \ref{tab:dims}, all points are taken into account for HR studies, so $N_{s, u_b} \times K = J \times K$.

\renewcommand{\arraystretch}{0.97}
\setlength{\tabcolsep}{3pt}

\begin{longtable}{|c|c|c|c|c|c|c|c|c|}
\hline
Dataset & $J \times K$ & Case & Method(s) & Noise (\%) & $N_{s,u_b} \times K$ & $N_{s,w} \times K$ & $C_{R, u_b}$ & $C_{R, w}$ \\
\hline
\endfirsthead

\multirow{17}{*}{$Re = 100$} & \multirow{17}{*}{$139300 \times 151$}
& 1 & HR & 2 & $139300 \times 151$ & $440 \times 151$ & $1$ & $317$ \\
\cline{3-9}
& & 2 & HR & 5 & $139300 \times 151$ & $440 \times 151$ & $1$ & $317$ \\
\cline{3-9}
& & 3 & lcSVD & 5 & $8 \times 151$ & $440 \times 151$ & $17413$ & $317$ \\
\cline{3-9}
& & 4 & lcSVD & 5 & $18 \times 151$ & $440 \times 151$ & $7739$ & $317$ \\
\cline{3-9}
& & 5 & lcSVD & 5 & $32 \times 151$ & $440 \times 151$ & $4353$ & $317$ \\
\cline{3-9}
& & 6 & lcSVD & 5 & $440 \times 151$ & $440 \times 151$ & $317$ & $317$ \\
\cline{3-9}
& & 7 & lcSVD & 5 & $1800 \times 151$ & $440 \times 151$ & $77$ & $317$ \\
\cline{3-9}
& & 8 & lcSVD & 5 & $2800 \times 151$ & $440 \times 151$ & $49.8$ & $317$ \\
\cline{3-9}
& & 9 & lcSVD & 5 & $11200 \times 151$ & $440 \times 151$ & $12.4$ & $317$ \\
\cline{3-9}
& & 10 & lcSVD & 5 & $45000 \times 151$ & $440 \times 151$ & $3.10$ & $317$ \\
\cline{3-9}
& & 11 & HR & 10 & $139300 \times 151$ & $440 \times 151$ & $1$ & $317$ \\
\cline{3-9}
& & 12 & HR & 30 & $139300 \times 151$ & $440 \times 151$ & $1$ & $317$ \\
\cline{3-9}
& & 13 & HR & 50 & $139300 \times 151$ & $440 \times 151$ & $1$ & $317$ \\
\cline{3-9}
& & 14 & lcSVD & 50 & $8 \times 151$ & $440 \times 151$ & $17413$ & $317$ \\
\cline{3-9}
& & 15 & lcSVD & 50 & $18 \times 151$ & $440 \times 151$ & $7739$ & $317$ \\
\cline{3-9}
& & 16 & lcSVD & 50 & $32 \times 151$ & $440 \times 151$ & $4353$ & $317$ \\
\cline{3-9}
& & 17 & lcSVD & 50 & $2800 \times 151$ & $440 \times 151$ & $49.8$ & $317$ \\
\cline{3-9}
& & 18 & lcSVD & 50 & $45000 \times 151$ & $440 \times 151$ & $3.10$ & $317$ \\
\hline

\multirow{17}{*}{$Re = 2600$} & \multirow{17}{*}{$66822 \times 151$}
& 19 & HR & 2 & $66822 \times 151$ & $660 \times 151$ & $1$ & $101$ \\
\cline{3-9}
& & 20 & HR & 5 & $66822 \times 151$ & $660 \times 151$ & $1$ & $101$ \\
\cline{3-9}
& & 21 & lcSVD & 5 & $8 \times 151$ & $660 \times 151$ & $8353$ & $101$ \\
\cline{3-9}
& & 22 & lcSVD & 5 & $18 \times 151$ & $660 \times 151$ & $3712$ & $101$ \\
\cline{3-9}
& & 23 & lcSVD & 5 & $32 \times 151$ & $660 \times 151$ & $2088$ & $101$ \\
\cline{3-9}
& & 24 & lcSVD & 5 & $42 \times 151$ & $660 \times 151$ & $1591$ & $101$ \\
\cline{3-9}
& & 25 & lcSVD & 5 & $150 \times 151$ & $660 \times 151$ & $445$ & $101$ \\
\cline{3-9}
& & 26 & lcSVD & 5 & $660 \times 151$ & $660 \times 151$ & $101$ & $101$ \\
\cline{3-9}
& & 27 & lcSVD & 5 & $1064 \times 151$ & $660 \times 151$ & $62.8$ & $101$ \\
\cline{3-9}
& & 28 & lcSVD & 5 & $4200 \times 151$ & $660 \times 151$ & $15.9$ & $101$ \\
\cline{3-9}
& & 29 & lcSVD & 5 & $16800 \times 151$ & $660 \times 151$ & $3.98$ & $101$ \\
\cline{3-9}
& & 30 & HR & 10 & $66822 \times 151$ & $660 \times 151$ & $1$ & $101$ \\
\cline{3-9}
& & 31 & HR & 30 & $66822 \times 151$ & $660 \times 151$ & $1$ & $101$ \\
\cline{3-9}
& & 32 & HR & 50 & $66822 \times 151$ & $660 \times 151$ & $1$ & $101$ \\
\cline{3-9}
& & 33 & lcSVD & 50 & $8 \times 151$ & $660 \times 151$ & $8353$ & $101$ \\
\cline{3-9}
& & 34 & lcSVD & 50 & $32 \times 151$ & $660 \times 151$ & $2088$ & $101$ \\
\cline{3-9}
& & 35 & lcSVD & 50 & $1064 \times 151$ & $660 \times 151$ & $62.8$ & $101$ \\
\cline{3-9}
& & 36 & lcSVD & 50 & $16800 \times 151$ & $660 \times 151$ & $3.98$ & $101$ \\
\hline

\multirow{16}{*}{Jet LES} & \multirow{16}{*}{$6825 \times 151$}
& 37 & HR & 2 & $6825 \times 151$ & $27 \times 151$ & $1$ & $253$ \\
\cline{3-9}
& & 38 & HR & 5 & $6825 \times 151$ & $27 \times 151$ & $1$ & $253$ \\
\cline{3-9}
& & 39 & lcSVD & 5 & $36 \times 151$ & $27 \times 151$ & $190$ & $253$ \\
\cline{3-9}
& & 40 & lcSVD & 5 & $64 \times 151$ & $27 \times 151$ & $107$ & $253$ \\
\cline{3-9}
& & 41 & lcSVD & 5 & $81 \times 151$ & $27 \times 151$ & $84$ & $253$ \\
\cline{3-9}
& & 42 & lcSVD & 5 & $100 \times 151$ & $27 \times 151$ & $68$ & $253$ \\
\cline{3-9}
& & 43 & lcSVD & 5 & $180 \times 151$ & $27 \times 151$ & $38$ & $253$ \\
\cline{3-9}
& & 44 & lcSVD & 5 & $220 \times 151$ & $27 \times 151$ & $31$ & $253$ \\
\cline{3-9}
& & 45 & lcSVD & 5 & $440 \times 151$ & $27 \times 151$ & $16$ & $253$ \\
\cline{3-9}
& & 46 & lcSVD & 5 & $1760 \times 151$ & $27 \times 151$ & $4$ & $253$ \\
\cline{3-9}
& & 47 & HR & 10 & $6825 \times 151$ & $27 \times 151$ & $1$ & $253$ \\
\cline{3-9}
& & 48 & HR & 30 & $6825 \times 151$ & $27 \times 151$ & $1$ & $253$ \\
\cline{3-9}
& & 49 & HR & 50 & $6825 \times 151$ & $27 \times 151$ & $1$ & $253$ \\
\cline{3-9}
& & 50 & lcSVD & 50 & $36 \times 151$ & $27 \times 151$ & $190$ & $253$ \\
\cline{3-9}
& & 51 & lcSVD & 50 & $100 \times 151$ & $27 \times 151$ & $68$ & $253$ \\
\cline{3-9}
& & 52 & lcSVD & 50 & $440 \times 151$ & $27 \times 151$ & $16$ & $253$ \\
\cline{3-9}
& & 53 & lcSVD & 50 & $1760 \times 151$ & $27 \times 151$ & $4$ & $253$ \\
\hline

\caption{From left to right columns: Dataset used, organized data reshaped into $J$ for all $K$ snapshots, computation case, methodology used for DA computation, shape of the data collected by the sensors selecting $N_{s,u_b}$ equidistant data points of $u_b$ for all $K$ snapshots, shape of the data collected by the sensors selecting $N_{s,w}$ equidistant data points of $w$ for all $K$ snapshots, and the data compression $C_R$ rate between the original data and the downsampled dataset (for both $u_b$ and $w$).}
\label{tab:dims}
\end{longtable}

The cases showcased in this results section were strategically chosen to illustrate three fundamental capabilities: (1) the method's robust performance under high noise conditions, (2) the significant computational advantages achievable through data compression, and (3) the practical trade-offs between accuracy and computational efficiency.

\subsection{Circular cylinder dataset at $Re = 100$}\label{Re100}

The cases presented were strategically selected to validate three key aspects of the study for the circular cylinder at $Re = 100$ presented in section \ref{testcase1}. Cases 12 and 13 demonstrate the method's noise resilience, while cases 7 and 10 establish the viability of computationally efficient LR computations that preserve solution quality. Furthermore, cases 14-18 in table \ref{tab:summarySVD_Re100} provide additional evidence of the LR method's accuracy retention under high noise levels, demonstrating consistent performance across various compression ratios even at 50\% noise.

\begin{figure}[H]
  \centering
  \includegraphics[width=1\textwidth]{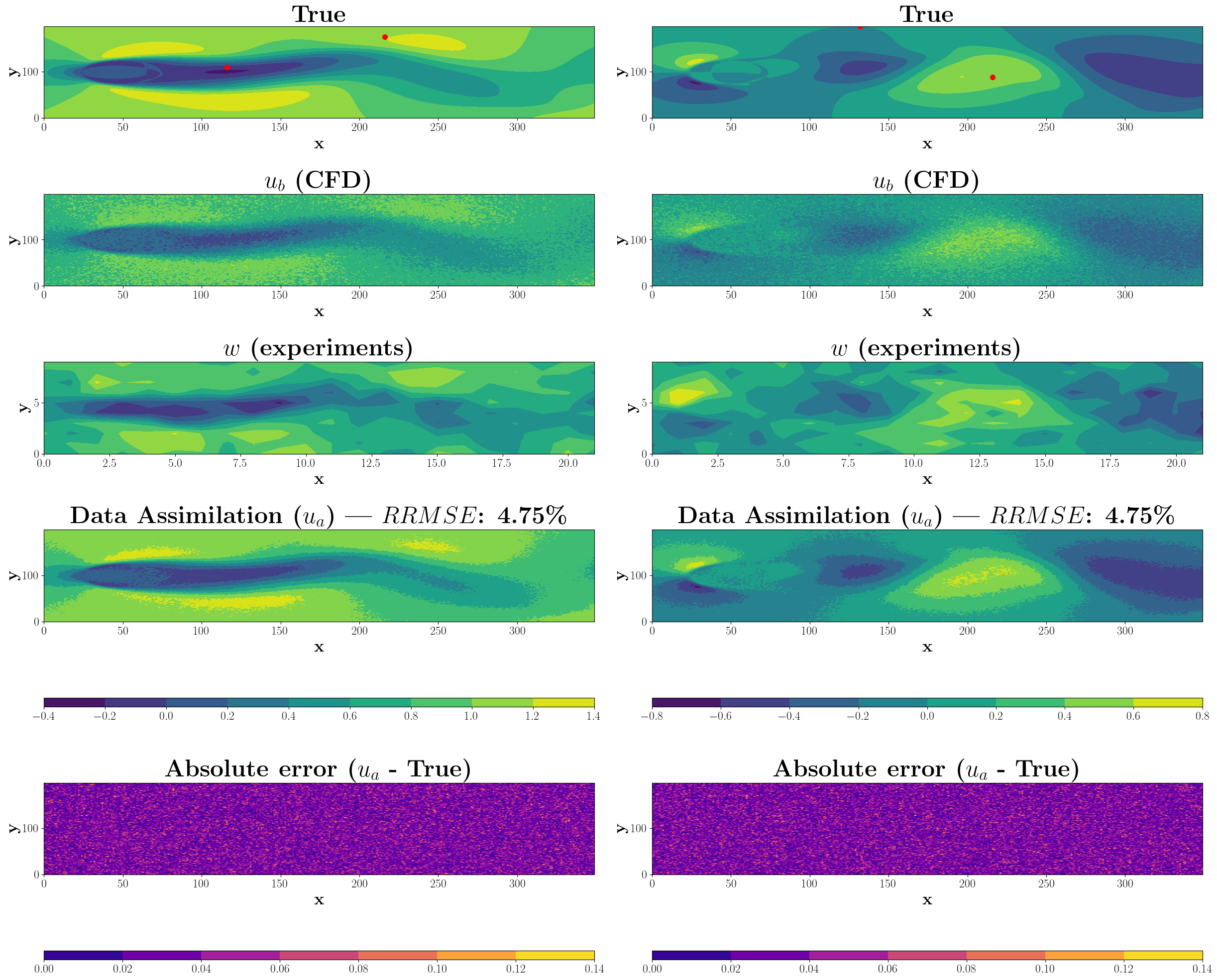}
  \caption{Case 12 in table \ref{tab:dims}. Cylinder dataset at $Re = 100$ results with 30\% of noise level applied on both $u_b$ and $w$, HR method. $N_{s, u_b} \times K = 139300 \times 151$ ($C_{R,u_b} = 1$) and $N_{s, w} \times K = 440 \times 151$ ($C_{R,w} = 317$). From the top to the bottom: True data, CFD data, experimental data, DA reconstruction, absolute error ($| u_a - \text{True}|$). The left column represents the streamwise velocity, and the right column represents the normal velocity.}
  \label{Re100ResNoise30}
\end{figure}

\begin{figure}[H]
  \centering
  \includegraphics[width=1\textwidth]{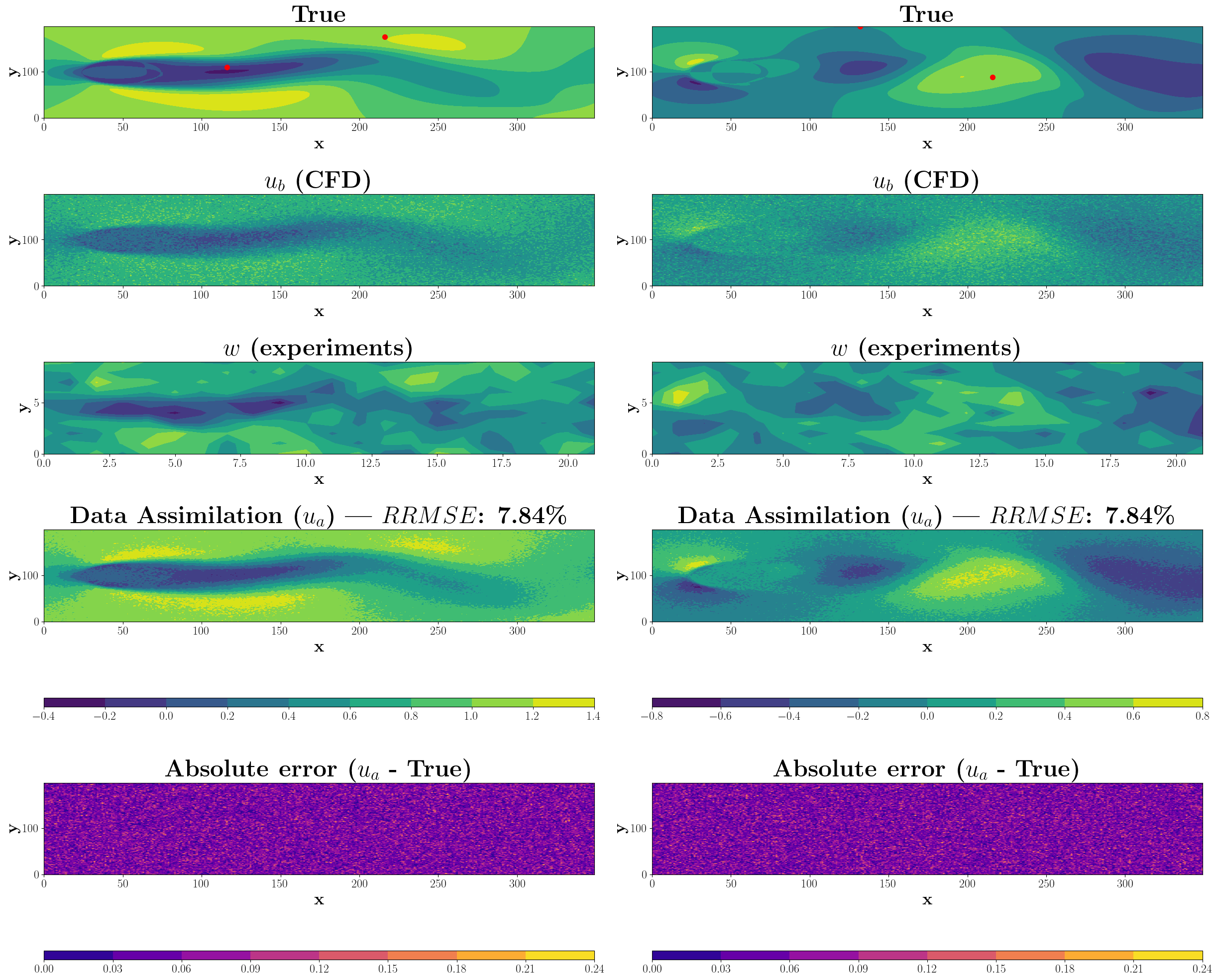}
  \caption{Case 13 in table \ref{tab:dims}. Cylinder dataset at $Re = 100$ results with 50\% of noise level applied on both $u_b$ and $w$, HR method. $N_{s, u_b} \times K = 139300 \times 151$ ($C_{R,u_b} = 1$) and $N_{s, w} \times K = 440 \times 151$ ($C_{R,w} = 317$). From the top to the bottom: True data, CFD data, experimental data, DA reconstruction, absolute error ($| u_a - \text{True}|$). The left column represents the streamwise velocity, and the right column represents the normal velocity.}
  \label{Re100ResNoise50}
\end{figure}

For the noise resilience demonstration, cases 12 and 13 in table \ref{tab:dims} are observed, with a number of points $N_{s, u_b} \times K = 139300 \times 151$ ($C_{R,u_b} = 1$) and $N_{s, w} \times K = 440 \times 151$ ($C_{R,w} = 317$), with 30\% and 50 \% of noise level applied on both $u_b$ and $w$ respectively, represented in figures \ref{Re100ResNoise30} and \ref{Re100ResNoise50} fixed at one time instant $t = 150$. The figures show the reference data (True), the CFD data with noise added, and then the assimilated data, where a first graphical result is given on the estimation of the True state. Experimental $(w)$ data is shown with a reduced dimension. Finally, the last figures show the absolute error between $u_a$ and $\text{True}$. As can be seen, the error is uniform and bounded throughout the domain, with no region having a higher error zones than others.

The results are particularly promising, with 30\% noise and an $RRMSE$ of 4.75\%, showing the strong performance of the EnKF method in high-resolution (HR) applications. Even with 50\% noise, the $RRMSE$ remains below 8\%, highlighting the method's ability to effectively reduce noise. These results emphasize the potential of EnKF for solving complex problems while maintaining accuracy, even in noisy environments.

\begin{figure}[H]
  \centering
  \includegraphics[width=1\textwidth]{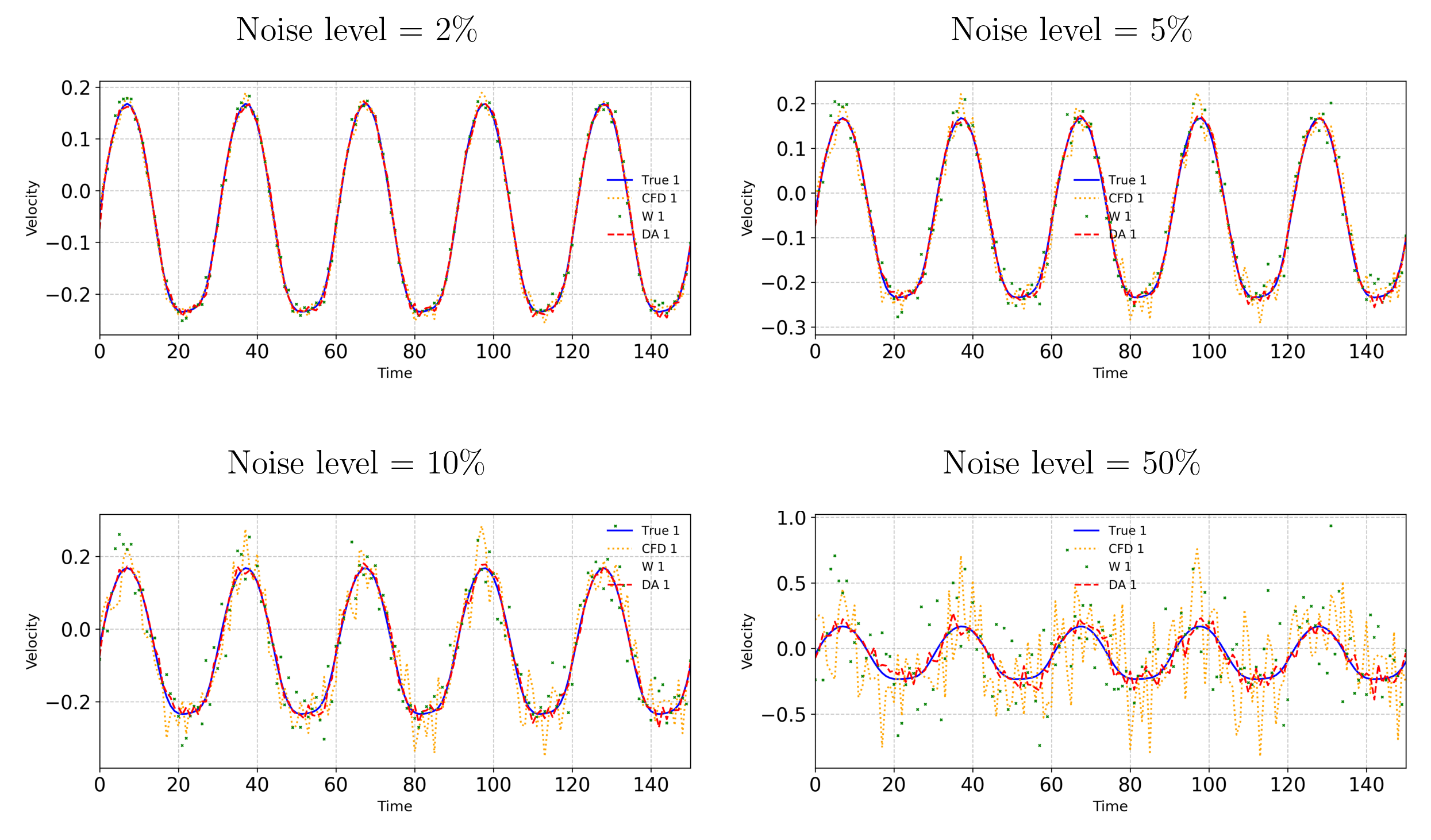}
  \caption{Velocity tracking point (1st red dot from the left side column 1 in figures \ref{Re100ResNoise30} and \ref{Re100ResNoise50}) circular cylinder dataset at $Re = 100$, with noise levels from 2\% to 50\%, and $C_{R,w}$ = 317. True data is shown in blue, $u_a$ in yellow, $w$ in green, and the assimilated data in red.}
  \label{2trackingVelRe100}
\end{figure}

\begin{figure}[H]
  \centering
  \includegraphics[width=1\textwidth]{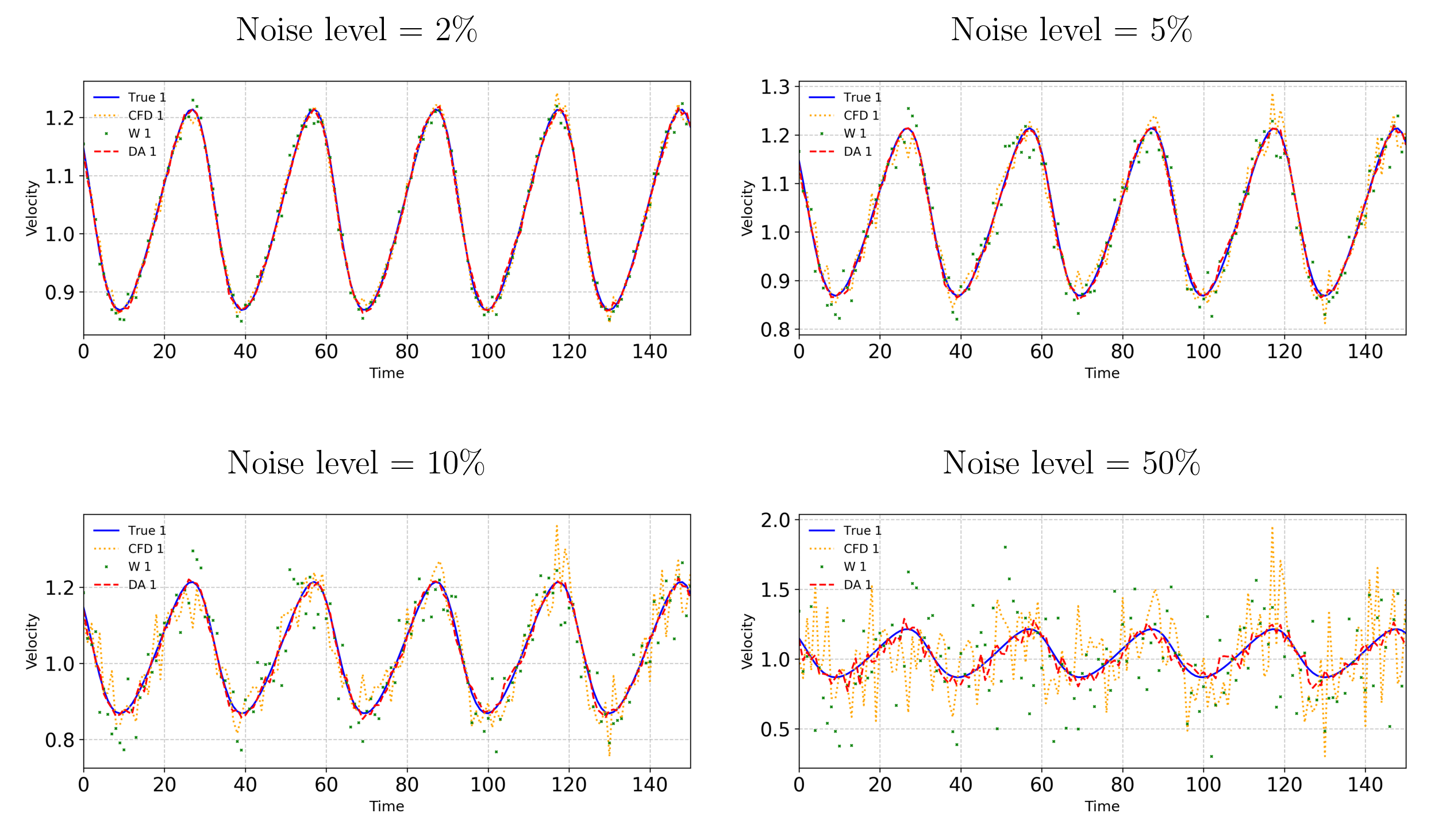}
  \caption{Velocity tracking point (2nd red dot from the left side column 1 in figures \ref{Re100ResNoise30} and \ref{Re100ResNoise50}) circular cylinder dataset at $Re = 100$, with noise levels from 2\% to 50\%, and $C_{R,w}$ = 317. True data is shown in blue, $u_a$ in yellow, $w$ in green, and the assimilated data in red.}
  \label{3trackingVelRe100}
\end{figure}

The red dots in the figures represent velocity tracking over time, providing a clearer view of how DA estimates the True state. Figures \ref{2trackingVelRe100} and \ref{3trackingVelRe100} show the velocity evolution at a representative point over time, comparing the True state, CFD data (\( u_b \)), experimental measurements (\( w \)), and assimilated data (\( u_a \)). These results demonstrate that the DA method achieves relatively accurate performance in estimating the True state across various noise levels.

The information in these figures can be cross-referenced with the table \ref{tab:summarySVD_Re100} to quantify the observations. As expected, increasing the noise level results in a corresponding increase in $MAE$ and $RRMSE$. The variation of $RRMSE$ is relatively low when the noise level is high. For noise level 2\%, 5\%, 10\%, 30\% and 50\%, the $RRMSE$ reaches 0.90\%, 0.97\%, 1.77\%, 4.75\%, 7.84\% respectively, and 0.45\%, 0.59\%, 0.93\%, 2.51\%, 4.14\% respectively for $MAE$. 

\begin{figure}[H]
  \centering
  \includegraphics[width=1\textwidth]{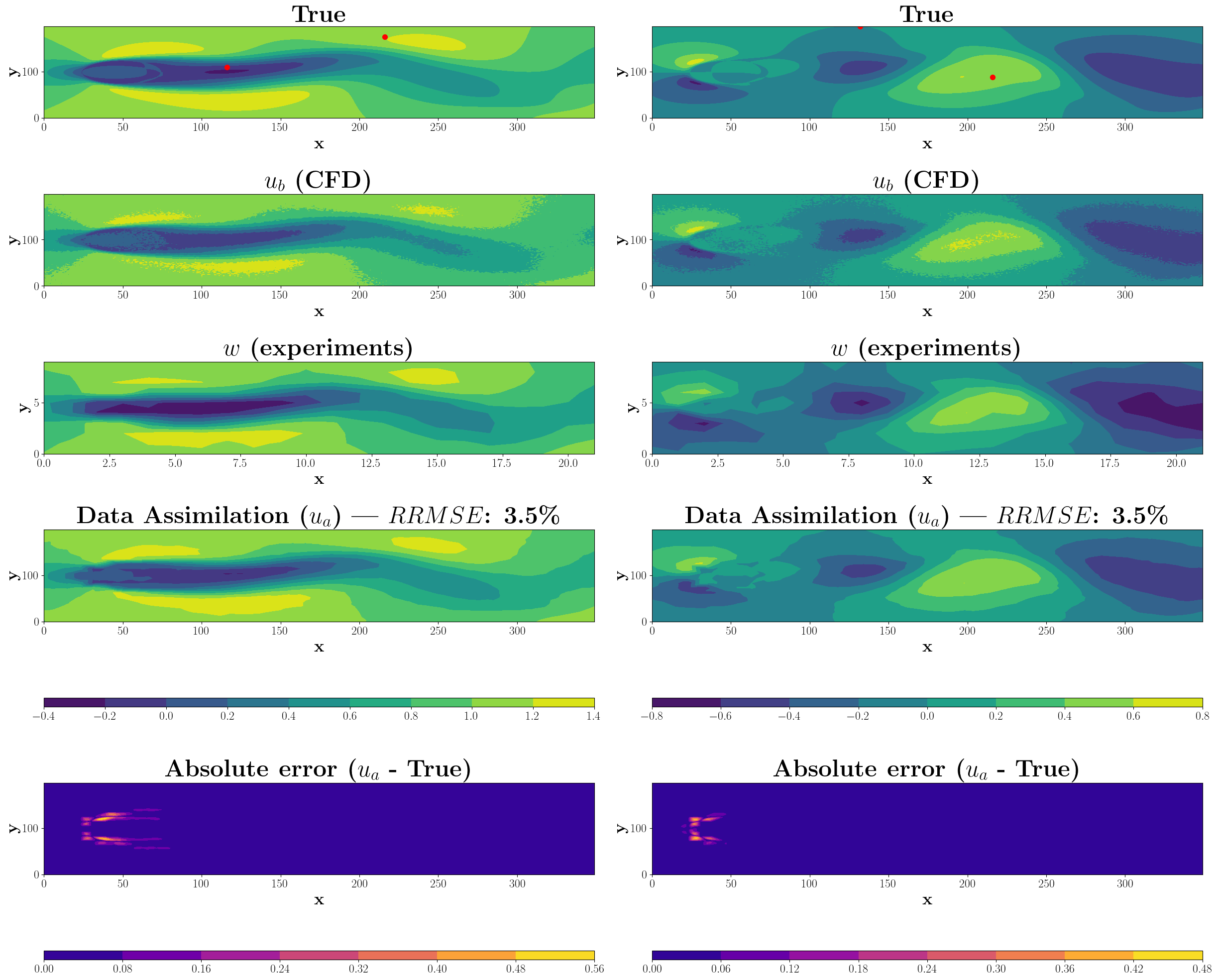}
  \caption{Case 7 in table \ref{tab:dims}. Cylinder dataset at $Re = 100$ results with 5\% of noise level applied on both $u_b$ and $w$, with lcSVD reconstruction method. $N_{s, u_b} \times K = 1800 \times 151$ ($C_{R,u_b} = 77$) and $N_{s, w} \times K = 440 \times 151$ ($C_{R,w} = 317$). From the top to the bottom: True data, CFD data, experimental data, DA reconstruction, absolute error ($| u_a - \text{True}|$). The left column represents the streamwise velocity, and the right column represents the normal velocity.}
  \label{Re100ResNoise5}
\end{figure}

\begin{figure}[H]
  \centering
  \includegraphics[width=1\textwidth]{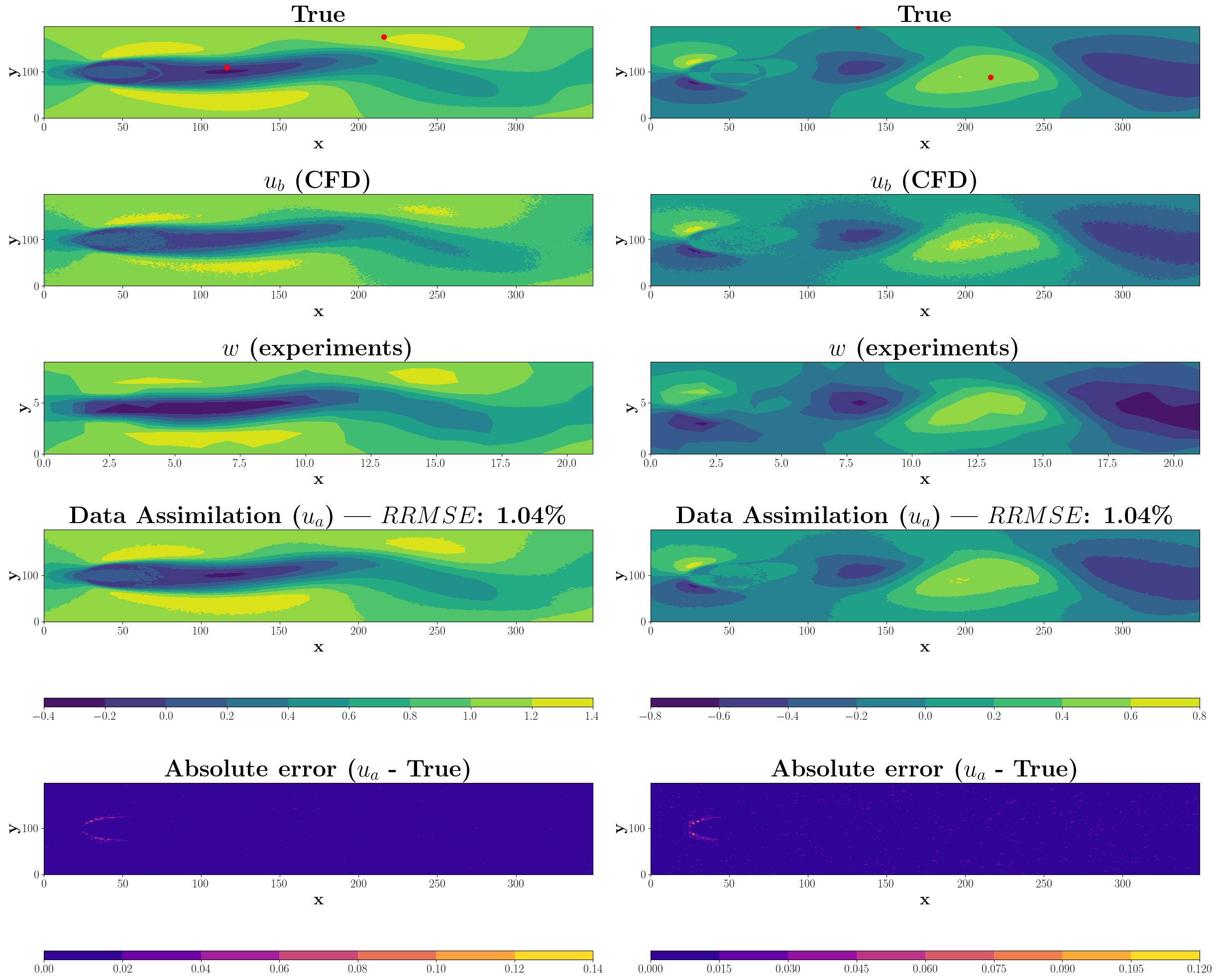}
  \caption{Case 10 in table \ref{tab:dims}. Cylinder dataset at $Re = 100$ results with 5\% of noise level applied on both $u_b$ and $w$, with lcSVD reconstruction method. $N_{s, u_b} \times K = 45000 \times 151$ ($C_{R,u_b} = 3.10$) and $N_{s, w} \times K = 440 \times 151$ ($C_{R,w} = 317$). From the top to the bottom: True data, CFD data, experimental data, DA reconstruction, absolute error ($| u_a - \text{True}|$). The left column represents the streamwise velocity, and the right column represents the normal velocity.}
  \label{Re100ResNoise5bis}
\end{figure}

Then, we compare the previous HR results with the LR computations. Figures \ref{Re100ResNoise5} and \ref{Re100ResNoise5bis} correspond to cases 7 and 10 in table \ref{tab:dims}, with the number of points given as $N_{s, u_b} \times K = 1800 \times 151$ ($C_{R,u_b} = 77$) and $N_{s, u_b} \times K = 45000 \times 151$ ($C_{R,u_b} = 3.10$) respectively. In both cases, $N_{s, w} \times K = 440 \times 151$ ($C_{R, w} = 317$), with noise levels of 5\% applied to both $u_b$ and $w$, respectively. The figures are shown at a single time instant, $t = 150$.

The results presented in table \ref{tab:summarySVD_Re100} offer a comprehensive comparison of computational efficiency and accuracy for the cylinder dataset at $Re = 100$. It provides results for the dataset with different levels of noise, comparing in detail the performance of lcSVD reconstruction with various downsampling configurations. For the HR method at 5\% of noise, the computation time is 1977 seconds, and the RAM usage is 297.33 GB. This high computational cost reflects the demands of the full dataset without dimensionality reduction or compression.  

LR methods drastically improve computational efficiency. For instance, using the $C_{R,u_b} = 317$ downsampling configuration, the computation time is reduced to 37.9 seconds, giving a speed-up $S = 53.2$ compared to the HR method. RAM usage drops to approximately 0.78 GB for the same case. With the $C_{R,u_b} = 3.10$ configuration, the computation time is 328.4 seconds, and RAM usage is around 31.90 GB and the speed-up $S=6.2$. This substantial decrease in computation time and memory usage shows the effectiveness of LR methods in managing computational resources.  

Despite these reductions, low-resolution methods maintain acceptable levels of accuracy. For the HR method, the $MAE$ and $RRMSE$ are 0.59\% and 0.97\%, respectively, for a noise level of 5\%. Table \ref{tab:summarySVD_Re100} shows that the LR method maintains low error metrics across its configurations. For example, the $C_{R,u_b} = 3.10$ configuration achieves a $MAE$ of 0.59\% and an $RRMSE$ of 1.04\%. These values are relatively close to those reported for the HR method, illustrating the ability of LR techniques to preserve or even enhance accuracy despite substantial data compression. The $C_{R,u_b} = 317$ downsampling configuration achieves a $MAE$ of 0.60\% and an $RRMSE$ of 8.56\%, indicating that this configuration considerably influences the DA results.  

Considering a good compromise between accuracy, computation time, and memory usage, the $C_{R,u_b} = 12.4$ downsampling configuration is particularly noteworthy. It achieves a computation time of 74.4 seconds with a RAM usage of 2.70 GB while maintaining high accuracy with a $MAE$ of 0.6\% and an $RRMSE$ of 1.26\%. This balance between speed, memory efficiency, and accuracy underscores the LR method as a powerful and efficient alternative for managing large datasets with reduced computational resources compared to the full dataset approach.  

At high compression rates, the LR method demonstrates robust results under high noise levels (50\%). This is especially relevant for large-scale noisy CFD applications where data compression is crucial. While the HR method achieves an $MAE$ of 4.14\% and an $RRMSE$ of 7.84\%, the highest compression rate ($C_{R,u_b} = 17413$) achieves an $MAE$ of 6.66\% and an $RRMSE$ of 51.31\%, which is on the same order of magnitude as the noise level. Similarly, at $C_{R,u_b} = 7739$, the $MAE$ is reduced to 6.37\% with an $RRMSE$ of 41.77\%, also in the same order of magnitude as the noise level. At $C_{R,u_b} = 4353$, the LR method achieves an $RRMSE$ of 39.04\%, again comparable to the noise level. At lower compression rates, such as $C_{R,u_b} = 3.10$, where more data is retained, the performance gap narrows significantly, with an $MAE$ of 6.29\% and an $RRMSE$ of 9.86\%.

These results confirm that even at 50\% noise, the LR method yields performance levels that are remarkably close to those observed under 5\% noise conditions. LR method with lcSVD reconstruction is a promising tool to reconstruct simple flows and noisy cases (by filtering out noise), while keeping the computational needs lower than HR method, making it a valuable option for large-scale applications.

\begin{table*}[h]
  \centering
  \adjustbox{valign=c, max width=\textwidth}{
    \begin{tabular}{|c|c|c|c|c|c|c|c|c|c|c|c|}
      \hline
      \multicolumn{1}{|c|}{Noise (\%)} &
      \multicolumn{1}{|c|}{Case} &
      \multicolumn{1}{c|}{$C_{R,u_b}$} &
      \multicolumn{1}{c|}{$C_{R,w}$} &
      \multicolumn{1}{c|}{$N_{s, u_b} \times K$} &
      \multicolumn{1}{c|}{$N_{s, w} \times K$} &
      \multicolumn{1}{c|}{$t_{comp}$ (s)} &
      \multicolumn{1}{c|}{$S$} &
      \multicolumn{1}{c|}{$MAE$ (\%)} &
      \multicolumn{1}{c|}{$RRMSE$ (\%)} &
      \multicolumn{1}{c|}{RAM (GB)} &
      \multicolumn{1}{c|}{$R_{\text{comp}} (\%)$} \\ \hline
      2  & 1 & 1 & 317 & $21034300$  & $66440$   & 1977 & 1 & 0.45 & 0.90 & 297.33 & -\\ \hline
      \multirow{5}{*}{5} 
        &2& 1  & 317   & $21034300$ & $66440$   & 1977  & 1  & 0.59  & 0.97 & 297.33 & -\\ 
        &3 & 17413  & 317 & $1208$  & $66440$   & 1583 & 1.25 & 0.60 & 51.1 & 0.61 & 99.79\\
        &4 & 7739 & 317   & $2718$  & $66440$   & 1581 & 1.25 & 0.60 & 41.3  & 0.67 & 99.78\\
        &5 & 4353  & 317  & $4832$  & $66440$   & 240 & 8.2 & 0.60  & 38.6 & 0.71 & 99.76\\
        &6& 317   & 317   & $66440$   & $66440$   & 37.9   & 53.2 & 0.60  & 8.56 & 0.78 & 99.74\\ 
        &7& 77  & 317   & $271800$  & $66440$   & 59.5   & 33.9 & 0.60  & 3.54   & 0.80 & 99.73\\ 
        &8& 49.8  & 317   & $422800$  & $66440$   & 62.9   & 32.1 & 0.60  & 2.33   & 0.85 & 99.71\\ 
        &9& 12.4  & 317   & $1691200$ & $66440$   & 74.4   & 27.2 & 0.60  & 1.26   & 2.70 & 99.09\\ 
        &10& 3.10  & 317   & $6795000$ & $66440$   & 328.4  & 6.2  & 0.59  & 1.04   & 31.90 & 89.27\\
         \hline
      10 &11& 1    & 317    & $21034300$   & $66440$   & 1977 & 1 & 0.93 & 1.77  & 297.33 & -\\ \hline
      30 &12& 1    & 317   & $21034300$   & $66440$   & 1977 & 1 & 2.51  & 4.75  & 297.33 & -\\ \hline
      \multirow{5}{*}{50}
        &13& 1    & 317   & $21034300$   & $66440$   & 1977 & 1 & 4.14  & 7.84  & 297.33 & -\\
        &14& 17413    & 317   & $1208$   & $66440$   & 1583 & 1.25 & 6.66  & 51.31  & 0.61 & 99.79\\
        &15& 7739    & 317   & $2718$   & $66440$   & 1581 & 1.25 & 6.37  & 41.77  & 0.67 & 99.78\\
        &16& 4353    & 317   & $4832$   & $66440$   & 240 & 8.2 & 6.03  & 39.04  & 0.71 & 99.76\\
        &17& 49.8    & 317   & $422800$   & $66440$   & 62.9 & 32.1 & 6.29  & 9.86  & 0.85 & 99.71\\
        &18& 3.10  & 317   & $6795000$ & $66440$   & 328.4  & 6.3  & 5.97  & 8.84 & 31.90 & 89.27\\
        \hline

    \end{tabular}
  }
  \caption{Merged results for the circular cylinder dataset at $Re = 100$, combining HR and LR (lcSVD) methods. From left to right: noise rate, case, data compression rates $C_{R,u_b}$ and $C_{R,w}$, shape of the data collected by the sensors selecting $N_{s,u_b}$ equidistant data points of $u_b$ multiplied by $K$ snapshots, shape of the data collected by the sensors selecting $N_{s,w}$ equidistant data points of $w$ multiplied by $K$ snapshots, computation time, speed-up, $MAE$, $RRMSE$, RAM usage, and RAM compression rate $R_{\text{comp}}$ (\%).}
  \label{tab:summarySVD_Re100}
\end{table*}

\begin{figure}[H]
  \centering
  \includegraphics[width=1\textwidth]{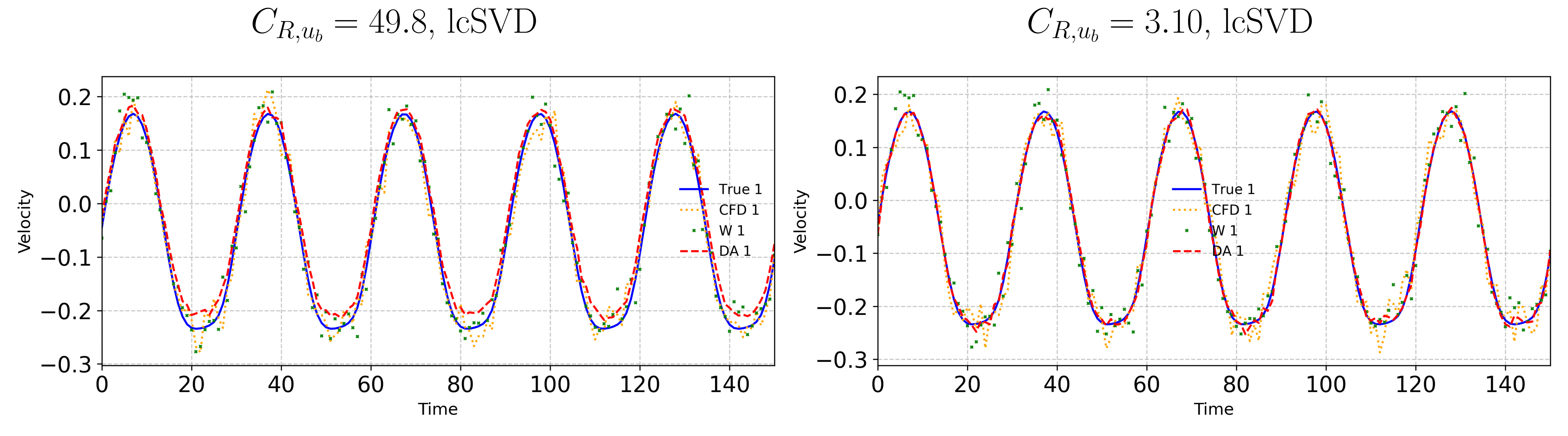}
  \caption{Velocity tracking point (1st red dot from the left side, column 1 in figures \ref{Re100ResNoise5} and \ref{Re100ResNoise5bis}) circular cylinder dataset at $Re = 100$, with 5\% of noise level, for $C_{R,u_b}$ = 3.10, $C_{R,u_b}$ = 49.8 with $C_{R,w}$ = 317. True data is shown in blue, $u_a$ in yellow, $w$ in green, and the assimilated data in red.}
  \label{2trackingVelRe100LR}
\end{figure}

\begin{figure}[H]
  \centering
  \includegraphics[width=1\textwidth]{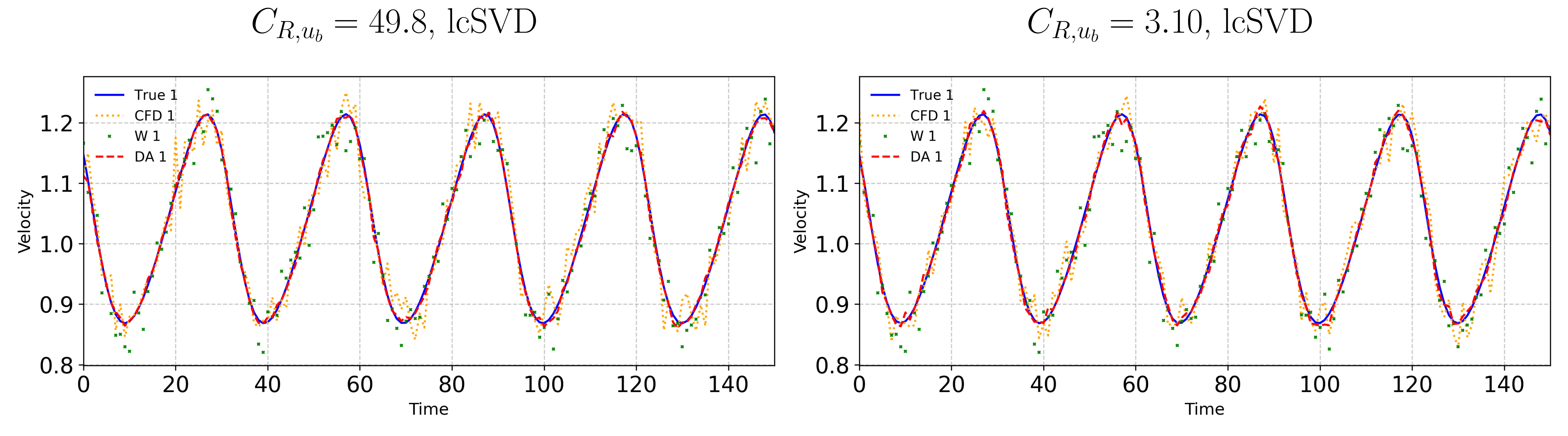}
  \caption{Velocity tracking point (2nd red dot from the left side, column 1 in figures \ref{Re100ResNoise5} and \ref{Re100ResNoise5bis}) circular cylinder dataset at $Re = 100$, with 5\% of noise level, for $C_{R,u_b}$ = 3.10, $C_{R,u_b}$ = 49.8 with $C_{R,w}$ = 317. True data is shown in blue, $u_a$ in yellow, $w$ in green, and the assimilated data in red.}
  \label{3trackingVelRe100LR}
\end{figure}

Figures \ref{2trackingVelRe100LR}, and \ref{3trackingVelRe100LR} demonstrate the influence of the compression rate on the accuracy of True state estimation. With a low $C_{R,u_b} = 3.10$, the DA estimate aligns much more closely with the True state, achieving an accuracy comparable to the HR method, as shown in figures \ref{2trackingVelRe100}, and \ref{3trackingVelRe100}.

The case of $C_{R,u_b} = 49.8$ is particularly noteworthy, as the estimation accuracy depends significantly on the distribution of tracking points. In figure \ref{2trackingVelRe100LR}, the estimated state is almost indistinguishable from the True state, with only minor discrepancies, such as slight distortions in the sinusoidal shape, primarily near the troughs. This performance is close to that achieved with $C_{R, u_b} = 3.10$. Additionally, in figure \ref{3trackingVelRe100LR}, the estimated states for both sets of tracking points closely match the True curve, demonstrating an accurate approximation of the True state.

\subsection{Turbulent circular cylinder at $Re = 2600$}\label{Re2600}

The next dataset studied is the turbulent circular cylinder at $Re = 2600$ detailed in section \ref{testcase3}. The cases presented were chosen to assess the main study's objectives. Cases 31 and 32 illustrate the method's robustness under significant noise contamination, while cases 25 and 29 confirm the effectiveness of the LR approach that maintain solution accuracy and increase computational efficiency. Additionally, cases 33-36 in table \ref{tab:summaryRe2600_SVD} offer further validation of the LR method's ability to preserve accuracy under elevated noise conditions, showing reliable performance across different compression ratios even at 50\% noise.

\begin{figure}[H]
  \centering
  \includegraphics[width=1\textwidth]{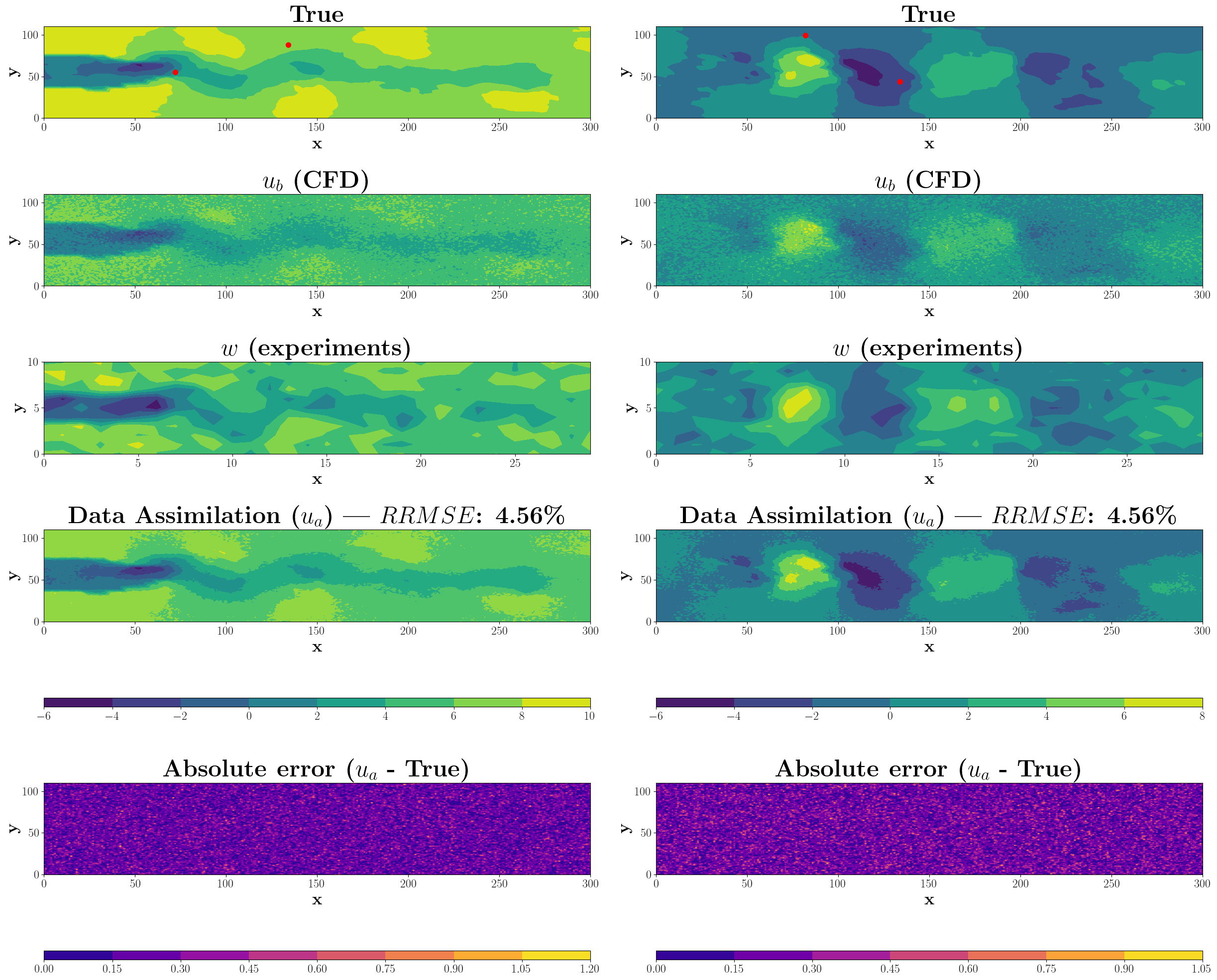}
  \caption{Case 31 in table \ref{tab:dims}. Turbulent cylinder dataset at $Re = 2600$ results with 30\% of noise level applied on both $u_b$ and $w$, HR method. $N_{s, u_b} \times K = 66822 \times 151$ ($C_{R,u_b} = 1$) and $N_{s, w} \times K = 660 \times 151$ ($C_{R,w} = 101$). From the top to the bottom: True data, CFD data, experimental data, DA, absolute error ($| u_a - \text{True}|$). The left column represents the streamwise velocity, and the right column represents the normal velocity.}
  \label{Re2600ResNoise30}
\end{figure}

\begin{figure}[H]
  \centering
  \includegraphics[width=1\textwidth]{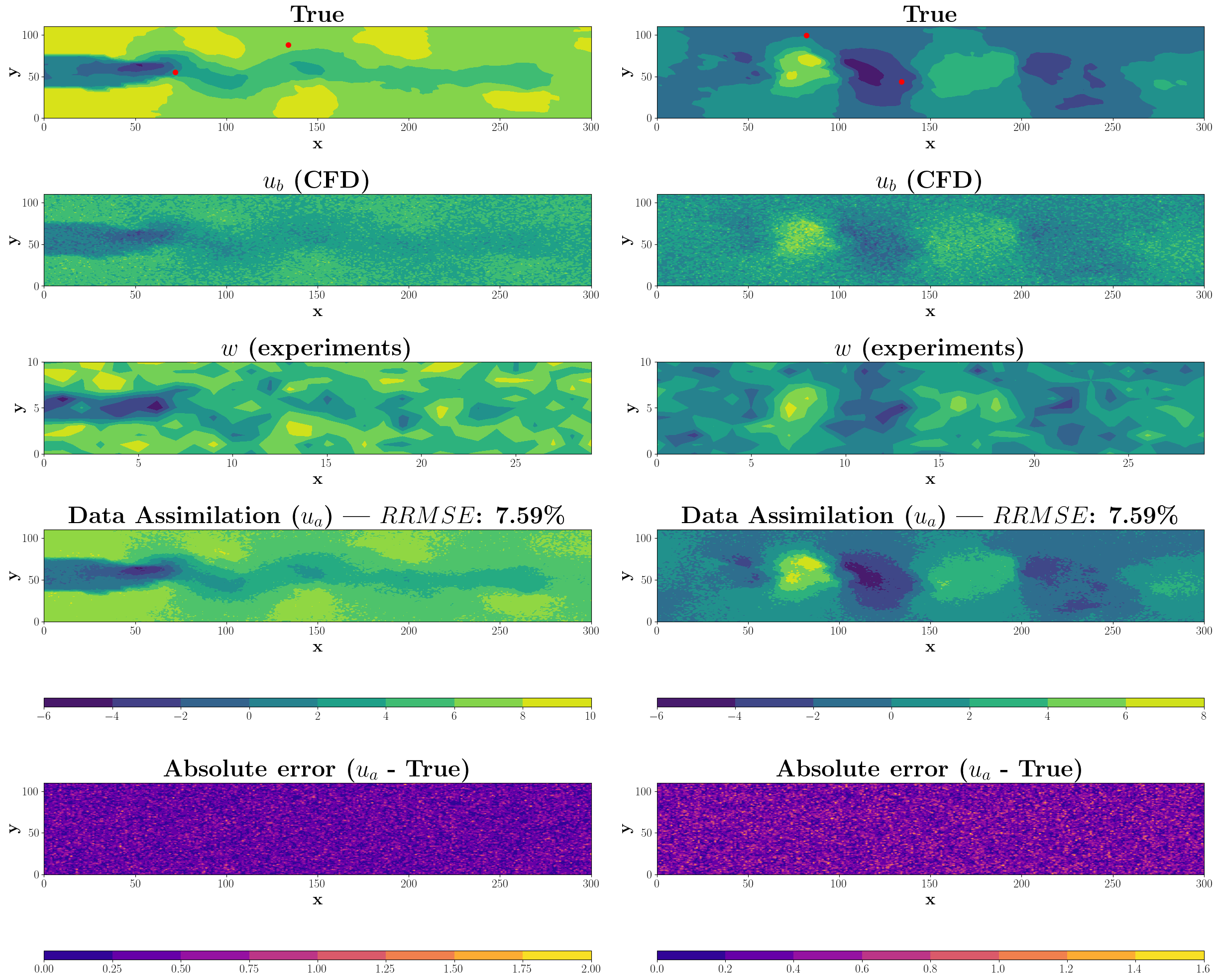}
  \caption{Case 32 in table \ref{tab:dims}. Turbulent cylinder dataset at $Re = 2600$ results with 50\% of noise level applied on both $u_b$ and $w$, HR method. $N_{s, u_b} \times K = 66822 \times 151$ ($C_{R,u_b} = 1$) and $N_{s, w} \times K = 660 \times 151$ ($C_{R,w} = 101$). From the top to the bottom: True data, CFD data, experimental data, DA, absolute error ($| u_a - \text{True}|$). The left column represents the streamwise velocity, and the right column represents the normal velocity.}
  \label{Re2600ResNoise50}
\end{figure}

\begin{figure}[H]
  \centering
  \includegraphics[width=1\textwidth]{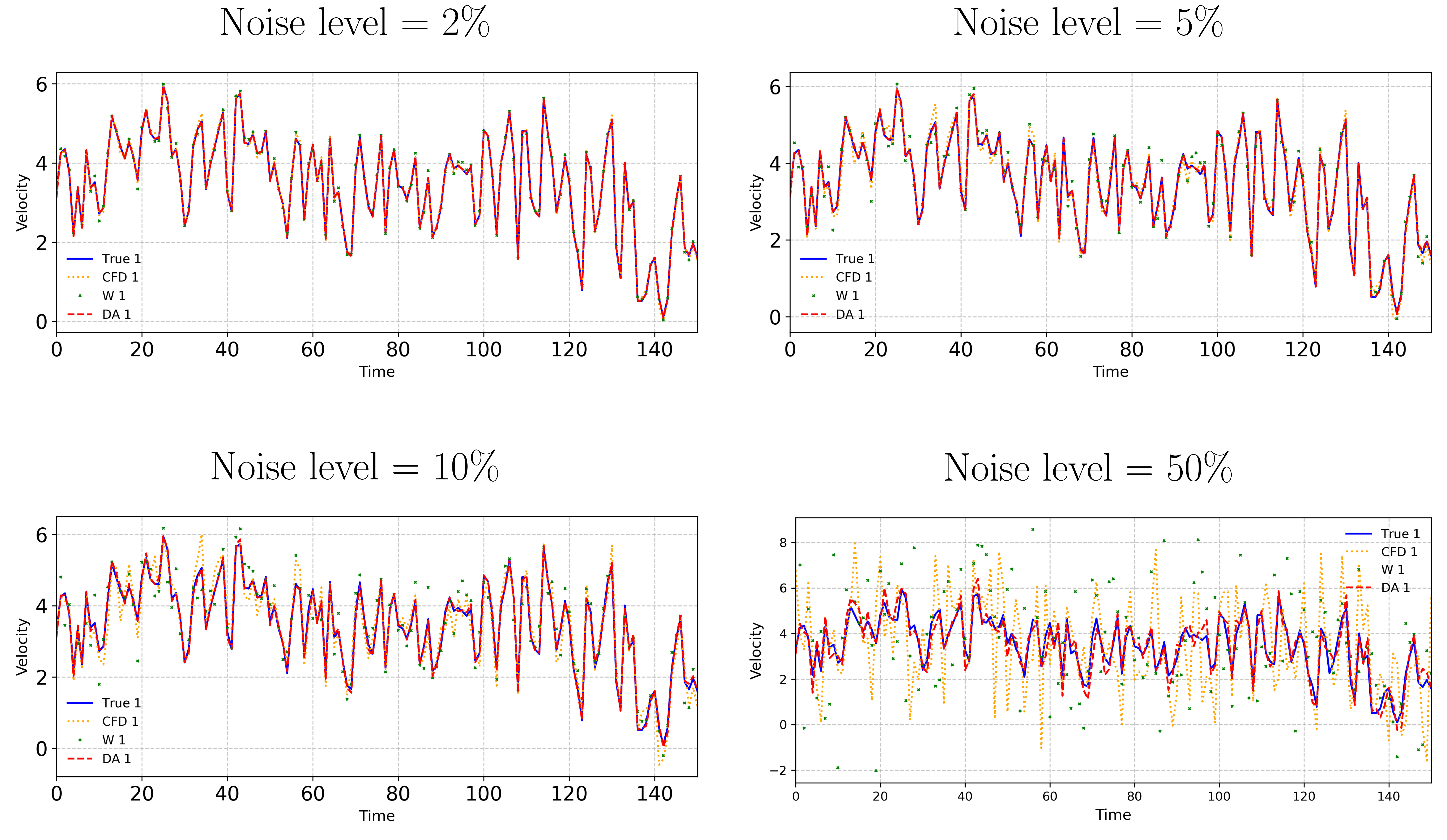}
  \caption{Velocity tracking point (1st red dot from the left side, column 1 in figures \ref{Re2600ResNoise30} and \ref{Re2600ResNoise50}) turbulent cylinder dataset at $Re = 2600$, with noise levels from 2\% to 50\%, and $C_{R,w}$ = 101. True data is shown in blue, $u_a$ in yellow, $w$ in green, and the assimilated data in red.}
  \label{2trackingVelRe2600}
\end{figure}

\begin{figure}[H]
  \centering
  \includegraphics[width=1\textwidth]{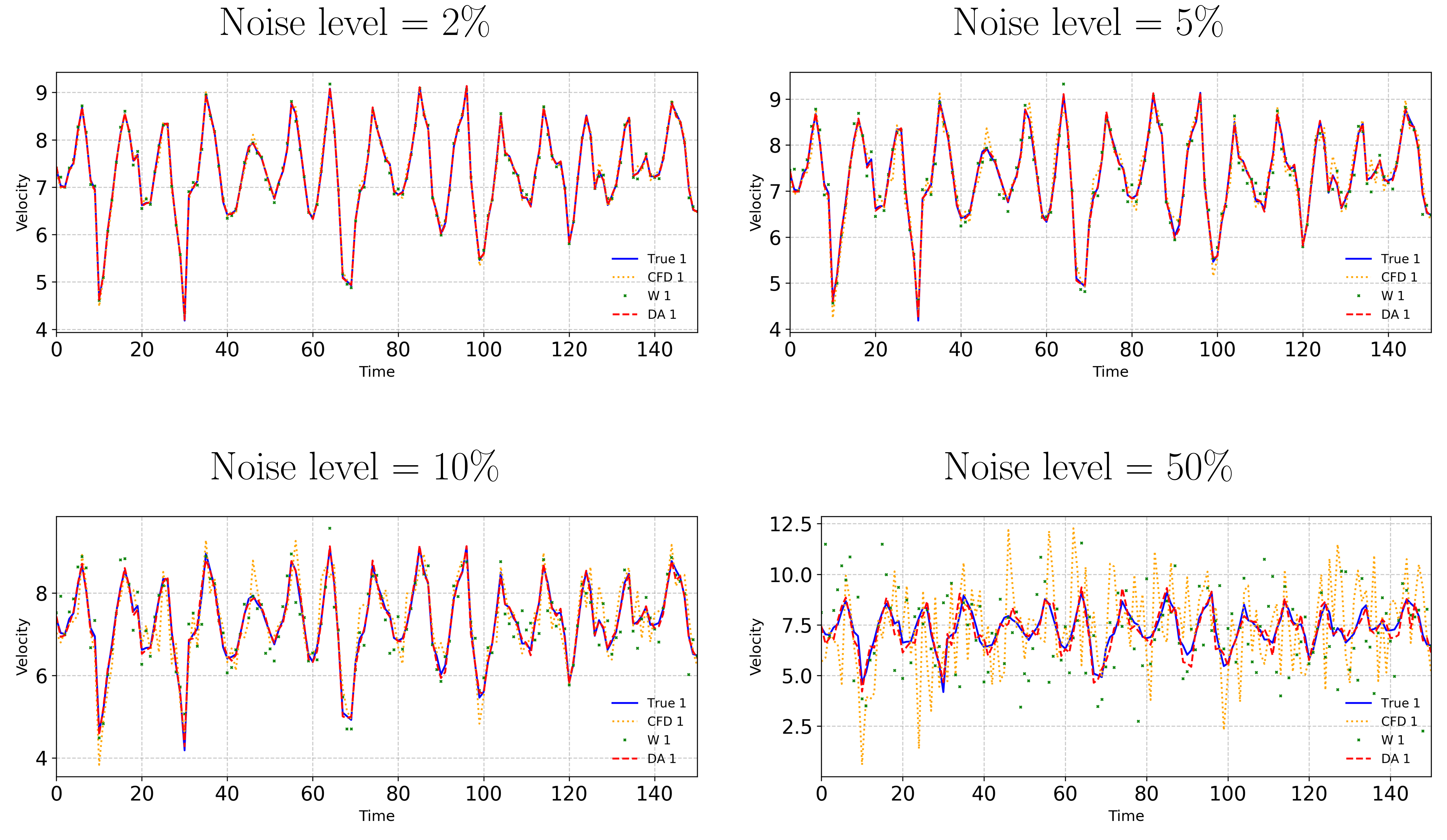}
  \caption{Velocity tracking point (2nd red dot from the left side, column 1 in figures \ref{Re2600ResNoise30} and \ref{Re2600ResNoise50}) turbulent cylinder dataset at $Re = 2600$, with noise levels from 2\% to 50\%, and $C_{R,w}$ = 101. True data is shown in blue, $u_a$ in yellow, $w$ in green, and the assimilated data in red.}
  \label{3trackingVelRe2600}
\end{figure}

The cases 31 and 32 in table \ref{tab:dims}, with a number of points $N_{s, u_b} \times K = 66822 \times 151$ ($C_{R,u_b} = 1$) and $N_{s, w} \times K = 660 \times 151$ ($C_{R,w} = 101$), with 30\% and 50 \% of noise level applied on both $u_b$ and $w$, respectively for figures \ref{Re2600ResNoise30} and \ref{Re2600ResNoise50} fixed at one time instant $t = 150$, are showed to demonstrate its estimation capabilities under high noise levels. This test case contains fewer points than the previous one in section \ref{Re100} (table \ref{tab:dims}). The goal of this dataset is to observe the performances with a turbulent test case. Also, here $N_{s, w} \times K = 660 \times 151$ ($C_{R, w} = 101$), which is higher than circular cylinder at $Re = 100$ ($N_{s, w} \times K = 440 \times 151$ with $C_{R, w} = 317$), so we have more information from $w$ in this test case. The results in the figures are cross-referenced with table \ref{tab:summaryRe2600_SVD} to quantify the results of this test case. As shown in figures \ref{2trackingVelRe2600} and \ref{3trackingVelRe2600}, the performances of the DA for estimating the True state is still relatively accurate when comparing it with the velocity tracking points, like with the results in section \ref{Re100}, regardless of the noise level applied to the original tensor. 

For noise levels of 2\%, 5\%, 10\%, 30\%, and 50\%, the $RRMSE$ reaches 0.45\%, 0.83\%, 1.55\%, 4.56\%, and 7.59\% respectively, while the $MAE$ is 1.35\%, 3.21\%, 6.31\%, 18.70\%, and 31.10\%, respectively.  The $RRMSE$ remains well below the noise level for all the cases, highlighting the strong denoising capabilities of the EnKF method to estimate the True state of the system even in high-noise regimes.

\begin{figure}[H]
  \centering
  \includegraphics[width=1\textwidth]{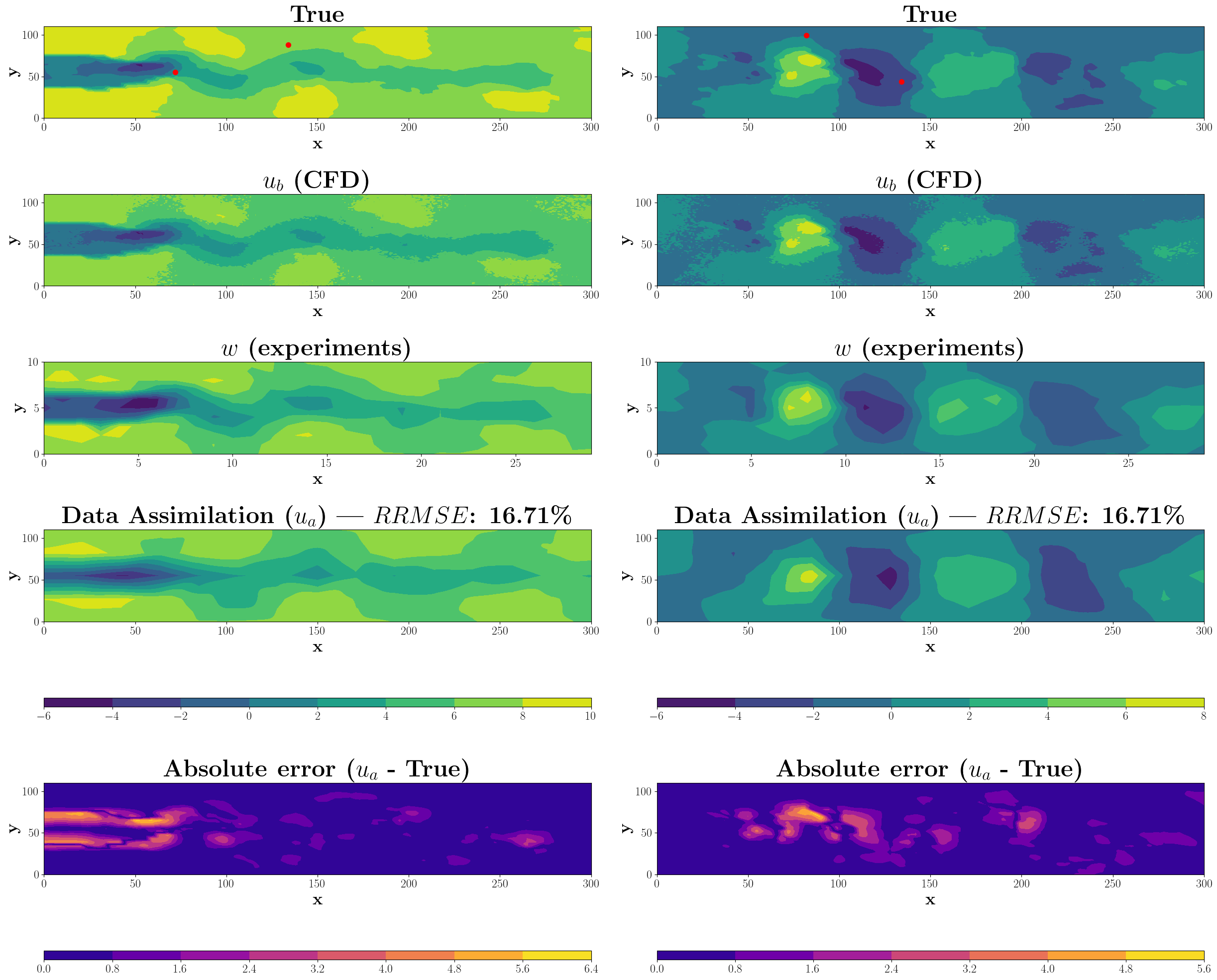}
  \caption{Case 25 in table \ref{tab:dims}. Turbulent cylinder dataset at $Re = 2600$ results with 5\% of noise level on both $u_b$ and $w$, lcSVD reconstruction method. $N_{s, u_b} \times K = 150 \times 151$ ($C_{R,u_b} = 445$) and $N_{s, w} \times K = 660 \times 151$ ($C_{R,w} = 110$). From the top to the bottom: True data, CFD data, experimental data, DA, absolute error ($u_a - \text{True}$). The left column represents the streamwise velocity, and the right column represents the normal velocity.}
  \label{Re2600ResNoise5}
\end{figure}

\begin{figure}[H]
  \centering
  \includegraphics[width=1\textwidth]{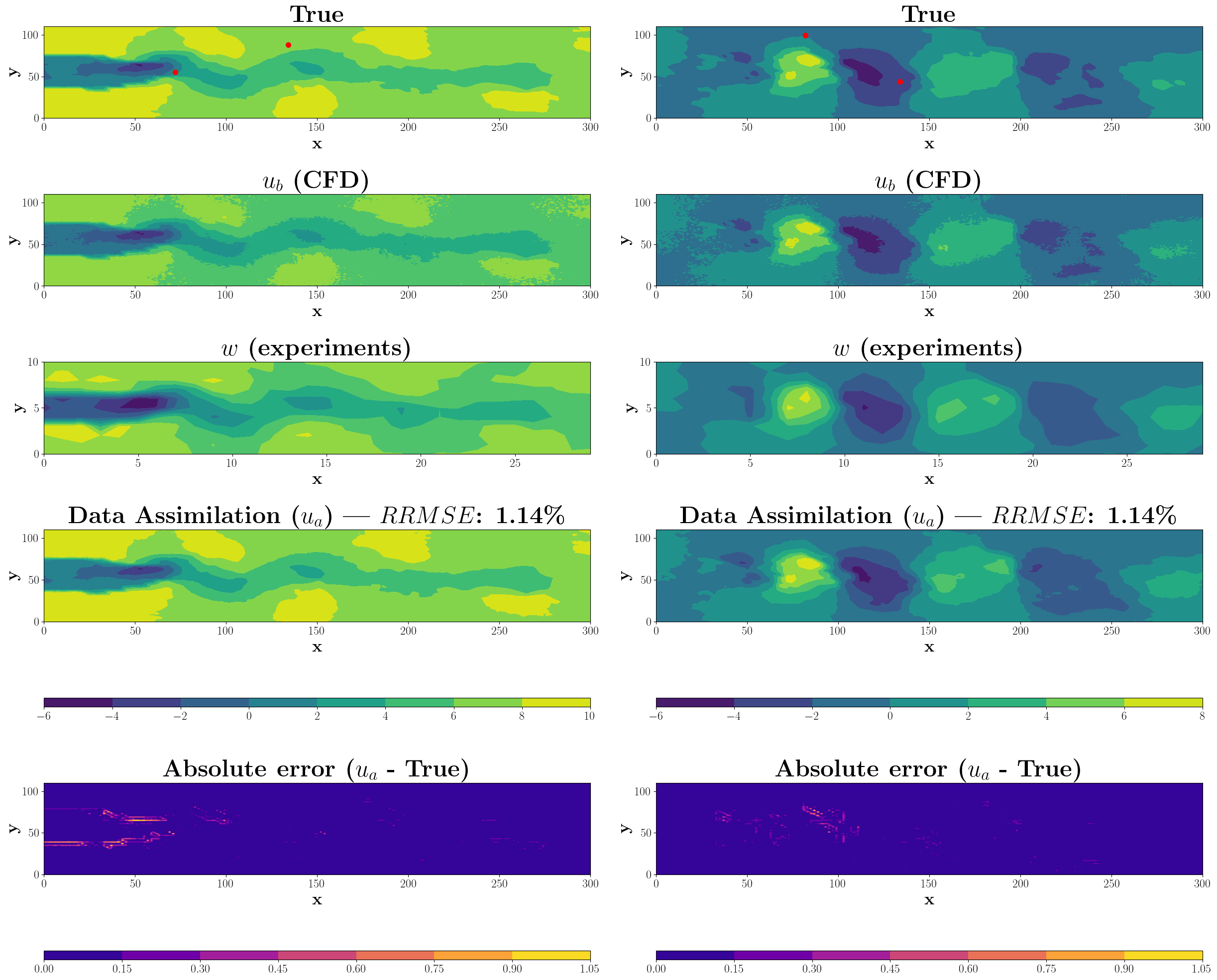}
  \caption{Case 29 in table \ref{tab:dims}. Turbulent cylinder dataset at $Re = 2600$ results with 5\% of noise level applied on both $u_b$ and $w$, lcSVD reconstruction method. $N_{s, u_b} \times K = 16800 \times 151$ ($C_{R,u_b} = 3.98$) and $N_{s, w} \times K = 660 \times 151$ ($C_{R,w} = 101$). From the top to the bottom: True data, CFD data, experimental data, DA, absolute error ($u_a - \text{True}$). The left column represents the streamwise velocity, and the right column represents the normal velocity.}
  \label{Re2600ResNoise5bis}
\end{figure}

We then contrast the HR results with the LR computations. Figures \ref{Re2600ResNoise5} and \ref{Re2600ResNoise5bis} correspond to cases 25 and 29 in table \ref{tab:dims}, with the number of points given as $N_{s, u_b} \times K = 150 \times 151$ ($C_{R,u_b} = 445$) and $N_{s, u_b} \times K = 16800 \times 151$ ($C_{R,u_b} = 3.98$), with a noise level of 5\% applied to both $u_b$ and $w$, respectively. The figures are shown at a single time instant, $t = 150$. This test case presents similar results compared to the circular cylinder at $Re = 100$ for LR computations.

The results in table \ref{tab:summaryRe2600_SVD} provide a detailed comparison of the performance of lcSVD reconstruction method compared to the HR for the cylinder dataset at \(Re = 2600\). Both methods exhibit significant improvements in computational efficiency compared to standard approaches without downsampling. The HR method at 5\% of noise shows a computation time $t_{comp} = 453.3$ seconds and a RAM usage of 37.35 GB.

LR methods significantly optimize computation time and memory usage. For instance, with a $C_{R,u_b} = 445$ downsampling configuration, lcSVD reduces the computation time to 21.3 seconds, resulting in a speed-up $S=21.3$ compared to the HR method. Similarly, RAM usage decreases notably, from over 37.35 GB for the full dataset to 3.27 GB. These reductions demonstrate the efficiency gains of LR methods when processing large datasets.

Despite these performance improvements, LR methods maintain a high level of accuracy. The $MAE$ and $RRMSE$ metrics confirm that the method preserves the accuracy of the original data despite significant compression. Again, with a $C_{R,u_b} = 445$ downsampling configuration, the $MAE$ and the $RRMSE$ reach 3.31\% and 16.71\%, respectively. For even more accuracy, the $C_{R,u_b} = 3.98$ downsampling configuration shows a $MAE$ of 3.21\% and an $RRMSE$ of 1.14\%, closely matching the results for the same configuration using the HR method, which has a $MAE$ of 3.21\% and an $RRMSE$ of 0.83\%. With a high compression rate, such as the $C_{R,u_b} = 1591$ configuration, the $MAE$ achieves 3.32\% and the $RRMSE$ is 31.17\%. However, the computation time for lcSVD goes up to 175.9 seconds for this $C_{R,u_b}$.

The $C_{R,u_b} = 15.9$ downsampling configuration is a particularly noteworthy trade-off case. It achieves a computation time of 33.1 seconds with a RAM usage of 3.40 GB while maintaining high accuracy with a $MAE$ of 3.23\% and an $RRMSE$ of 2.61\%. This trade-off between computational speed, memory savings, and predictive accuracy highlights the LR method as an effective and resource-efficient solution for handling large-scale datasets compared to the full-resolution approach.

For this turbulent dataset, lcSVD provides promising results under a 50\% noise level. With the HR method, we achieve an $MAE$ of 31.10\% and an $RRMSE$ of 7.59\%. At the highest compression rate ($C_{R,u_b} = 8353$), the LR method yields an $MAE$ of 33.31\% and an $RRMSE$ of 41.71\%, which is close to the noise magnitude. Similarly, at $C_{R,u_b} = 2088$, we obtain an $MAE$ of 31.30\% and an $RRMSE$ of 38.16\%, also within the same order of magnitude as the imposed noise level. For a lower compression rate ($C_{R,u_b} = 62.8$), the method yields an $MAE$ of 32.14\% and $RRMSE$ of 8.31\%. Finally, with a $C_{R,u_b} = 3.98$, we obtain an $MAE$ of 32.13\% and $RRMSE$ of 6.66\%. These findings highlight that, for this turbulent dataset, lcSVD performs consistently across all compression rates, and can effectively handle high noise levels. The results confirm that even at 50\% noise level, the LR method is showing performance levels that are in the same order to the one observed at 5\% noise level and with HR results at 50\% of noise level. This has been achieved thanks to the noise filtering capabilities of the lcSVD reconstruction, which allows for accurate state estimation even in high-noise regimes.

The LR method combined with lcSVD reconstruction proves to be a promising approach for estimating the True state of turbulent flows from noisy data, while significantly reducing computational costs compared to the HR method. This makes it particularly well-suited for large-scale applications.

\begin{table*}[h]
  \centering
  \adjustbox{valign=c, max width=\textwidth}{
    \begin{tabular}{|c|c|c|c|c|c|c|c|c|c|c|c|}
      \hline
      \multicolumn{1}{|c|}{Noise (\%)} &
      \multicolumn{1}{|c|}{Case} &
      \multicolumn{1}{c|}{$C_{R,u_b}$} &
      \multicolumn{1}{c|}{$C_{R,w}$} &
      \multicolumn{1}{c|}{$N_{s, u_b} \times K$} &
      \multicolumn{1}{c|}{$N_{s, w} \times K$} &
      \multicolumn{1}{c|}{$t_{comp}$ (s)} &
      \multicolumn{1}{c|}{$S$} &
      \multicolumn{1}{c|}{$MAE$ (\%)} &
      \multicolumn{1}{c|}{$RRMSE$ (\%)} &
      \multicolumn{1}{c|}{RAM (GB)} &
      \multicolumn{1}{c|}{$R_{\text{comp}} (\%)$} \\ \hline
      2  & 19 & 1  & 101   & 10090122    & 99660   & 453.3  & 1  & 1.35 & 0.45 & 37.35 & -\\ \hline
      \multirow{6}{*}{5}
        & 20& 1   & 101   & 10090122 & 99660   & 453.3  & 1     & 3.21 & 0.83 & 37.35 & -\\
        & 21& 8353    & 101   & 1208   & 99660   & 2209   & 0.21  & 3.33 & 40.93 & 3.19 & 91.47\\
        &22& 3712    & 101   & 2718   & 99660   & 2102   & 0.22  & 3.33 & 37.62 & 3.20 & 91.42\\
        & 23& 2088    & 101   & 4832   & 99660   & 1628   & 0.28  & 3.33 & 36.93 & 3.22 & 91.38\\
        & 24& 1591    & 101   & 6342   & 99660   & 175.9   & 2.58  & 3.32 & 31.17 & 3.24 & 91.32\\
        & 25& 445    & 101   & 22650   & 99660   & 21.3   & 21.3  & 3.31 & 16.71 & 3.27 & 91.24\\
        & 26& 101  & 101   & 99660   & 99660   & 22.3   & 20.3  & 3.31 & 7.72 & 3.32 & 91.12\\ 
        & 27& 62.8    & 101   & 160664   & 99660   & 24.6   & 18.5  & 3.28 & 5.89 & 3.33 & 91.09\\
        & 28& 15.9   & 101   & 634200  & 99660   & 33.1   & 13.7  & 3.23 & 2.61 & 3.40 & 90.89\\ 
        & 29& 3.98   & 101   & 2536800 & 99660   & 85.9   & 5.3  & 3.21 & 1.14 & 5.51 & 85.25\\ \hline
      10 & 30& 1   & 101   & 10090122 & 99660   & 453.3  & 1     & 6.31 & 1.55  & 37.35 & -\\ \hline
      30 & 31& 1   & 101   & 10090122 & 99660   & 453.3  & 1     & 18.70 & 4.56  & 37.35 & -\\ \hline
      \multirow{5}{*}{50}
        &32& 1   & 101   & 10090122 & 99660   & 453.3  & 1     & 31.10 & 7.59  & 37.35 & -\\
        &33& 8353    & 101   & $1208$   & $99660$   & 2209 & 0.21 & 33.31  & 41.71  & 3.19 & 91.47\\
        &34& 2088    & 101   & $4832$   & $99660$   & 1628 & 0.28 & 31.30  & 38.16  & 3.22 & 91.38\\
        &35& 62.8    & 101   & $160664$   & $99660$   & 24.6 & 18.5 & 32.14  & 8.31  & 3.33 & 91.09\\
        &36& 3.98   & 101   & 2536800 & 99660   & 85.9   & 5.3  & 32.13 & 6.66 & 5.51 & 85.25\\
        \hline
      
    \end{tabular}
  }
  \caption{Merged results for turbulent cylinder dataset at $Re = 2600$ dataset, combining HR and LR (lcSVD) methods. From left to right: noise rate, case, data compression rates $C_{R,u_b}$ and $C_{R,w}$, shape of the data collected by the sensors selecting $N_{s,u_b}$ equidistant data points of $u_b$ multiplied by $K$ snapshots, shape of the data collected by the sensors selecting $N_{s,w}$ equidistant data points of $w$ multiplied by $K$ snapshots, computation time, speed-up, $MAE$, $RRMSE$, RAM usage, and RAM compression rate $R_{\text{comp}}$ (\%).}
  \label{tab:summaryRe2600_SVD}
\end{table*}

\begin{figure}[H]
  \centering
  \includegraphics[width=1\textwidth]{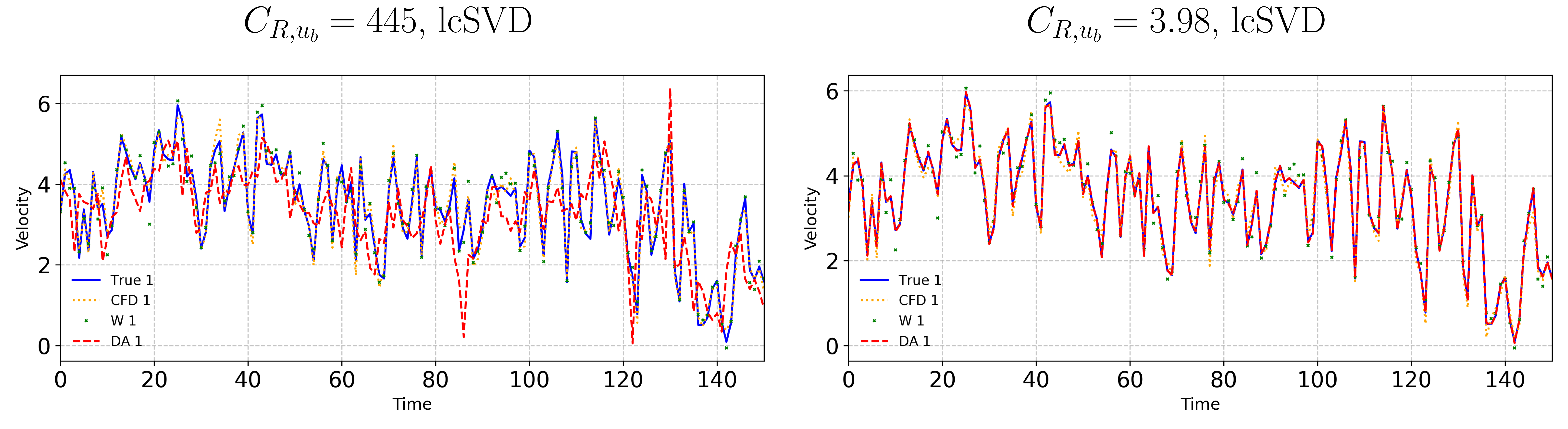}
  \caption{Velocity tracking point (1st red dot from the left side, column 1 in figures \ref{Re2600ResNoise5} and \ref{Re2600ResNoise5bis}) turbulent cylinder dataset at $Re = 2600$, with 5\% of noise level, for $C_{R,u_b}$ = 3.98, $C_{R,u_b}$ = 445 with $C_{R,w}$ = 101. True data is shown in blue, $u_a$ in yellow, $w$ in green, and the assimilated data in red.}
  \label{2trackingVelRe2600LR}
\end{figure}

\begin{figure}[H]
  \centering
  \includegraphics[width=1\textwidth]{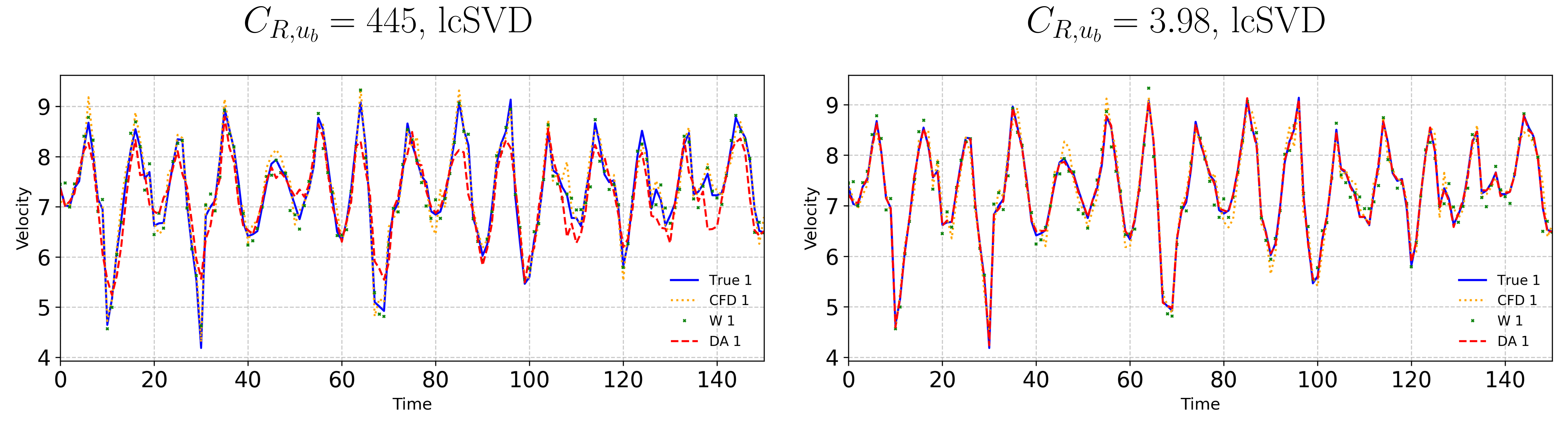}
  \caption{Velocity tracking point (2nd red dot from the left side, column 1 in figures \ref{Re2600ResNoise5} and \ref{Re2600ResNoise5bis}) turbulent cylinder dataset at $Re = 2600$, with 5\% of noise level, for $C_{R,u_b}$ = 3.98, $C_{R,u_b}$ = 445 with $C_{R,w}$ = 101. True data is shown in blue, $u_a$ in yellow, $w$ in green, and the assimilated data in red.}
  \label{3trackingVelRe2600LR}
\end{figure}

Figures \ref{2trackingVelRe2600LR}, and \ref{3trackingVelRe2600LR} illustrate how the compression rate $C_{R, u_b}$ affects the accuracy of True state estimation. When $C_{R, u_b} = 445$, the DA estimate (\( u_b \), red curve) significantly deviates from the True state (\( u_a \), blue curve), indicating poor estimation accuracy, especially when compared to the HR method. However, increasing the number of tracking points to $C_{R, u_b} = 3.98$ results in the DA estimate aligning much closer with the True state, reaching an accuracy comparable to the HR method, as demonstrated in figures \ref{2trackingVelRe2600}, and \ref{3trackingVelRe2600}. Based on both the previous and current cases, it appears that for the cylinder configuration, an optimal compression rate can be found. Specifically, for $C_{R, u_b}$ values close to 4, the LR method achieves results comparable to the HR method, while significantly reducing computational costs.

\subsection{Turbulent jet large eddy}\label{JetLES}

The last dataset studied is the turbulent jet large eddy simulation (presented in section \ref{testcase2}). Cases 48 and 49 validate the method's resilience against substantial noise levels, while cases 45 and 46 establish the viability of LR computations that maintain accuracy while improving computational performance. Moreover, cases 50--53 in table \ref{tab:summarySVD_JetLES} provide supplementary confirmation of the LR method's capacity to retain precision under high noise environments, exhibiting consistent behavior across various compression ratios even at 50\% noise.

\begin{figure}[H]
  \centering
  \includegraphics[width=1\textwidth]{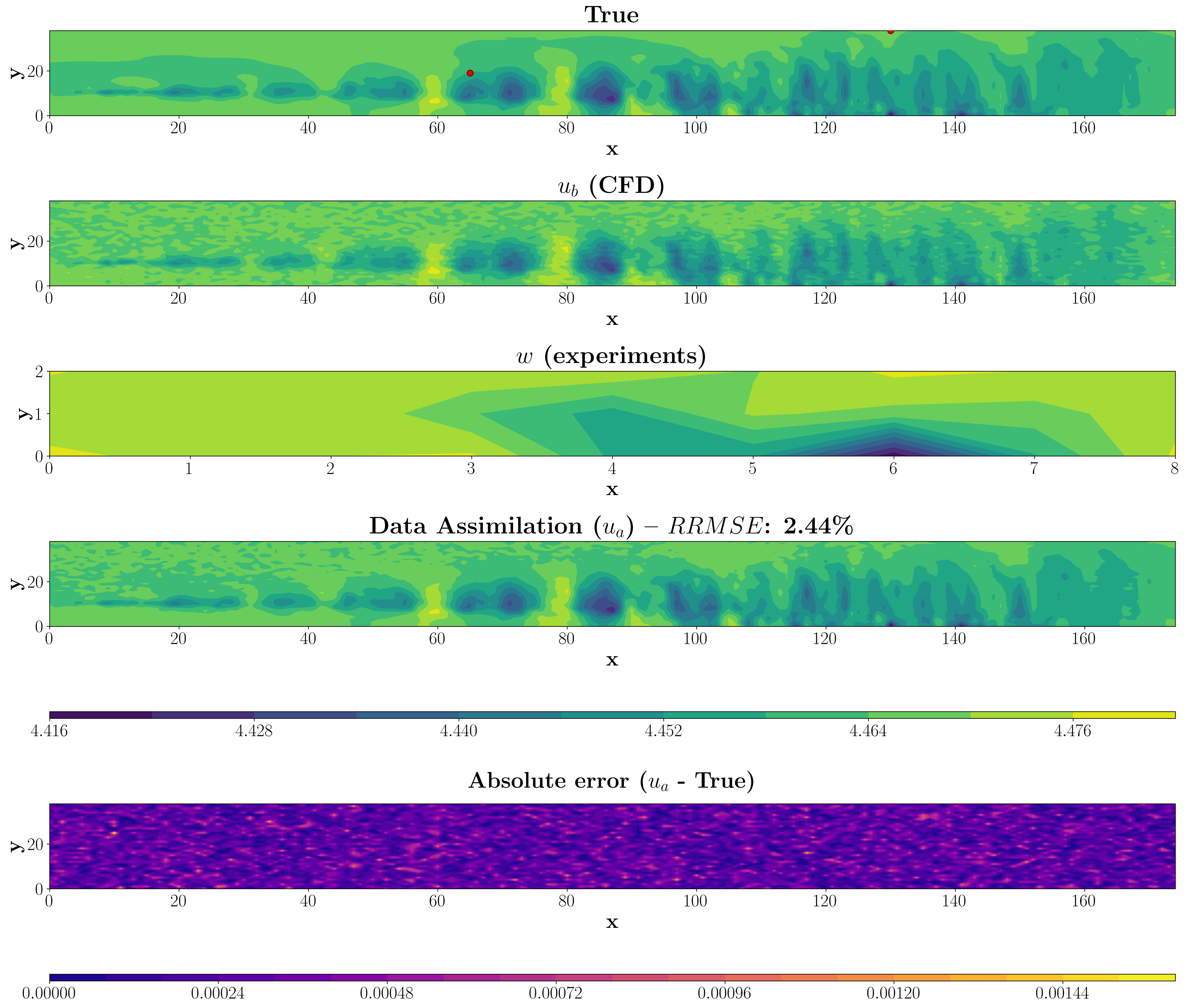}
  \caption{Case 48 in table \ref{tab:dims}. Turbulent jet large eddy (streamwise velocity) results with 30\% of noise level applied on both $u_b$ and $w$, HR method. $N_{s, u_b} \times K = 6825 \times 151$ ($C_{R,u_b} = 1$) and $N_{s, w} \times K = 27 \times 151$ ($C_{R,w} = 253$). From the top to the bottom: True data, CFD data, experimental data, DA, absolute error ($| u_a - \text{True}|$).}
  \label{JetLESnoise30}
\end{figure}

\begin{figure}[H]
  \centering
  \includegraphics[width=1\textwidth]{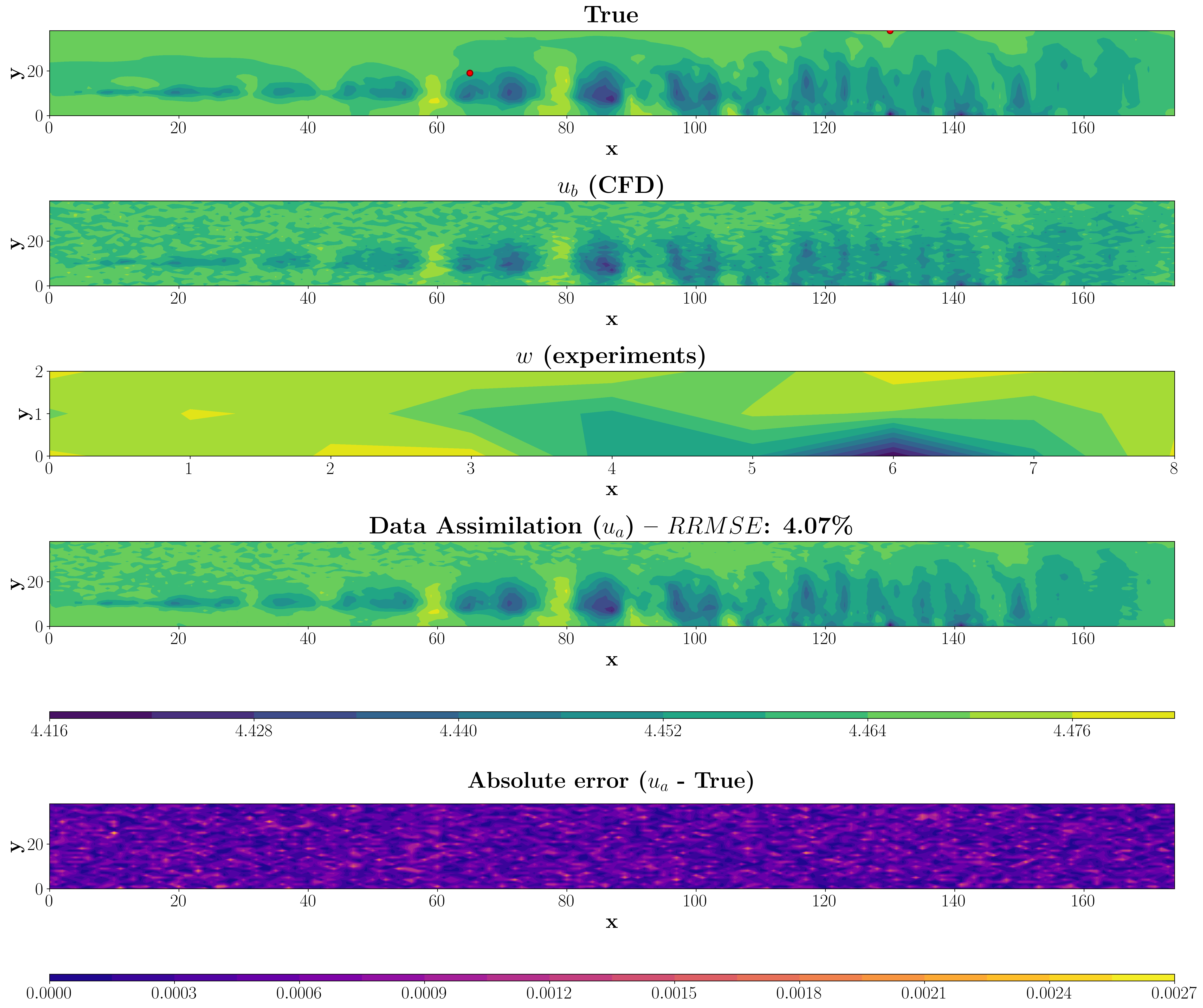}
  \caption{Case 49 in table \ref{tab:dims}. Turbulent jet large eddy (streamwise velocity) results with 50\% of noise level applied on both $u_b$ and $w$, HR method. $N_{s, u_b} \times K = 6825 \times 151$ ($C_{R,u_b} = 1$) and $N_{s, w} \times K = 27 \times 151$ ($C_{R,w} = 253$). From the top to the bottom: True data, CFD data, experimental data, DA, absolute error ($| u_a - \text{True}|$).}
  \label{JetLESnoise50}
\end{figure}

The results of the test cases applied are shown in figures \ref{JetLESnoise30} and \ref{JetLESnoise50}, corresponding to the case 48 and 49 from table \ref{tab:dims}. The difference here, is that the number of points ($N_{s, u_b} \times K$, table \ref{tab:dims}) is very much lower than in the previous cases, as well as the number of experimental points ($N_{s,w} \times K$ = $27 \times 151$). It is indeed interesting to see how the DA method is estimating the True state with a low number of points. We can see that the absolute error ($| u_a - \text{True}|$) is distributed on the whole field, showing that the EnKF is correcting on the whole field at available points and time instants.

By analyzing the results in table \ref{tab:summarySVD_JetLES}, we can quantify how the density of data points significantly shapes the system’s performance.  

While a dataset with fewer points might seem to limit the amount of available information, in cases where noise and uncertainty are low, this data density reduction can actually reduce the system's sensitivity to noise. In this context, the smaller number of points in the experimental dataset ($w$) reduces the impact of random noise, leading to lower uncertainty and a stable error probability. This low number of points significantly decreases the computation time compared to the HR case. This consistency in the error distribution highlights the robustness of the system. While fewer data points might traditionally suggest a higher risk of large errors, in this case, the opposite occurs. The limited number of points leads to reduced exposure to noise, as noise gets distributed more evenly across fewer data points. Consequently, the modifications introduced by data assimilation are moderate, and the system maintains low error probabilities. The Gaussian distribution reinforces this, as the narrow and stable error range, centered around a small peak, shows that the model continues to perform well even under substantial noise.

It is evident that higher noise levels lead to increased values of both $MAE$ and $RRMSE$, although these remain significantly lower than those obtained in the previously studied HR case. For noise levels of 2\%, 5\%, 10\%, 30\%, and 50\%, the $RRMSE$ values are 0.16\%, 0.41\%, 0.81\%, 2.44\%, and 4.07\%, respectively. Correspondingly, the $MAE$ values are $0.05$\%, $0.12$\%, $0.24$\%, $0.73$\%, and $1.21$\%. These results confirm the impressive robustness and efficiency of the EnKF-based estimation, even under substantial noise levels.

\begin{figure}[H]
  \centering
  \includegraphics[width=1\textwidth]{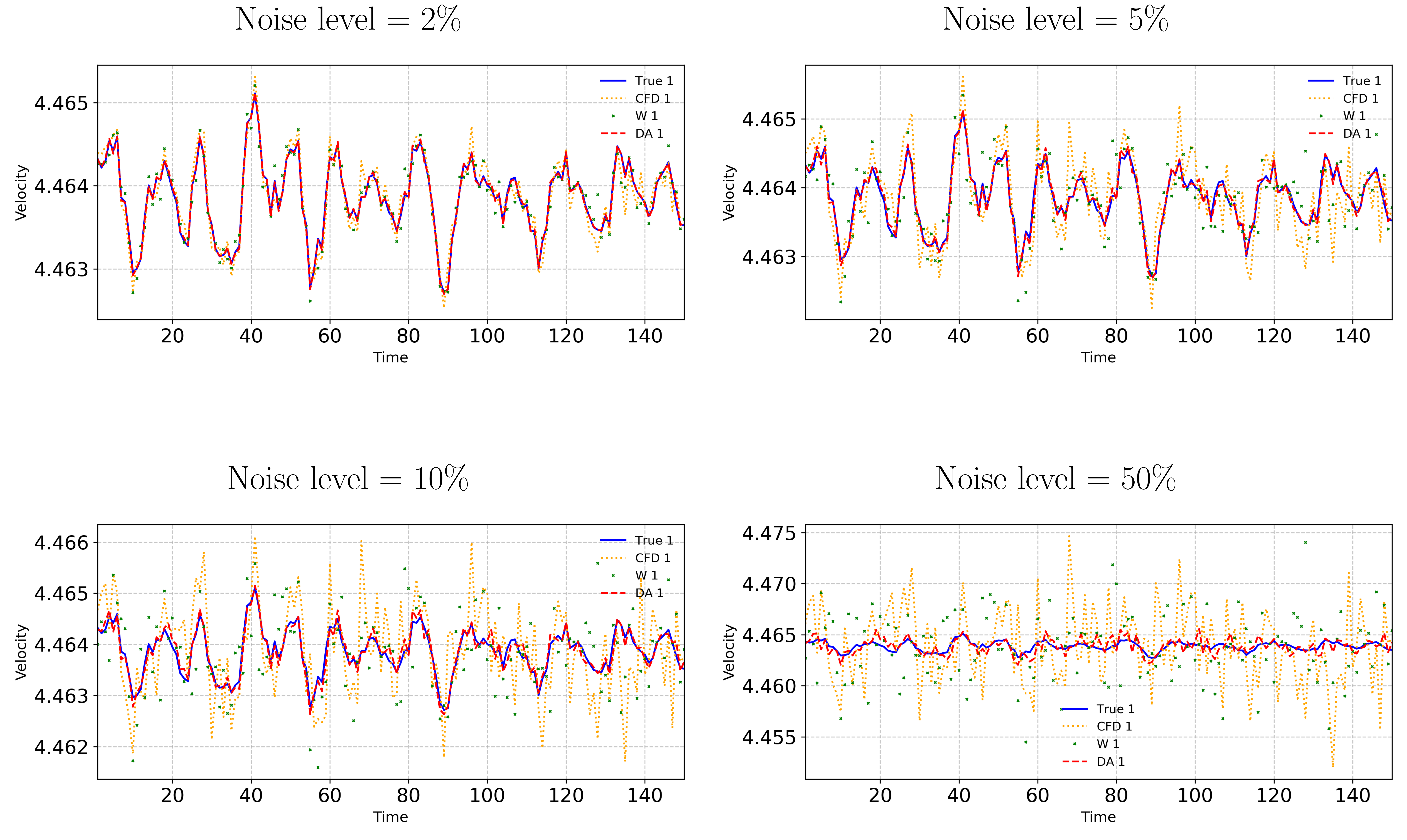}
  \caption{Velocity tracking point (1st red dot from the left side in figures \ref{JetLESnoise30} and \ref{JetLESnoise50}) turbulent jet large eddy, with noise levels from 2\% to 50\%, with $C_{R,w}$ = 253. True data is shown in blue, $u_a$ in yellow, $w$ in green, and the assimilated data in red.}
  \label{1trackingVelJetLES}
\end{figure}

\begin{figure}[H]
  \centering
  \includegraphics[width=1\textwidth]{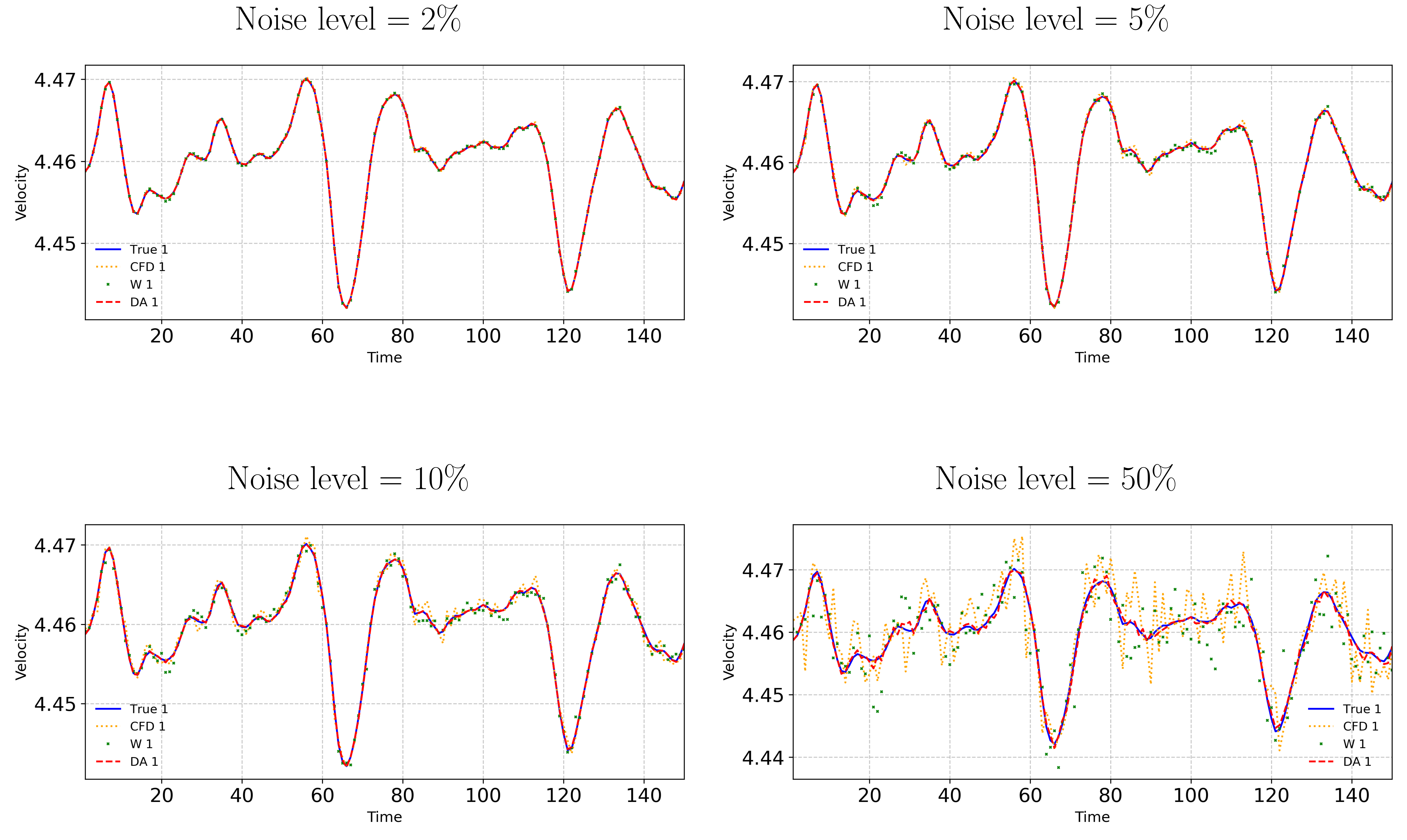}
  \caption{Velocity tracking point (2nd red dot from the left side in figures \ref{JetLESnoise30} and \ref{JetLESnoise50}) turbulent jet large eddy, with noise levels from 2\% to 50\%, with $C_{R,w}$ = 253. True data is shown in blue, $u_a$ in yellow, $w$ in green, and the assimilated data in red.}
  \label{2trackingVelJetLES}
\end{figure}

\begin{figure}[H]
  \centering
  \includegraphics[width=1\textwidth]{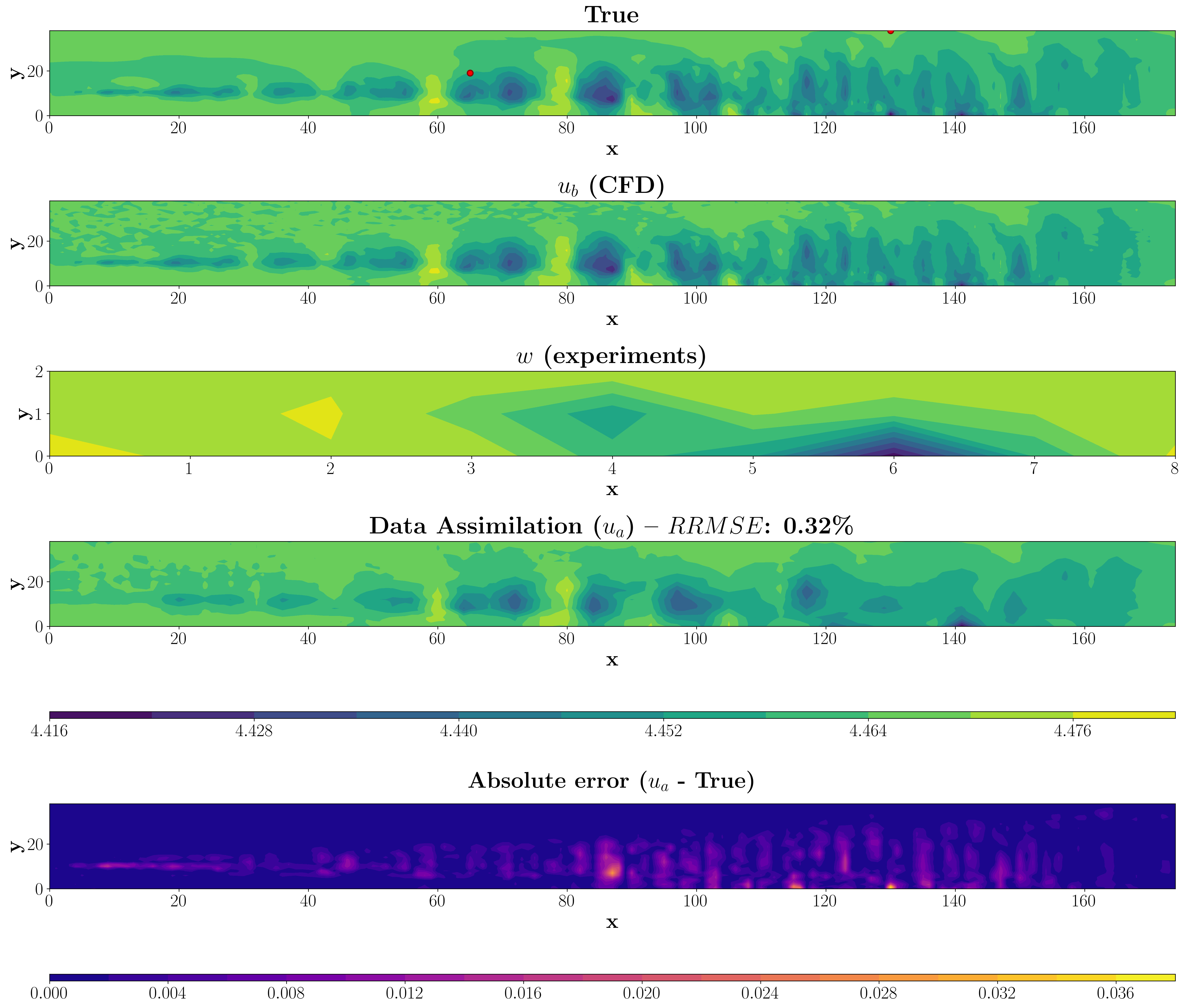}
  \caption{Case 45 in table \ref{tab:dims}. Turbulent jet large eddy results with 5\% of noise level, $C_{R,u_b} = 16$, with lcSVD reconstruction method. $N_{s, u_b} \times K = 440 \times 151$ ($C_{R,u_b} = 16$) and $N_{s, w} \times K = 27 \times 151$ ($C_{R,w} = 253$). From the top to the bottom: True data, CFD data, DA, experimental data, absolute error ($u_a - \text{True}$). The left column represents the streamwise velocity, and the right column represents the normal velocity.}
  \label{JetLESResNoise5}
\end{figure}

\begin{figure}[H]
  \centering
  \includegraphics[width=1\textwidth]{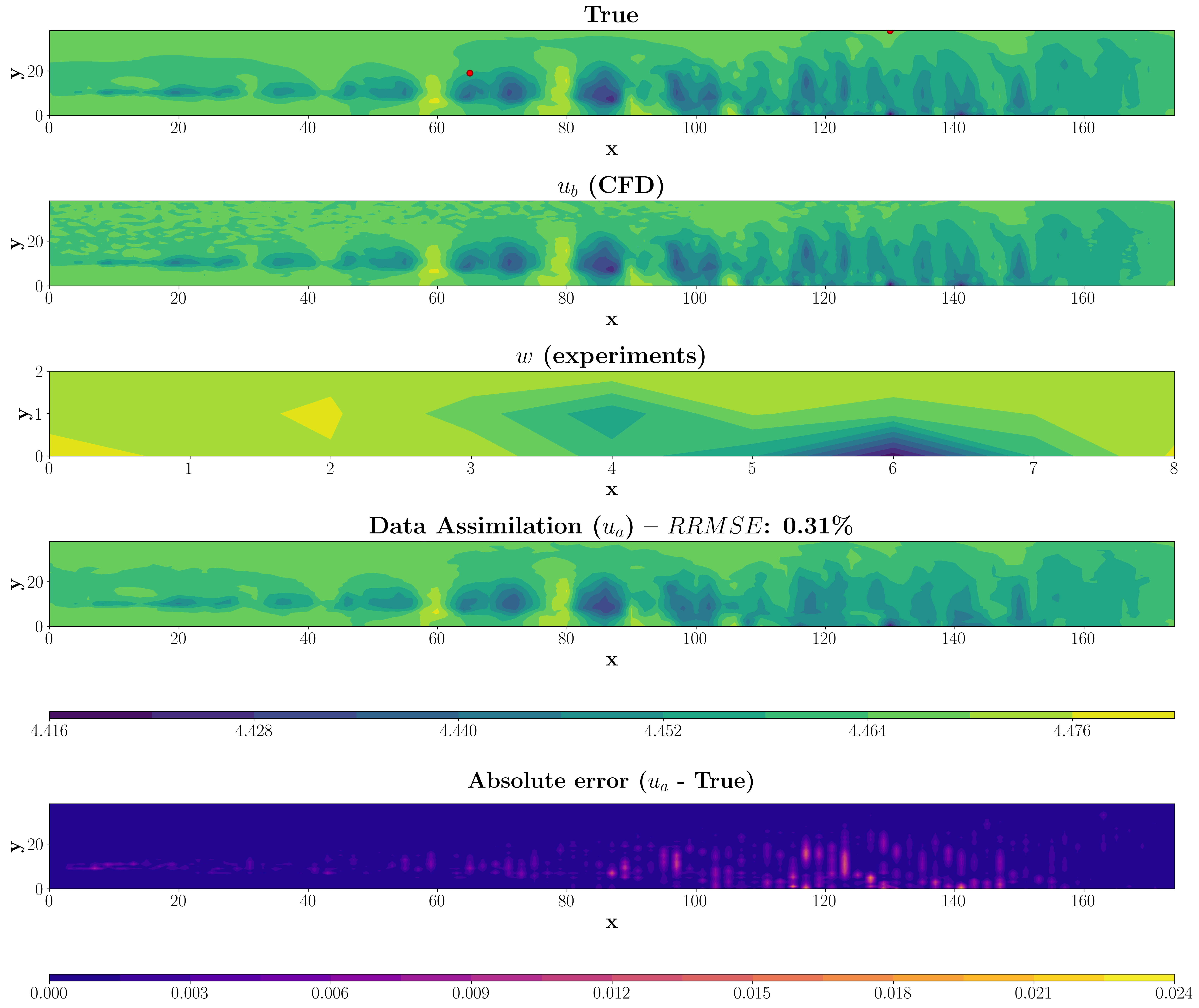}
  \caption{Case 46 in table \ref{tab:dims}. Turbulent jet large eddy results with 5\% of noise level, $C_{R,u_b} = 4$, with lcSVD reconstruction method. $N_{s, u_b} \times K = 1760 \times 151$ ($C_{R,u_b} = 16$) and $N_{s, w} \times K = 27 \times 151$ ($C_{R,w} = 253$). From the top to the bottom: True data, CFD data, DA, experimental data, absolute error ($u_a - \text{True}$). The left column represents the streamwise velocity, and the right column represents the normal velocity.}
  \label{JetLESResNoise5bis}
\end{figure}

Then, we compare the performance LR and HR methods, keeping in mind that this test case is influenced by a lower number of points compared to the other test cases. Figures \ref{JetLESResNoise5} and \ref{JetLESResNoise5bis} give a first graphical appreciation of the low-resolution results applied to turbulent jet large eddy test case.

The results presented in table \ref{tab:summarySVD_JetLES} provides a detailed comparison of the computational efficiency and accuracy. For the HR method, the computation time reaches 10.70 seconds, and the RAM usage 0.90 GB.

LR method improves computational efficiency, but this time not for every case. Among the most efficient cases, the $C_{R,u_b} = 84$ configuration achieves almost the same computation time than the HR method, 9.80 seconds ($S = 1.09$), with around 0.22 GB of RAM usage, standing out as an effective compromise between accuracy, computation time, and memory usage. With the $C_{R,u_b} = 4$ configuration, the computation time is reduced to 7.67 seconds, with a RAM usage of 0.28 GB, 0.62 GB less than the HR method.

LR method also leads to higher computation times for some configurations within this test case. For example, for $C_{R,u_b} = 107, 68, 16$, the computation times reach 64.12, 71.24, and 83.92 seconds respectively. Nevertheless, LR methods still maintain a satisfactory level of accuracy in this test case. For the HR method, $MAE$ and $RRMSE$ are $0.15$\%, $0.14$\%, $0.13$\% and 0.35\%, 0.35\%, 0.32\% respectively, for a noise level of 5\%. Table \ref{tab:summarySVD_JetLES} indicates that with lower downsampling configurations, the LR methods continue to provide low error metrics: the $C_{R,u_b} = 4$ configuration yields an $MAE$ of $0.12$\% and an $RRMSE$ of 0.31\%. Remarkably, the $RRMSE$ values for all cases are lower than those obtained using the HR method, highlighting the ability of LR techniques to enhance noisy, low-fidelity data. This improvement is largely due to lcSVD’s ability to extract and reconstruct only the most coherent flow structures. In the context of this jet turbulent case, where small-scale structures are numerous but often dominated by noise or less relevant dynamics, lcSVD naturally filters out these fine scales by retaining only a low number of dominant modes. As a result, the method reduces reconstruction error by focusing on the main features that show the essential flow information.

At 50\% noise level and for the highest compression rate ($C_{R,u_b} = 190$), the method achieves a $MAE$ of 1.43\% and a $RRMSE$ of 3.46\%. At $C_{R,u_b} = 68$, the $MAE$ drop to 1.39\% and the $RRMSE$ is 3.22\%. Finally, for the lowest compression rate ($C_{R,u_b} = 4$), the $MAE$ decreases to 1.23\% and the $RRMSE$ to 2.91\%. When comparing these values to the results at 5\% noise level, we observe that they remain within the same range as those obtained at 50\% noise for both LR and HR approaches. Moreover, as seen for the 5\% noise level cases applied to the JetLES, the LR method further decreases the $RRMSE$ at 50\% noise level compared to the HR method. These results further confirm that, for this low-fidelity turbulent dataset, lcSVD effectively reduces the error across various compression rates, demonstrating its robustness, reliability, and ability to enhance the reconstruction thanks to its noise-filtering capabilities.

Given the characteristics of this test case, namely the relatively low number of computational points and the presence of moderate velocity gradients, the LR approach demonstrates satisfactory performance in terms of memory compression, computational speed-up, and accuracy. Notably, even under these turbulent conditions where traditional HR simulations are computationally affordable, the LR method achieves substantial gains in efficiency achieving a similar or better quality of the results than for the HR. This highlights the robustness and broad applicability of the LR technique: it not only performs well in complex, high-resolution scenarios but also remains effective and reliable in regimes where the need for compression is less critical. Furthermore, the observed ability of LR to filter out noise while maintaining high accuracy at strong compression rates underscores its potential as a versatile and efficient tool for fluid flow simulations.

\begin{table*}[h]
  \centering
  \adjustbox{valign=c, max width=\textwidth}{
    \begin{tabular}{|c|c|c|c|c|c|c|c|c|c|c|c|}
      \hline
      \multicolumn{1}{|c|}{Noise (\%)} &
      \multicolumn{1}{|c|}{Case} &
      \multicolumn{1}{c|}{$C_{R,u_b}$} &
      \multicolumn{1}{c|}{$C_{R,w}$} &
      \multicolumn{1}{c|}{$N_{s, u_b} \times K$} &
      \multicolumn{1}{c|}{$N_{s, w} \times K$} &
      \multicolumn{1}{c|}{$t_{comp}$ (s)} &
      \multicolumn{1}{c|}{$S$} &
      \multicolumn{1}{c|}{$MAE$ (\%)} &
      \multicolumn{1}{c|}{$RRMSE$ (\%)} &
      \multicolumn{1}{c|}{RAM (GB)} &
      \multicolumn{1}{c|}{$R_{\text{comp}} (\%)$} \\ \hline
      2 & 37 & 1 & 253 & 1030575   & 4077   & 10.70  & 1 & $0.05$ & 0.16 & 0.90 & -\\ \hline
      \multirow{6}{*}{5} 
      &38& 1     & 253   & 1030575 & 4077   & 10.70    & 1  & $0.12$ & 0.41   & 0.90 & -\\
         &39& 190   & 253   & $5436$   & $4077$   & 35.35   & 0.30  & $0.15$ & 0.37 & 0.19 & 78.89\\
         &40& 107   & 253   & $9664$   & $4077$   & 64.72   & 0.17 & $0.15$  & 0.36 & 0.21 & 76.67\\
         &41& 84   & 253   & $12231$   & $4077$   & 9.80  & 1.09  & $0.15$ & 0.35 & 0.22 & 75.56\\ 
         &42& 68    & 253    & $15100$  & $4077$   & 71.24   & 0.15 & $0.14$  & 0.35 & 0.22 & 75.32\\
         &43& 38    & 253   & 27180  & 4077   & 71.77   & 0.15 & $0.14$ & 0.34 & 0.22 & 75.18\\
         &44& 31    & 253   & 33220  & 4077   & 76.31   & 0.14  & $0.13$   & 0.34 & 0.22 & 75.03\\ 
         &45& 16    & 253   & 66440  & 4077   & 83.92   & 0.13  & $0.13$ & 0.32   & 0.23 & 74.44\\ 
         &46& 4     & 253   & 265760 & 4077   & 7.67    & 1.40  & $0.12$ & 0.31   & 0.28 & 68.89\\\hline
      10 &47& 1  & 253   & 1030575   & 4077   & 10.70 & 1  & $0.24$ & 0.81 & 0.90 & -\\ \hline
      30 &48& 1  & 253   & 1030575   & 4077   & 10.70 & 1  & $0.73$  & 2.44 & 0.90 & -\\ \hline
      \multirow{5}{*}{50}
      &49& 1  & 253   & 1030575   & 4077   & 10.70 & 1 & $1.21$  & 4.07  & 0.90 & -\\ 
      &50& 190 & 253   & $5436$   & $4077$ & 35.35 & 0.30 & $1.43$  & 3.46  & 0.19 & 78.89\\
      &51& 68 & 253   & $15100$   & $4077$ & 71.24 & 0.15 & $1.39$ & 3.22  & 0.22 & 75.32\\
      &52& 16    & 253 & $66440$   & $4077$& 83.92 & 0.13 & $1.30$ & 3.13  & 0.23 & 74.44\\
      &53& 4     & 253   & 265760 & 4077   & 7.67  & 1.40  & $1.23$ & 2.91   & 0.28 & 68.89\\ 
      \hline
    \end{tabular}
  }
  \caption{Merged results for turbulent jet large eddy, combining HR and LR (lcSVD) methods. From left to right: noise rate, case, data compression rates $C_{R,u_b}$ and $C_{R,w}$, shape of the data collected by the sensors selecting $N_{s,u_b}$ equidistant data points of $u_b$ multiplied by $K$ snapshots, shape of the data collected by the sensors selecting $N_{s,w}$ equidistant data points of $w$ multiplied by $K$ snapshots, computation time, speed-up, $MAE$, $RRMSE$, RAM usage, and RAM compression rate $R_{\text{comp}}$ (\%).}
  \label{tab:summarySVD_JetLES}
\end{table*}

\begin{figure}[H]
  \centering
  \includegraphics[width=1\textwidth]{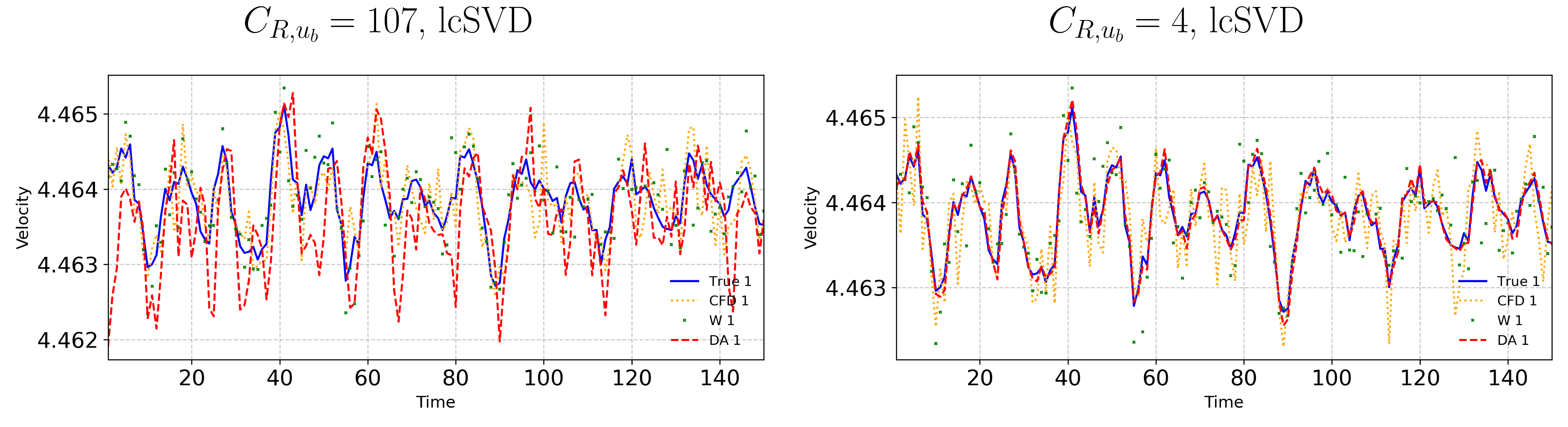}
  \caption{Velocity tracking point (1st red dot from the left side in figures \ref{JetLESResNoise5} and \ref{JetLESResNoise5bis}) turbulent jet large eddy, with 5\% of noise level, for $C_{R,u_b}$ = 107 and $C_{R,u_b}$ = 4. True data is shown in blue, $u_a$ in yellow, $w$ in green, and the assimilated data in red.}
  \label{1trackingVelJetLESLR}
\end{figure}

\begin{figure}[H]
  \centering
  \includegraphics[width=1\textwidth]{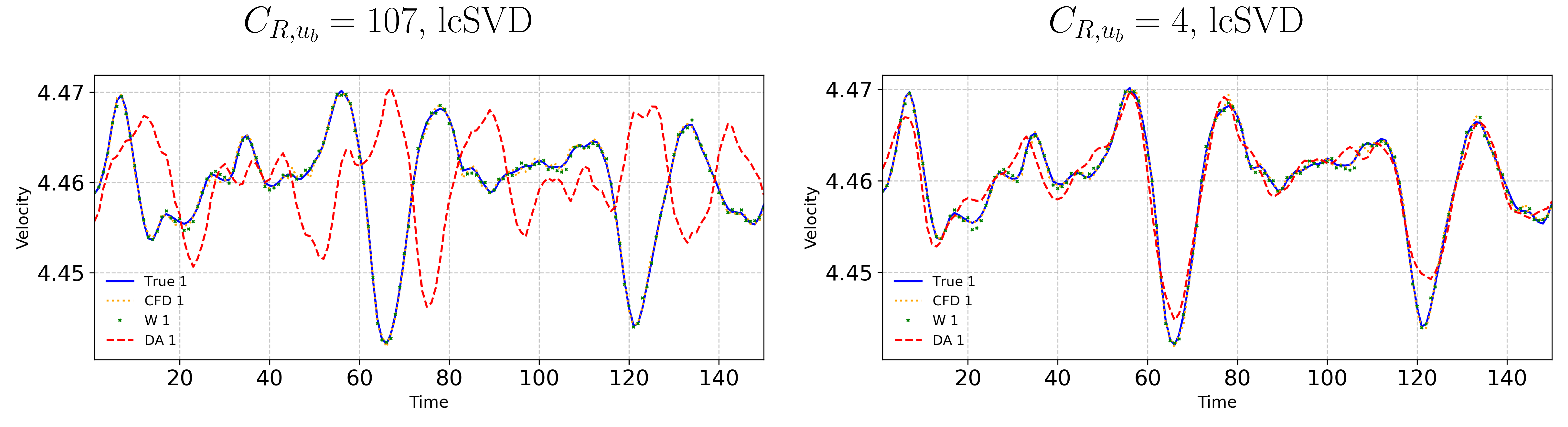}
  \caption{Velocity tracking point (2nd red dot from the left side in figures \ref{JetLESResNoise5} and \ref{JetLESResNoise5bis}) turbulent jet large eddy, with 5\% of noise level, for $C_{R,u_b}$ = 107 and $C_{R,u_b}$ = 4. True data is shown in blue, $u_a$ in yellow, $w$ in green, and the assimilated data in red.}
  \label{2trackingVelJetLESLR}
\end{figure}

Figures \ref{1trackingVelJetLESLR} and \ref{2trackingVelJetLESLR} demonstrate the influence of $C_{R,u_b}$, on the accuracy of True state estimation. For $C_{R,u_b} = 107$, the DA estimate (\( u_b \), red curve) shows a significant deviation from the True state (\( u_a \), blue curve), indicating a relatively poor estimation, especially when compared to the HR method. However, with an increased number of tracking points, $C_{R,u_b} = 4$, the DA estimate aligns much more closely with the True state, achieving an accuracy comparable to the HR method, but couldn't achieve a strong estimation of the figure \ref{2trackingVelJetLESLR}, where some differences can be observed close around the peaks mainly. 

The previous results show how lcSVD reconstruction method provides similar results for $RRMSE$ with a lower RAM memory usage, but not always a decreased computation time.

\section{Conclusion}\label{conclusion}
\small

In this paper, we explored a Reduced-Order Model (ROM) approach designed to improve the estimation of fluid flow states by combining simulation and experimental data through Data Assimilation (DA) techniques. Our method builds on the Ensemble Kalman Filter (EnKF), adapted here to operate efficiently within a lower-dimensional framework. By working with downsampled data and applying a tailored reconstruction strategy using low-cost Singular Value Decomposition (lcSVD), we were able to cut down both computational time and memory usage significantly while keeping high accuracy. 

Through extensive testing, this DA approach demonstrated consistent and reliable performance across a variety of datasets and noise levels. At a 50\% noise level, the Relative Root Mean Square Error ($RRMSE$) was observed to be 7.84\% for the circular cylinder dataset at \( Re = 100 \), 7.59\% for the turbulent circular cylinder at \( Re = 2600 \), and 4.07\% for the turbulent jet LES dataset, highlighting the model's robustness in managing significant noise. As the noise level decreased, the model's accuracy improved notably. At a noise level of 5\%, the $RRMSE$ values reduced to 0.97\%, 0.83\%, and 0.41\%, respectively, for the three test cases, demonstrating high accuracy. These results confirm the capability of the method to maintain performance across varying levels of data noise, making it suitable for diverse fluid dynamics applications where data uncertainty is prevalent.

A critical element of this ROM approach is the data compression achieved through strategic downsampling. By optimizing the number of data points within each dataset, we effectively reduced the database size without significantly impacting the accuracy of the model. The compression ratio was significant, while the reconstruction errors remained low, thus confirming the reliability of the employed compression techniques in accurately reconstructing the data. This compression not only facilitates faster processing but also enables the application of DA techniques to larger datasets and more complex flow fields. In terms of computation, reductions in processing times were achieved with a speed-up ($S$) up to 53.20, 21.30, and 1.40 for the circular cylinder at \( Re = 100 \), the turbulent circular cylinder at \( Re = 2600 \), and the turbulent jet LES, respectively, when compared to high-resolution simulations. This substantial reduction in computation time makes it feasible to perform real-time DA in fluid dynamics applications. Furthermore, memory usage was optimized, with reductions from over 297.33 GB in high-resolution configurations to 0.85 GB at $C_{R,u_b}$ = 49.8 for the circular cylinder at \( Re = 100 \), from 37.35 GB to 3.40 GB at $C_{R,u_b}$ = 15.9 for the turbulent circular cylinder at \( Re = 2600 \), and from 0.90 GB to 0.22 GB at $C_{R,u_b}$ = 68 for the turbulent jet LES, demonstrating the ROM's potential for use in computational environments with limited resources, within an acceptable level of accuracy.

Additionally, when comparing the $RRMSE$ across various flow scenarios, it becomes evident that the methods exhibit different performance levels at varying compression rates. The ROM showed promising estimations at higher compression rates, leading to a noticeable reduction in $RRMSE$, particularly under high noise conditions. 

For instance, for high compression rates, $C_{R,u_b}$ = 17413, 8353, and 190, the $RRMSE$ are 51.10\%, 40.93\% and 0.37\% at 5\% of noise, and 51.31\%, 41.71\%, and 3.46\% at 50\% of noise respectively for the 3 test cases. These cases are achieving low computation time and RAM needs, but the accuracy compared to the HR method is not sufficient. At low compressed cases, $C_{R,u_b}$ = 3.10, 3.98, and 4, we obtain an $RRMSE$ of 1.04\%, 1.14\% and 0.31\% at 5\% of noise, and 8.84\%, 6.66\%, and 2.91\% for 50\% of noise respectively for the 3 test cases. 

These results also demonstrate the capability of lcSVD to reconstruct very noisy data (50\%) close to the HR computations, and in some cases, approaching the performance obtained with only 5\% noise for both high and low compression rates. This robustness is attributed to the noise filtering capabilities of the lcSVD reconstruction, which enables accurate state estimation in low and high noise-contaminated simple and complex flow scenarios. In particular, for the JetLES case, the lcSVD-based LR reconstruction achieves $RRMSE$ values even lower than those obtained with the HR method under equivalent noise levels. This result is explained by the nature of the JetLES turbulent flow, which contains a wide range of flow structures across scales, including numerous small-scale features. lcSVD exploits this by retaining only a limited number of dominant modes, effectively filtering out the small-scale turbulent fluctuations that are often contaminated by noise. As a result, the reconstruction not only compresses the data but also enhances it, yielding more accurate estimates than those produced by the unfiltered high-resolution solution.

In summary, this ROM framework with DA presents a robust and practical solution for fluid dynamics simulations. By balancing computational efficiency, data compression, and accuracy, it offers new opportunities for DA in large-scale and complex fluid dynamics scenarios. The flexibility in downsampling configurations makes this approach suitable for a variety of applications, including environmental monitoring and engineering design, where real-time or resource-efficient simulations are crucial. Future work may focus on exploring optimal sensor placement for data reconstruction methods and the development of parallel computing implementations to further reduce computation time, enabling the processing of even larger datasets and expanding the utility of the framework in complex fluid dynamics applications.

This ROM framework demonstrates considerable promise, particularly for physics-based applications such as Computational Fluid Dynamics (CFD). Its ability to balance computational efficiency with accuracy makes it especially suitable for scenarios that require real-time or resource-efficient simulations. One such application is urban CFD, where the ROM can enhance the modeling of complex fluid interactions in densely built environments. This framework will be incorporated into the next version of ModelFLOWs-app \cite{Modelflow}, facilitating practical implementations in fluid dynamics, alongside tools for data post-processing, pattern identification, and the development of reduced-order models using modal decomposition and machine learning techniques. By integrating this ROM into future CFD workflows, the field can progress toward more efficient and scalable solutions for large-scale, real-world problems.

\section{Acknowledgements}

The MODELAIR and ENCODING project has received funding from the European Union’s Horizon Europe research and innovation programme under the Marie Sklodowska-Curie grant agreement No. 101072559 and 101072779, respectively. The results of this publication reflect only the author(s) view and do not necessarily reflect those of the European Union. The European Union can not be held responsible for them. The authors gratefully acknowledge the Universidad Politécnica de Madrid (\url{https://www.upm.es/}) for providing computing resources on Magerit Supercomputer.

The work of SEA is supported by the U.S. Department of Energy (DOE) Office of Advanced Scientific Computing Research (ASCR) through the Pacific Northwest National Laboratory Distinguished Fellowship in Scientific Computing (Project No. 71268). Pacific Northwest National Laboratory is operated by Battelle Memorial Institute for DOE under Contract DE-AC05-76RL01830.


\newpage
\begin{thebibliography}{9}

\bibitem{boundariesCFD}
Oberkampf, W. L., \& Trucano, T. G. (2002). Verification and validation in computational fluid dynamics. \textit{Progress in Aerospace Sciences}, 38(3), 209-272.

\bibitem{expCFD}
Gargiulo, A., Duetsch-Patel, J. E., Borgoltz, A., Devenport, W. J., Roy, C. J., \& Lowe, K. T. (2023). Strategies for computational fluid dynamics validation experiments. \textit{Journal of Verification, Validation and Uncertainty Quantification}, 8(3), 031004. \url{https://doi.org/10.1115/1.4063639}

\bibitem{boundaryAircraft}
Wu, X., Zhang, W., \& Song, S. (2017). Uncertainty quantification and sensitivity analysis of transonic aerodynamics with geometric uncertainty. \textit{International Journal of Aerospace Engineering}, 2017(1), 8107190. \url{https://doi.org/10.1155/2017/8107190}

\bibitem{boundaryComb}
Barzegar, R., Shafee, S., \& Khalilarya, S. (2013). Computational fluid dynamics simulation of the combustion process, emission formation and the flow field in an in-direct injection diesel engine. \textit{Thermal Science}, 17(1), 11-23. \url{https://doi.org/10.2298/TSCI111218108B}

\bibitem{boundariesWind}
Khan, M., Odemark, Y., \& Fransson, J. H. M. (2017). Effects of inflow conditions on wind turbine performance and near wake structure. \textit{Open Journal of Fluid Dynamics}, 7(1), 105-129. \url{https://doi.org/10.4236/ojfd.2017.71008}

\bibitem{boundariesUrban1}
Blocken, B. (2015). Computational fluid dynamics for urban physics: Importance, scales, possibilities, limitations and ten tips and tricks towards accurate and reliable simulations. \textit{Building and Environment}, 91, 219-245. \url{https://doi.org/10.1016/j.buildenv.2015.02.015}

\bibitem{boundariesUrban2}
Parente, A., Gorlé, C., van Beeck, J., \& Benocci, C. (2011). A comprehensive modelling approach for the neutral atmospheric boundary layer: Consistent inflow conditions, wall function and turbulence model. \textit{Boundary-Layer Meteorology}, 140(3), 411-428. \url{https://doi.org/10.1007/s10546-011-9621-5}

\bibitem{boundariesUrban3}
García-Sánchez, C., Vitalis, S., Paden, I., \& Stoter, J. (2021). The impact of level of detail in 3D city models for CFD-based wind flow simulations. \textit{The International Archives of the Photogrammetry, Remote Sensing and Spatial Information Sciences}, 46, 67-72. \url{https://doi.org/10.5194/isprs-archives-XLVI-4-W4-2021-67-2021}

\bibitem{DA5}
Meldi, M., \& Poux, A. (2017). A reduced order model based on Kalman filtering for sequential data assimilation of turbulent flows. \textit{Journal of Computational Physics}, 347, 207-234. \url{https://doi.org/10.1016/j.jcp.2017.06.042}

\bibitem{DAlargeScale}
Moldovan, G., Lehnasch, G., Cordier, L., \& Meldi, M. (2021). A multigrid/ensemble Kalman filter strategy for assimilation of unsteady flows. \textit{Journal of Computational Physics}, 443, 110481. \url{https://doi.org/10.1016/j.jcp.2021.110481}

\bibitem{DA_CFD_LES}
Aristodemou, E., Arcucci, R., Mottet, L., Robins, A., Pain, C., \& Guo, Y.-K. (2019). Enhancing CFD-LES air pollution prediction accuracy using data assimilation. \textit{Building and Environment}, 165, 106383. \url{https://doi.org/10.1016/j.buildenv.2019.106383}

\bibitem{DA}
Gillijns, S., Barrero Mendoza, O., Chandrasekar, J., De Moor, B., Bernstein, D. S., \& Ridley, A. (2006). What is the ensemble Kalman filter and how well does it work? In \textit{Proceedings of the 2006 American Control Conference} (pp. 6-pp). IEEE. \url{https://doi.org/10.1109/ACC.2006.1657419}

\bibitem{DA1}
Nino-Ruiz, E. D., Sandu, A., \& Deng, X. (2018). An ensemble Kalman filter implementation based on modified Cholesky decomposition for inverse covariance matrix estimation. \textit{SIAM Journal on Scientific Computing}, 40(2), A867-A886. \url{https://doi.org/10.1137/16M1097031}

\bibitem{DA2}
Mons, V., Chassaing, J.-C., Gomez, T., \& Sagaut, P. (2016). Reconstruction of unsteady viscous flows using data assimilation schemes. \textit{Journal of Computational Physics}, 316, 255-280. \url{https://doi.org/10.1016/j.jcp.2016.04.022}

\bibitem{DA3}
Colburn, C. H., Cessna, J. B., \& Bewley, T. R. (2011). State estimation in wall-bounded flow systems. Part 3. The ensemble Kalman filter. \textit{Journal of Fluid Mechanics}, 682, 289-303. \url{https://doi.org/10.1017/jfm.2011.222}

\bibitem{DA4}
Villanueva, L., Valero, M. M., Glumac, A. Š., \& Meldi, M. (2024). Augmented state estimation of urban settings using on-the-fly sequential data assimilation. \textit{Computers \& Fluids}, 269, 106118. \url{https://doi.org/10.1016/j.compfluid.2023.106118}

\bibitem{DA6}
Chandramouli, P., Mémin, E., \& Heitz, D. (2020). 4D large scale variational data assimilation of a turbulent flow with a dynamics error model. \textit{Journal of Computational Physics}, 412, 109446. \url{https://doi.org/10.1016/j.jcp.2020.109446}

\bibitem{CFDcost}
Brodtkorb, A. R., \& Holm, H. H. (2019). Real-world oceanographic simulations on the GPU using a two-dimensional finite-volume scheme. \textit{arXiv preprint arXiv:1912.02457}. \url{https://doi.org/10.48550/arXiv.1912.02457}

\bibitem{CFDcost2}
Mateevitsi, V. A., Bode, M., Ferrier, N., Fischer, P., Göbbert, J. H., Insley, J. A., Lan, Y.-H., Min, M., Papka, M. E., Patel, S., Rizzi, S., \& Windgassen, J. (2023). Scaling computational fluid dynamics: In situ visualization of NekRS using SENSEI. In \textit{Proceedings of the SC '23 Workshops of The International Conference on High Performance Computing, Network, Storage, and Analysis} (pp. 862-867). \url{https://doi.org/10.1145/3624062.3624159}

\bibitem{DAcost}
Majda, A. J., \& Tong, X. T. (2018). Performance of ensemble Kalman filters in large dimensions. \textit{Communications on Pure and Applied Mathematics}, 71(5), 892-937. \url{https://doi.org/10.1002/cpa.21722}

\bibitem{ashton2020low}
Hetherington, A., \& Le Clainche, S. (2023). Low-cost singular value decomposition with optimal sensor placement. \textit{arXiv preprint arXiv:2311.09791}. \url{https://doi.org/10.48550/arXiv.2311.09791}

\bibitem{Sirovich87}
Sirovich, L. (1987). Turbulence and the dynamics of coherent structures. I. Coherent structures. \textit{Quarterly of Applied Mathematics}, 45(3), 561-571. \url{https://doi.org/10.1090/qam/910462}

\bibitem{MODELFLOW}
Hetherington, A., Corrochano, A., Abadía-Heredia, R., Lazpita, E., Muñoz, E., Díaz, P., Corrochano, S., García-Vallejo, D., Ferrer, E., \& Le Clainche, S. (2024). ModelFLOWs-app: Data-driven post-processing and reduced order modelling tools. \textit{Computer Physics Communications}, 301, 109217. \url{https://doi.org/10.1016/j.cpc.2024.109217}

\bibitem{shady}
Ahmed, S. E., Pawar, S., \& San, O. (2020). PyDA: A hands-on introduction to dynamical data assimilation with Python. \textit{Fluids}, 5(4), 225. \url{https://doi.org/10.3390/fluids5040225}

\bibitem{lewis}
Lewis, J. M., Lakshmivarahan, S., \& Dhall, S. (2006). \textit{Dynamic data assimilation: A least squares approach} (Vol. 13). Cambridge University Press.

\bibitem{evensen}
Evensen, G. (1994). Sequential data assimilation with a nonlinear quasi-geostrophic model using Monte Carlo methods to forecast error statistics. \textit{Journal of Geophysical Research: Oceans}, 99(C5), 10143-10162. \url{https://doi.org/10.1029/94JC00572}

\bibitem{burgers}
Burgers, G., van Leeuwen, P. J., \& Evensen, G. (1998). Analysis scheme in the ensemble Kalman filter. \textit{Monthly Weather Review}, 126(6), 1719-1724. \url{https://doi.org/10.1175/1520-0493(1998)126%3C1719:ASITEK%3E2.0.CO;2}

\bibitem{evensen2}
Evensen, G. (2003). The ensemble Kalman filter: Theoretical formulation and practical implementation. \textit{Ocean Dynamics}, 53, 343-367. \url{https://doi.org/10.1007/s10236-003-0036-9}

\bibitem{Lumley}
Lumley, J. L. (1967). The structure of inhomogeneous turbulent flows. \textit{Atmospheric turbulence and radio wave propagation}, 166-178. \url{https://cir.nii.ac.jp/crid/1574231874542771712}

\bibitem{LeClaincheetalJFM20}
Le Clainche, S., Izbassarov, D., Rosti, M., Brandt, L., \& Tammisola, O. (2020). Coherent structures in the turbulent channel flow of an elastoviscoplastic fluid. \textit{Journal of Fluid Mechanics}, 888, A5. \url{https://doi.org/10.1017/jfm.2020.31}

\bibitem{LeClaincheVega17}
Le Clainche, S., \& Vega, J. M. (2017). Higher order dynamic mode decomposition. \textit{SIAM Journal on Applied Dynamical Systems}, 16(2), 882-925. \url{https://doi.org/10.1137/15M1054924}

\bibitem{PCA}
Parente, A., \& Sutherland, J. C. (2013). Principal component analysis of turbulent combustion data: Data pre-processing and manifold sensitivity. \textit{Combustion and Flame}, 160(2), 340-350. \url{https://doi.org/10.1016/j.combustflame.2012.09.016}

\bibitem{Lupod}
Rapún, M. L., Terragni, F., \& Vega, J. M. (2017). LUPOD: Collocation in POD via LU decomposition. \textit{Journal of Computational Physics}, 335, 1-20. \url{https://doi.org/10.1016/j.jcp.2017.01.005}

\bibitem{turbulentDA}
Deng, Z., He, C., Wen, X., \& Liu, Y. (2018). Recovering turbulent flow field from local quantity measurement: Turbulence modeling using ensemble-Kalman-filter-based data assimilation. \textit{Journal of Visualization}, 21, 1043-1063. \url{https://doi.org/10.1007/s12650-018-0508-0}

\bibitem{vK vortex}
Jackson, C. P. (1987). A finite-element study of the onset of vortex shedding in flow past variously shaped bodies. \textit{Journal of Fluid Mechanics}, 182, 23-45. \url{https://doi.org/10.1017/S0022112087002234}

\bibitem{3dim floquet}
Barkley, D., \& Henderson, R. D. (1996). Three-dimensional Floquet stability analysis of the wake of a circular cylinder. \textit{Journal of Fluid Mechanics}, 322, 215-241. \url{https://doi.org/10.1017/S0022112096002777}

\bibitem{nek5000}
Nek5000. (2023). \textit{Nek5000: Open source spectral element CFD solver}. \url{https://nek5000.mcs.anl.gov/}

\bibitem{VKIcylinder}
Mendez, M. A., Hess, D., Watz, B. B., \& Buchlin, J.-M. (2020). Multiscale proper orthogonal decomposition (mPOD) of TR-PIV data—A case study on stationary and transient cylinder wake flows. \textit{Measurement Science and Technology}, 31(9), 094014. \url{https://doi.org/10.1088/1361-6501/ab82be}

\bibitem{JetLES}
Brès, G. A., Jordan, P., Jaunet, V., Le Rallic, M., Cavalieri, A. V., Towne, A., Lele, S. K., Colonius, T., \& Schmidt, O. T. (2018). Importance of the nozzle-exit boundary-layer state in subsonic turbulent jets. \textit{Journal of Fluid Mechanics}, 851, 83-124. \url{https://doi.org/10.1017/jfm.2018.476}

\bibitem{mdpi2020}
Jain, N., Bravo, L., Kim, D., Murugan, M., Ghoshal, A., Ham, F., \& Flatau, A. (2020). Massively parallel large eddy simulation of rotating turbomachinery for variable speed gas turbine engine operation. \textit{Energies}, 13(3), 703. \url{https://doi.org/10.3390/en13030703}

\bibitem{CharLES}
Brès, G. A., Bose, S. T., Emory, M., Ham, F. E., Schmidt, O. T., Rigas, G., \& Colonius, T. (2018). Large-eddy simulations of co-annular turbulent jet using a Voronoi-based mesh generation framework. In \textit{2018 AIAA/CEAS Aeroacoustics Conference} (p. 3302). \url{https://doi.org/10.2514/6.2018-3302}

\bibitem{Vreman}
Vreman, A. W. (2004). An eddy-viscosity subgrid-scale model for turbulent shear flow: Algebraic theory and applications. \textit{Physics of Fluids}, 16(10), 3670-3681. \url{https://doi.org/10.1063/1.1785131}

\bibitem{BCLES}
Freund, J. B. (1997). Proposed inflow/outflow boundary condition for direct computation of aerodynamic sound. \textit{AIAA Journal}, 35(4), 740-742. \url{https://doi.org/10.2514/2.167}

\bibitem{Sponge_layers}
Mani, A. (2012). Analysis and optimization of numerical sponge layers as a nonreflective boundary treatment. \textit{Journal of Computational Physics}, 231(2), 704-716. \url{https://doi.org/10.1016/j.jcp.2011.10.017}

\bibitem{JetExperiments}
Zaman, K. B. M. Q. (1998). Asymptotic spreading rate of initially compressible jets—Experiment and analysis. \textit{Physics of Fluids}, 10(10), 2652-2660. \url{https://doi.org/10.1063/1.869778}

\bibitem{JetNozzles}
Zaman, K. B. M. Q. (1999). Spreading characteristics of compressible jets from nozzles of various geometries. \textit{Journal of Fluid Mechanics}, 383, 197-228. \url{https://doi.org/10.1017/S0022112099003833}

\bibitem{Modelflow}
ModelFLOWs Research Group. (2023). \textit{ModelFLOWs-app}. \url{https://modelflows.github.io/modelflowsapp/}

\end{thebibliography}
\end{document}